\documentclass[journal,draftclsnofoot,onecolumn,12pt,twoside]{IEEEtran}

\ifCLASSINFOpdf

\else

\fi

\usepackage{url}
\usepackage{amsmath}
\usepackage{amsthm}
\usepackage{enumitem}

\newtheorem{lemma}{Lemma}
\newtheorem{remark}{Remark}

\newtheorem{auxiliary code}{Auxiliary Code}

\newtheorem{definition}{Definition}

\usepackage{algorithm}
\usepackage[noend]{algpseudocode}

\algrenewcommand\algorithmicindent{1em}%

\usepackage{tikz}
\usepackage{graphicx}
\usetikzlibrary{positioning}
\usepackage[font={small}]{caption}
\usepackage{subcaption}
\usepackage{array}
\usepackage{multirow}
\usepackage{setspace}%
\usepackage{optidef}
\usepackage[scientific-notation=true]{siunitx}
\usepackage{amsbsy}
\usepackage[noadjust]{cite}

\def \talpha {\alpha_{\ttg}}

\def \tA {\mathrm{A}}

\def \tCMT {\mathcal{T}}

\def \tT {\mathcal{T} }
\def \tsymbol {\tau}
\def \tproof {{\fontfamily{qcr}\selectfont Proof}}
\def \troot {{\fontfamily{qcr}\selectfont Root}}
\def \tverify {{\fontfamily{qcr}\selectfont Verify-Inclusion}}

\def \talpha {\alpha_{\min}}

\def \tP {\mathbf{F}}
\def \ttT {\mathbf{T}}
\def \tFG {\mathcal{G}}
\def \tpFG {\widehat{\mathcal{G}}}

\def \tI {\mathcal{A}}
\def \tF {\mathcal{F}}

\def \ttN {\overline{C}}

\def \tSS {\Psi}
\def \tssingle {\psi}
\def \tssingletree {ST}
\def \tLS {{\fontfamily{qcr}\selectfont Leaf-Set}}

\def \tlast {\delta}
\def \tlastVN {\mathcal{V}}

\def \tttN {C}
\def \tttC {\text{{\fontfamily{qcr}\selectfont CodeSym}}}
\def \tttS {\text{{\fontfamily{qcr}\selectfont data}}}
\def \tttP {\text{{\fontfamily{qcr}\selectfont parity}}}
\def \tparity {{\fontfamily{qcr}\selectfont encodeParity}}

\def \tparent {{\fontfamily{qcr}\selectfont formParentIn}}

\def \tdecodelayer {\text{{\fontfamily{qcr}\selectfont decodeLayer}}}

\def \tdropped {\text{{\fontfamily{qcr}\selectfont dropped}}}

\def \tfrozen {\text{{\fontfamily{qcr}\selectfont frozen}}}

\def \tdropindex {\text{{\fontfamily{qcr}\selectfont dI}}}

\def \ttotVN {\text{{\fontfamily{qcr}\selectfont TVN}}}

\def \tIm {\widehat{\mathcal{A}}}
\def \tFm {\widehat{\mathcal{F}}} 
\def \tNm {\widehat{N}}  
 
\def \tNmm {{N}}  
 
\def \tImm {{\mathcal{A}}}
\def \tFmm {{\mathcal{F}}} 

\def \tNoracle {\theta}
\def \tbetaoracle {\beta}

\def \tkoracle {g}
\def \tgammaoracle {\gamma}
\def \tCdispersal {\mathcal{C}}
\def \tAdispersal {\mathcal{A}}
\def \tmu {\mu}
\def \tteta {\eta}
\def \tcomm {\text{{\fontfamily{qcr}\selectfont CommCost}}}
\def \tVNi {\lambda}
\def \tCNi {z}

\def \tfrozenprune {\text{{\fontfamily{qcr}\selectfont pruneFrozenVN}}}
\def \tremovedegreoneCN {\text{{\fontfamily{qcr}\selectfont pruneDeg1CN}}}
\def \tmergedegretwoCN {\text{{\fontfamily{qcr}\selectfont mergeDeg2CN}}}
\def \tremoveemptyCN {\text{{\fontfamily{qcr}\selectfont pruneEmptyCN}}}
\def \tptotVN
{\text{{\fontfamily{qcr}\selectfont totVN}}}

\def \tname {SEF}

\def\closed#1#2#3{#1 \in [#2,#3]}
\def\openleft#1#2#3{#1 \in (#2,#3]}
\def\closedN#1#2{#1 \in [#2]}

\newcolumntype{L}{>{\centering\arraybackslash}m{4.5cm}}    
\newcolumntype{M}{>{\centering\arraybackslash}m{5cm}} 
\newcolumntype{N}{>{\centering\arraybackslash}m{5.5cm}} 
\newcolumntype{O}{>{\centering\arraybackslash}m{5.5cm}}

\usepackage{anyfontsize}

\usepackage{multicol}
\usepackage[T1]{fontenc}

\newcommand\deb[1]{{\color{black}#1}}
\newcommand\debb[1]{{\color{black}#1}}
\newcommand\redtext[1]{{\color{black}#1}}
\newcommand\greentext[1]{{\color{black}#1}}
\newcommand\browntext[1]{{\color{black}#1}}
\newcommand\bluetext[1]{{\color{black}#1}}

\usepackage{accents}
\def\symaccent#1{\widetilde{#1}}

\urldef{\bitcoinsize}\url{https://www.blockchain.com/charts/avg-block-size}

\urldef{\bitcoincash}\url{https://bitinfocharts.com/comparison/bitcoin%20cash-size.html#3y}
	
	\urldef{\bitcoinsv}\url{https://bitinfocharts.com/comparison/bitcoin%20sv-size.html#3y}

\usepackage{multirow}

\begin{document}

\title{Polar Coded Merkle Tree: Mitigating Data Availability Attacks in Blockchain Systems Using Informed Polar Code Design \vspace{-0.2cm} }

\author{Debarnab~Mitra,~\IEEEmembership{Student Member,~IEEE,}
Lev~Tauz,~\IEEEmembership{Student Member,~IEEE}
        and~Lara~Dolecek,~\IEEEmembership{Senior Member,~IEEE}%
\thanks{D.~Mitra, L.~Tauz, and L.~Dolecek  are with the ECE Department, UCLA, Los Angeles, CA 90095 USA (e-mail: debarnabucla@ucla.edu, levtauz@ucla.edu, dolecek@ee.ucla.edu). A part of this paper was presented at the IEEE International Symposium of Information Theory 2022 \cite{PCMT}.}%
\vspace{-1.2cm}}

\maketitle

\vspace{-1cm}
\begin{abstract}
\vspace{-0.3cm}
\emph{Data availability (DA) attack} is a well-known problem in certain blockchains where users accept an invalid block with unavailable portions. Previous works have used LDPC and 2-D Reed Solomon (2D-RS) codes with Merkle trees to mitigate DA attacks. These codes perform well across various metrics such as DA detection probability and communication cost. However, \redtext{these codes} are difficult to apply to blockchains with large blocks due to large decoding complexity and coding fraud proof size (2D-RS codes), and intractable code guarantees for large code lengths (LDPC codes). In this paper, we focus on large block size applications and address the above challenges by proposing the novel \emph{Polar Coded Merkle Tree} (PCMT): a Merkle tree encoded using the encoding graph of polar codes. We provide a specialized polar code design algorithm called \emph{Sampling Efficient Freezing} and an algorithm to prune the polar encoding graph. We demonstrate that the PCMT built using the above techniques results in a better DA detection probability and communication cost compared to LDPC codes, has a lower coding fraud proof size compared to LDPC and 2D-RS codes, provides tractable code guarantees at large code lengths \deb{(similar to 2D-RS codes)}, and has comparable decoding complexity to 2D-RS and LDPC codes. 

\end{abstract}

 \vspace{-0.6cm}
\begin{IEEEkeywords}
\vspace{-0.35cm}
Blockchain Systems, Data Availability Attacks, Polar codes, Coded Merkle Tree
\end{IEEEkeywords}

\IEEEpeerreviewmaketitle

\vspace{-0.6cm}
\section{Introduction}
\vspace{-0.2cm}
Blockchain is a tamper-proof ledger of transaction blocks connected together by hashes in the form of a chain. The ledger is maintained by a network of nodes in a decentralized manner. The decentralization and security properties of blockchains have led to their application in diverse fields such as cryptocurrencies, medical services, supply chains, copyright protection, and Internet of Things \cite{ECCSurvey}.
However, blockchains provide the security properties at the expense of poor performance in terms of storage overhead of nodes and transaction throughput. For instance, full nodes in blockchains such as Bitcoin and Ethereum are required to store the entire blockchain ledger whose sizes at the time of writing  are around 450GB \cite{bitcoinsize} and 1000GB \cite{ethereumsize}, respectively. The significant storage  overhead for full nodes prevents nodes with limited resources to join the blockchain network, which leads to the centralization of the blockchain \cite{DynamicDistStorage}. At the same time, Bitcoin and Ethereum have a poor throughput of only a few transactions per second \cite{LowThroughput}, which is significantly lower than well-known payment processing system like VISA, with a throughput of thousands of transactions per second. The design of high-performance blockchains with low storage overhead and high transaction throughput without sacrificing the security properties has been a major research area in recent years \deb{\cite{ECCSurvey}.} 

\begin{figure}[t]
    \centering
\begin{subfigure}{0.38\linewidth}
\begin{minipage}{0.99\linewidth}
\begin{tikzpicture}
  \node (img) {\includegraphics[scale=0.37]{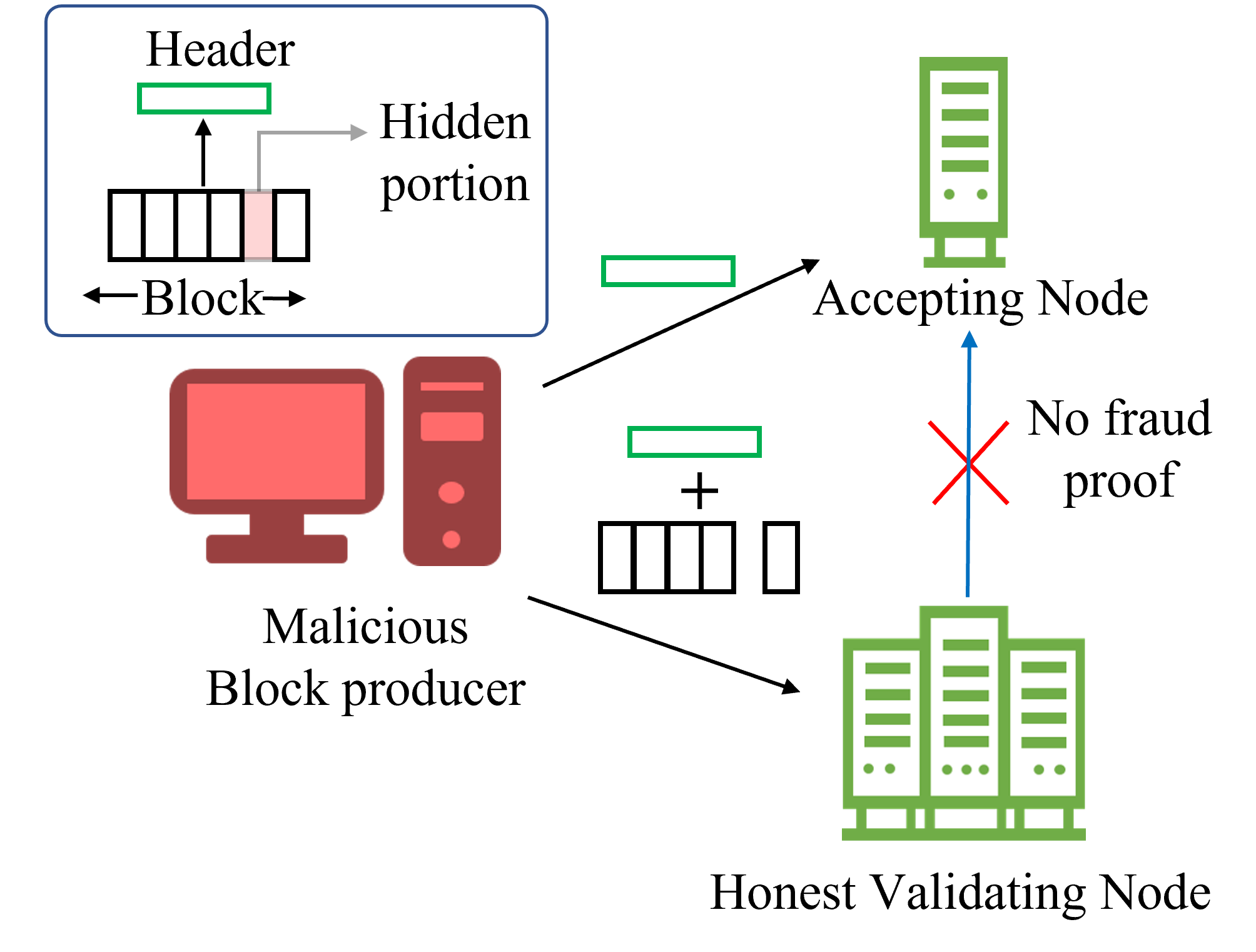}};
 \end{tikzpicture}
 \end{minipage}
\end{subfigure}%
    \begin{subfigure}{0.3\linewidth}
\begin{minipage}{0.99\linewidth}
\begin{tikzpicture}
  \node (img) {\includegraphics[scale=0.37]{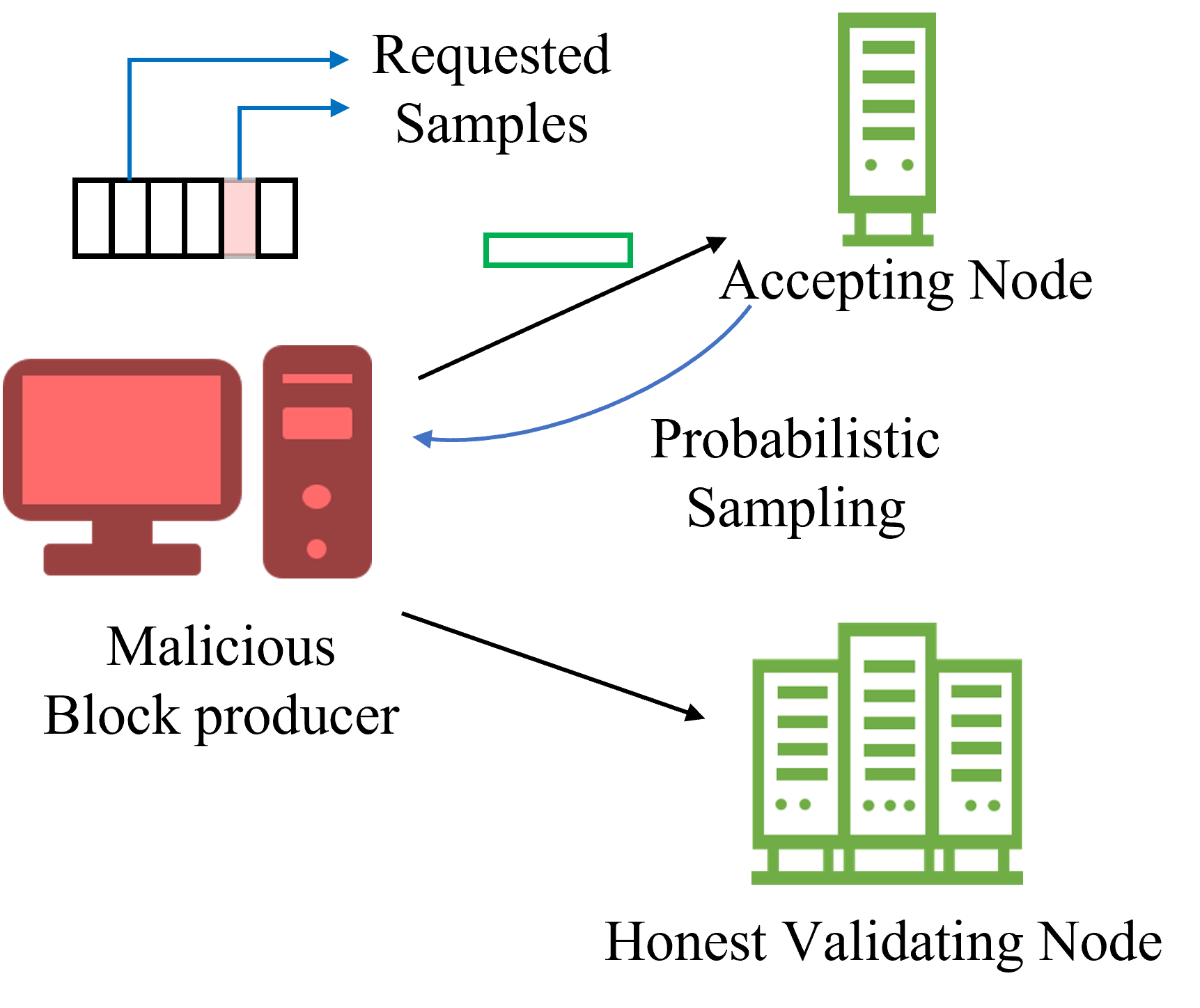}};
 \end{tikzpicture}
 \end{minipage}
    \end{subfigure}%
\begin{subfigure}{0.33\linewidth}
\begin{minipage}{0.99\linewidth}
\begin{tikzpicture}
  \node (img) {\includegraphics[scale=0.37]{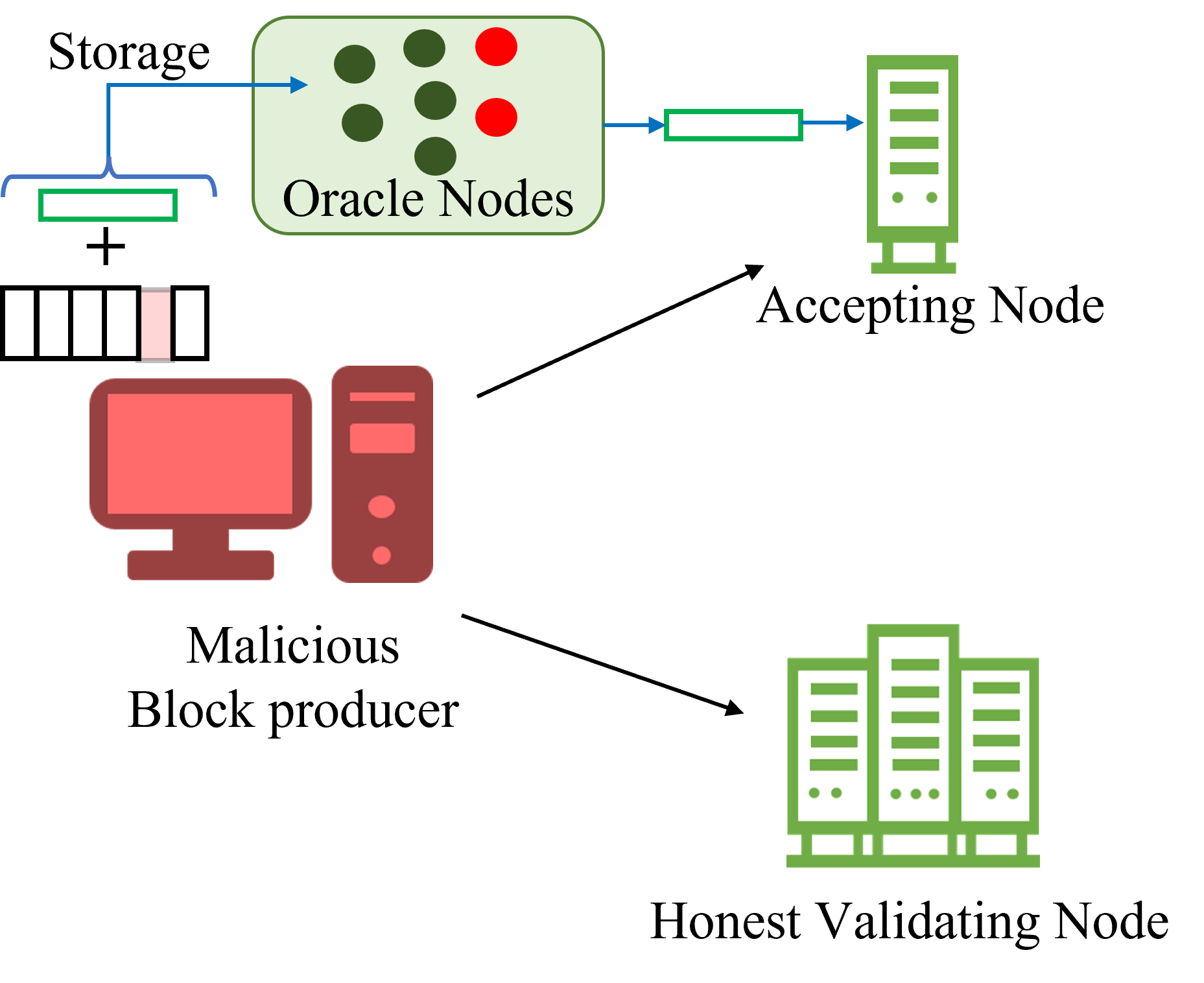}};
 \end{tikzpicture}
 \end{minipage}
\end{subfigure}
    \vspace{-14pt}
    \caption{Left panel: Data availability (DA) attack; Middle panel: Detection of DA attack using probabilistic sampling; Right panel: Preventing DA attacks by storing the block at certain intermediary oracle nodes.}
    \label{fig:DA_attack}
\end{figure}

\debb{A popular approach for improving the storage and throughput performance of blockchains is by allowing certain nodes called \emph{accepting nodes} to not store or validate the blockchain blocks. Instead, these nodes only store a small fraction of each block (called its \emph{header}) and rely on verifiable \emph{fraud-proofs} \cite{dataAvailOrg}  sent out by other nodes called \emph{validating nodes} (that store/validate the full blocks) to reject invalid blocks. Examples for the above model include blockchains with \emph{light nodes} \cite{dataAvailOrg} and \emph{side blockchains} \cite{AceD} that respectively improve the storage and throughput of blockchains. However,  only storing the header and relying on fraud proofs to reject invalid blocks make the accepting nodes vulnerable to \emph{data availability (DA) attacks} \cite{dataAvailOrg}, \cite{CMT} when the majority of validating nodes are malicious. In this attack, as illustrated in Fig. \ref{fig:DA_attack} left panel, a malicious block producer generates a block with invalid transactions, publishes its header to the accepting nodes, and hides the invalid portion of the block from the validating nodes. The validating nodes cannot validate the missing
portion of the block and are unable to generate fraud proofs. In the absence of fraud proofs, the invalid block is accepted by the accepting nodes. 
}

\debb{\redtext{A DA attack can be modeled as an adversarial erasure channel where the malicious node decides the positions in the transaction block to inject erasures into. Channel coding has been proposed in literature to mitigate DA attacks \cite{dataAvailOrg,CMT,SSskewITW,TCOMLDPC,AceD,DE-PEG}. 
In particular, the transaction block is encoded using an erasure code and the erasure coded} symbols of the block are either i) probabilistically sampled by each accepting node, i.e., they randomly request different coded symbols  and reject the block if a certain requested symbol is not returned by the block producer \cite{dataAvailOrg, CMT, SSskewITW, TCOMLDPC}; or ii) stored at certain intermediate nodes (called \emph{oracle nodes} in \cite{AceD}) to provably ensure that the original block can be decoded back from the symbols stored at the oracle nodes even if a fraction of them are malicious \cite{AceD},\cite{DE-PEG}. Important performance metrics in the above cases (which we call \emph{system specific metrics}) are the \emph{probability of failure} of detection of the hidden block portions in the case of probabilistic sampling and the \emph{communication costs} in the case of storage at oracle nodes. Encoding the blocks using erasure codes helps improve the system specific metrics in both scenarios.
In particular, the improvement in the system specific metrics depends on the  \emph{undecodable threshold} of the code which is defined as the minimum number of coded symbols the malicious block producer must hide to prevent validating nodes from generating fraud proofs. 
Despite improved system specific performance, erasure coding still allows the malicious block producer to carry out an \emph{incorrect-coding (IC) attack} where it incorrectly encodes the block such that the original block cannot be decoded back by the validating nodes. In this case, honest validating nodes can broadcast \emph{IC proofs} that allow accepting nodes to reject the header \cite{dataAvailOrg,CMT}. The IC proof size is another important metric that depends on the choice of the code. For linear codes, the IC proof size is  proportional to the degree of the parity check equations \bluetext{of the channel code used for encoding the transaction block}. }

\debb{ Previous works have used 2D Reed-Solomon (RS) \cite{dataAvailOrg, RSoptimize} and Low-Density Parity-Check (LDPC) \cite{CMT,SSskewITW,TCOMLDPC, AceD, DE-PEG} codes to encode the transaction block. However, as with classical applications of channel coding, each choice of code comes with its own set of trade-offs (see Section \ref{sec:prev_work}).
\greentext{Polar codes \cite{PolarCodesErikan}
have seen an enormous  success in the past decade including the pioneering work of Vardy and co-authors \cite{PolarCodesVardy,VaryListDecoding,VardyDeletion}. \browntext{As we demonstrate in this paper, these foundational results have far-reaching consequences} even in emerging technologies such as blockchains. In particular, we show a non-trivial application of  polar codes to mitigate DA attacks and demonstrate that it offers improved trade-offs in various performance metrics relevant to this application.}
}

\debb{The application of channel coding has to be carefully considered depending on the size of the transaction blocks in blockchains. They can range from a few MBs (small block size), e.g., Bitcoin \cite{BitcoinSize}, Bitcoin Cash \cite{BitcoinCash}, to hundreds of MBs (large block size), e.g., Bitcoin SV \cite{BitcoinSV}. Large blocksize applications require large code lengths since large code lengths allow for smaller partitioning of the block, thereby reducing the load on the network
bandwidth. In the context of DA attacks, authors in \cite{CMT} used random LDPC codes for large code lengths. However, as pointed out in \cite{TCOMLDPC}, random LDPC codes undermine the security of the system due to having a non-negligible probability of generating bad codes. \debb{At the same time, works in \cite{SSskewITW}, \cite{TCOMLDPC}, \cite{DE-PEG}   provide deterministic LDPC codes for short code lengths that result in good performance of the system specific metrics.
However, the undecodable threshold for deterministic LDPC codes is NP-hard to determine \cite{SSNP-hard} making it difficult to extend the techniques of \cite{SSskewITW}, \cite{TCOMLDPC}, \cite{DE-PEG} to large code lengths. Due to the above limitations of prior literature, we consider the problem of designing channel codes at large code lengths to mitigate DA attacks in large blocksize applications. An important performance metric at large code lengths is the \emph{threshold complexity} of the code which is defined as the complexity of finding the undecodable threshold. The threshold complexity affects the system design complexity in blockchains, and hence, \bluetext{needs to be small.}  } }

\debb{In general, for a given code, the following metrics are of importance at large code lengths: i)  undecodable threshold, ii) IC proof
size, iii) threshold complexity, and iv) decoding complexity. To result in good performance of the system specific metrics, the undecodable threshold must be large. The IC-proof size must be small since this proof is communicated to all nodes and can be used to congest the blockchain network.  As mentioned before, the threshold complexity must be small. Finally, the decoding complexity should also be small for high throughput blockchains.}

\vspace{-0.6cm}
\subsection{Contributions}
\vspace{-0.2cm}
\debb{In order to mitigate DA attacks while ensuring the integrity of the information communicated by different system entities,
instead of only encoding the transaction block, 
a cryptographic data structure called a \emph{Coded Merkle Tree (CMT)} (\bluetext{introduced in \cite{CMT} using sparse parity checks}) is used. A CMT is a Merkle tree \cite{Bitcoin} where each
layer of the tree is encoded using the choice of channel code used in the system.}
\greentext{Our contributions in this paper are listed as follows:}
\vspace{-0.1cm}
\begin{enumerate}[wide, labelwidth=!, labelindent=10pt]
    \item %
    We propose the \emph{Polar Coded Merkle Tree (PCMT)}, a CMT that is built using the encoding graph of polar codes. Although polar codes have dense parity check equations \cite{polarhighdensity}, they have sparse encoding graphs. Thus, we propose a novel technique for building a CMT using the encoding graph of polar codes and demonstrate that it results in small IC proof sizes. The IC proof size for the PCMT is around 30-60\% smaller compared to a CMT which uses LDPC codes. Note that the earlier CMT construction provided in \cite{CMT} uses the parity check matrix of a
    code for its construction which is unlike the PCMT that is built using the encoding graph.
    \item %
    We provide a specialized polar code design algorithm for the PCMT called \emph{Sampling}
    \emph{Efficient Freezing (SEF)}. The SEF algorithm has the following properties: i) it results in polar codes that have large undecodable thresholds and hence good performance of the system specific metrics, ii) it allows flexibility in designing polar codes of any length instead of just power of 2, and
    iii) polar codes designed using the SEF algorithm have an efficiently computable analytical expression for the undecodable threshold. Thus, \deb{the SEF algorithm} results in a very low threshold complexity, simplifying system design at large code lengths. SEF polar codes result in half an order of magnitude reduction in the probability of failure and around 8-10\% reduction
    in the communication cost compared to a CMT built using LDPC codes.
    Additionally, property iii) allows the derivation of the scaling law for the probability of failure for SEF polar codes. We show that SEF polar codes 
    \bluetext{have exponentially better probability of failure, with a factor $\Omega(\sqrt{K})$, 
    where $K$ is the information length, compared to the method that does not involve channel coding. }   
    \vspace{-0.7cm}
    \item %
    \debb{We provide a pruning algorithm that reduces the size of the polar encoding graph without changing its undecodable threshold.} Pruning helps further improve the performance of the system specific metrics by around 5-10\%, IC-proof size by around 3-10\%, and decoding complexity by more than 50\% compared to SEF polar codes. 
    \item We provide an extensive performance comparison of a PCMT and its pruned version with LDPC and 2D-RS codes to demonstrate the advantages of the techniques proposed in this paper. 
\end{enumerate}

\vspace{-0.6cm}
\subsection{Previous Work}\label{sec:prev_work}
\vspace{-0.15cm}

The first application of channel codes \bluetext{for overcoming DA attacks} was in \cite{dataAvailOrg} where 2D-RS codes were utilized.  
The techniques of \cite{dataAvailOrg} were recently optimized in \cite{RSoptimize}. \deb{Due to their algebraic constructions, 2D-RS codes provide large undecodable thresholds that can be easily calculated.}
However, 2D-RS codes
result in large IC-proof sizes and decoding complexity \cite{CMT,TCOMLDPC}. 
The above limitations of 2D-RS codes were overcome in \cite{CMT} where the authors proposed the CMT. In the CMT in \cite{CMT}, each layer is encoded using an LDPC code. 
\debb{The sparse parity check equations in LDPC codes result in small IC proofs. At the same time, LDPC codes also
allow the use of a low complexity peeling decoder \cite{ModernCodingTheory} for decoding each CMT layer.}
\debb{For LDPC codes, the undecodable threshold is the minimum stopping set size \cite{ModernCodingTheory} of the LDPC codes.
 } 

Authors in \cite{CMT} used codes from a random LDPC ensemble to construct the CMT. However, 
as pointed out in \cite{TCOMLDPC}, random LDPC codes undermine the security of the system due to a non-negligible probability of generating bad codes. In \cite{SSskewITW, TCOMLDPC, DE-PEG}, authors proposed specialized LDPC codes  for the CMT to mitigate DA attacks based on the PEG algorithm \cite{PEG} and demonstrate good system specific performance. 
\bluetext{However, the works in \cite{SSskewITW, TCOMLDPC,DE-PEG} were designed for short code lengths. The NP-hardness of computing the minimum stopping set size of LDPC codes \cite{SSNP-hard} makes it difficult to extend the techniques of \cite{SSskewITW, TCOMLDPC,DE-PEG} to large code lengths, which is the focus of this paper.}
Recently, \cite{InformationDispersal} proposed a technique to mitigate DA attacks without requiring IC proofs.  However, \cite{InformationDispersal} requires complex cryptographic computations at each block generation, which is infeasible in blockchains where nodes have low compute abilities, such as Proof of Stake \cite{SnowWhite} or Proof of Space \cite{PoSpace} blockchains. 

While we focus on mitigating DA attacks in this paper, channel coding has been extensively used to mitigate various other scalability issues in blockchain systems. For example,  \cite{downsampling} 
reduces the storage cost at full nodes using erasure coding and downsampling; 
\cite{CodedCompBlockchain} utilizes the concepts of coded computation, information dispersal, state machine replication, and two-dimensional sharding to reduce the storage, communication costs in sharded blockchains. We refer the reader to \cite{ECCSurvey} for
an extensive survey on  works that utilize channel coding for scaling blockchain systems. 

The rest of this paper is organized as follows. In Section \ref{sec:prelims}, we provide the preliminaries and the system model. In Section \ref{sec:pcmt}, we provide our novel construction of the Polar Coded Merkle Tree (PCMT). In Section \ref{sec:SEF}, we provide the SEF algorithm to design polar codes for the PCMT. In Section \ref{sec:pruned_construction}, we provide the pruning algorithm to reduce the size of the factor graphs used in the PCMT. We provide
simulation results in Section \ref{sec:sims} and concluding remarks in Section \ref{sec:conclusion}.

\vspace{-0.45cm}
\section{Preliminaries and System Model}\label{sec:prelims}
\vspace{-0.2cm}
We use the following notation for the rest of this paper. 
For a vector $\mathbf{a}$, let
$\mathbf{a}(i)$ denote the
$i$th element of $\mathbf{a}$ and let $\min(\mathbf{a};k)$ denote the $k$th smallest value of $\mathbf{a}$.  Let $Z^{\otimes n}$ denote the $n$th
Kronecker power of matrix/vector $Z$. 
Let $\vert S \vert$ be the cardinality of set $S$. Unless specified otherwise, all logarithms are with base 2 in this paper. Let $\tP_2 = \begin{bmatrix}
1 & 0 \\
1 & 1 
\end{bmatrix}$
and $\ttT_{N} = \begin{bmatrix}
1 \\
2 
\end{bmatrix}^{\bigotimes \lceil \log N\rceil}$, for positive integer $N$.
For integers $a$ and $b$ define $[a,b] = \{i \;\vert\; a \leq i \leq b\}$, $(a,b] = \{i \;\vert\; a < i \leq b\}$, and $[a] = \{i \;\vert\; 1 \leq i \leq a\}$, where elements in the three sets are integers. Let $ (x)_p := x \bmod {p}$. 
Also let ${\fontfamily{qcr}\selectfont \text{Hash}}$ and ${\fontfamily{qcr}\selectfont \text{concat}}$ represent the hash and string concatenation functions, respectively. \greentext{For functions $f$ and $g$, $f = \Omega(g)$ means $ \exists$ $n_0$ and a constant $e > 0$ such that
for all $n > n_0$, $eg(n) \leq  |f(n)|$. In other words, $f$ grows at least as fast as $g$.}

\vspace{-0.6cm}
\subsection{Coded Merkle Tree (CMT) Preliminaries}\label{sec:general-CMT-framework}
\vspace{-0.2cm}
Like a Merkle tree \cite{Bitcoin}, a CMT is a cryptographic commitment generator that is built from 
the transactions present in the block and is used to check the integrity of the transactions \cite{CMT}. Additionally, each CMT layer is encoded using an erasure code which allows us to mitigate DA attacks that may occur in any CMT layer.  In this section, we first provide a general framework for the CMT construction that captures its key properties. Later in Sections \ref{sec:pcmt} and \ref{sec:pruned_construction}, we present the construction of the PCMT and its pruned version within the general CMT framework.

A CMT is parameterized by $\tCMT = (K, R, q, l)$, where $K$ is the number of information symbols obtained by partitioning the transaction block, $R$ is the rate of the code used to encode each layer of the CMT, $q$ is a combining parameter which determines how many hashes are combined to form information symbols of the intermediate layers, and $l$ is the number of layers (excluding the CMT root). At a high level, the CMT is constructed as follows:  the transaction block is first partitioned into $K$ data symbols and a rate $R$ systematic channel code is applied to generate $N_l$ coded symbols. These $N_l$ coded symbols form the base layer of the CMT. The $N_l$
coded symbols are then hashed and the hashes of these $N_l$ coded symbols are combined to get  data symbols
of the parent layer. The data symbols of this layer are again coded using a rate $R$ systematic code and the coded symbols are further
hashed and combined to get data symbols of its parent layer. This iterative process is continued until we get 
$l$ layers. The hashes of the coded symbols in the final layer form the CMT root.  \debb{The CMT root is part of the header of each transaction block.}

\begin{figure}
    \centering
    \includegraphics[scale=0.33]{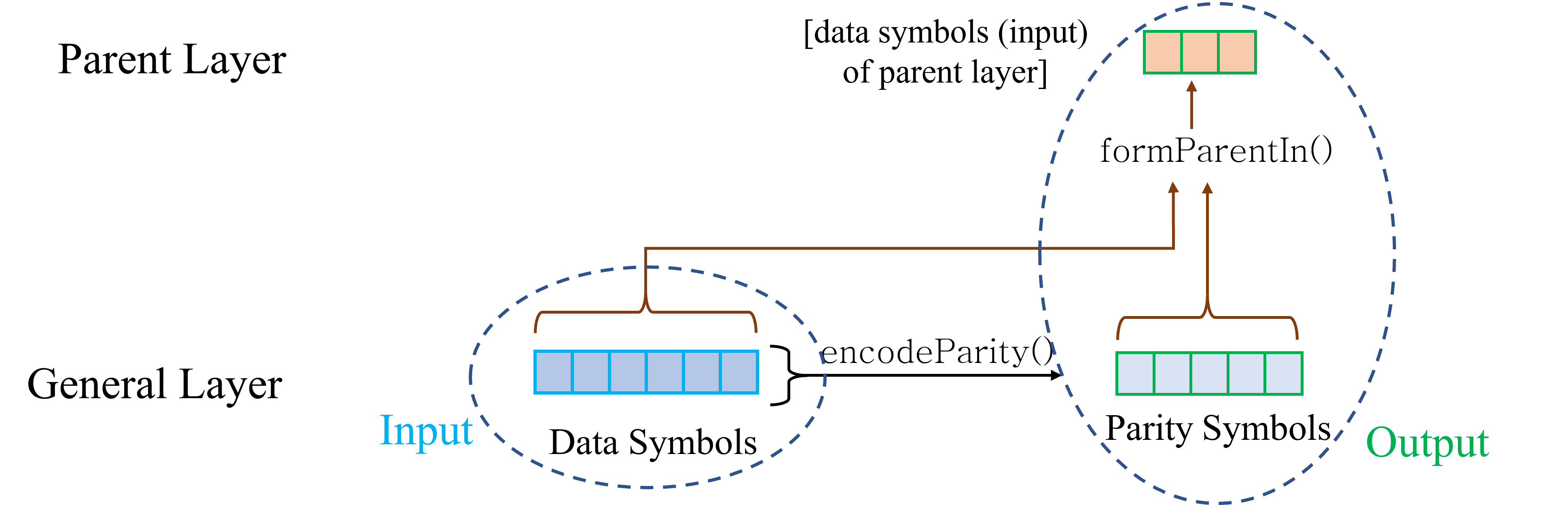}
    \vspace{-14pt}
    \caption{Recursive construction of the CMT. The input to the general layer is the data symbols. The outputs are the parity symbols (that become part of the general layer) and the data symbols of the parent layer. }
    \label{fig:recursiveCMT}
\end{figure}

Let the layers of the CMT be $L_0, L_1, \ldots, L_l$ where $L_l$ is the base layer and $L_0$ is the CMT root. For $j= 1,2,\ldots l $, let the code length and information length of $L_j$ be $N_j$ and $k_j$, respectively, where $k_j = \frac{K}{(qR)^{l-j}}$. Note that in prior work \deb{\cite{CMT, TCOMLDPC, AceD}}, $N_j = \frac{k_j}{R}$ but it need not be in general. \redtext{As explained above, the formation of the CMT is an iterative process.
Thus, throughout this paper, without loss of generality, we only consider layers $L_j$ and $L_{j-1}$ of the CMT which we call the \emph{general layer} and the \emph{parent layer}, respectively, and drop the subscript $j$ for the layer index \greentext{when the context is clear} (an illustration is shown in Fig. \ref{fig:recursiveCMT}).  
} 
The general layer gets as input $k$ data symbols. The outputs for this layer are $N$ coded symbols (that constitute the general layer) and $\symaccent{k}$ data symbols that form input to the parent layer.
All variables defined henceforth do not contain subscript $j$ \greentext{(unless necessary)} and are assumed to belong to the general or parent layer, with variables having a tilde accent belonging to the parent layer (otherwise the general layer). 
Note that for the defined variables, we bring back subscript $j$ for the layer index whenever necessary. 
Let $\tttN[i]$ be the $i$th coded symbol of the general layer\footnote{Thus, $\tttN_j[i]$ is the $i$th coded symbol of $L_j$ according to our notation.} and let $\tttC = \{\tttN[i] \;\vert\; \closedN{i}{N}\}$ be the set of all coded symbols of the general layer, where $\tttS = \{\tttN[i] \; \vert\; \closedN{i}{k}\}$ and $\tttP = \{\tttN[i] \;\vert\; \openleft{i}{k}{N}\}$ are the set of data and parity symbols of the general layer, respectively. 
 The input to the general layer is $\tttS$ and the outputs are $\tttP$ and $\symaccent{\tttS} = \{\symaccent{\tttN}[i] \; \vert\; \closedN{i}{\symaccent{k}}\}$. Outputs are formed as follows in the general CMT framework: 1) Form the parity symbols $\tttP$ from the data symbols $\tttS$ using a rate $R$ systematic linear code  via a procedure $\tttP = \text{\tparity}(\tttS)$; 2) Form the data symbols $\symaccent{\tttS}$ from the coded symbols $\tttC$ by a procedure  $\symaccent{\tttS} = \text{\tparent}(\tttC)$. Note that for initialization, when the general layer is the base layer, $\tttS$ is set to the $K$ partitions of the transaction block. 
 Generally in the \tparent$()$ procedure, hashes of a certain number of symbols of $\tttC$ are concatenated into  each data symbol $\symaccent{\tttN}[i]$ in $\symaccent{\tttS}$. The number of symbols  concatenated into each data symbol
 depends on the parameter $q$. After forming $l$ layers, the hashes of all the symbols in the final layer forms the root {\troot } of the CMT.

\vspace{-0.2cm}
Merkle proofs are an important construct that helps generate verifiable proofs that a malicious entity has altered the value of symbols, encoding rule, etc. Every coded symbol $\tsymbol$ in the CMT has a Merkle proof \tproof($\tsymbol$) that can be used to check the integrity of $\tsymbol$ with respect to the CMT root {\troot } using a procedure \tverify($\tsymbol$,  \tproof($\tsymbol$), \troot). The CMT is decoded using a hash-aware decoder (for example the hash-aware peeling decoder in \cite{CMT}) that decodes the CMT sequentially from layer $L_0$ to $L_l$. Due to the top-down decoding, when the decoder decodes the general layer, it already has the hashes of the symbols of this layer (provided by the parent layer) to compare against. Using sufficient symbols of each layer $L_j$, the hash-aware decoder decodes the layer using a procedure $\tdecodelayer(L_j)$. The hashes of all the decoded symbols are matched with their hashes provided in the parent layer. 

The hash-aware decoder via the above procedure of matching hashes allows to detect IC attacks and generate IC proofs. Let the decoded CMT symbols  $\tsymbol_1, \ldots, \tsymbol_{d}$ satisfy a degree $d$ parity check equation (of the erasure code used for encoding). From amongst these symbols, if there exists a symbol $\tsymbol_e$ whose hash does not match the hash provided by the parent of $\tsymbol_e$ in the CMT, an IC attack is detected. In this case, the IC proof consists of the following data: the CMT symbols $\{\tsymbol_1, \ldots, \tsymbol_{d}\} \setminus\tsymbol_e $ along with their Merkle proofs, and the Merkle proof of $\tsymbol_e$. The IC proof is verified using the following procedure: i) verify that each symbol $\tsymbol_i$, $\closedN{i}{d}, i \neq e$, satisfies \tverify($\tsymbol_i$,  \tproof($\tsymbol_i$), \troot), ii) decode ${\tsymbol}_e$ from the remaining symbols and check that  ${\tsymbol}_e$ does not satisfy \tverify(${\tsymbol}_e$,  \tproof(${\tsymbol}_e$), \troot).

 We now describe the method of \cite{CMT} of building the CMT from  the parity check matrix of any erasure code. 
For the CMT,
the $\text{\tparity}$() procedure is performed via systematic encoding using the parity check matrix.
The \tparent$()$ procedure is as follows:
\begin{align}\label{eqn:formparent-CMT}
     \symaccent{\tttN}[i] =  &{\fontfamily{qcr}\selectfont \text{concat}}(\{ {\fontfamily{qcr}\selectfont \text{Hash}}(\tttN[x]) \:\vert \:\closedN{x}{N}, \; i = 1 + (x-1)_{\symaccent{k}}\})\; \forall \closedN{i}{\symaccent{k}},
\end{align}

\noindent
where $\symaccent{\tttS} = \{\symaccent{\tttN}[i] \;\vert \; \closedN{i}{\symaccent{k}}\}$. The above method of grouping was chosen in the \tparent$()$ procedure because it allows an easy method to describe the Merkle proof of the CMT symbols as well as it satisfies a key property called the repetition property that we explain shortly. The Merkle proof of the CMT symbol $\tttN_{j}[i]$, $\closedN{i}{N_{j}}$ consists of a data symbol and a parity symbol from each intermediate layer of the tree that is above $L_j$ \cite{CMT,TCOMLDPC}. In particular, for $\closed{j}{2}{l}$, \tproof$(\tttN_{j}[i]) = \{\tttN_{j'}[\;1+(i-1)_{k_{j'}}]$, $\tttN_{j'}[\;1+k_{j'}+ (i-1)_{N_{j'} - k_{j'}}] \:\vert\: \closedN{j'}{j-1}\}$.
The coded symbols 
built using the $\text{\tparent}$ procedure in \eqref{eqn:formparent-CMT} satisfy an important property called the \emph{repetition} property \cite{AceD}. According to this property, the Merkle proofs of $\tteta$ fraction of distinct base layer coded symbols of the CMT contain
at least $\tteta$ fraction of distinct coded symbols from each CMT layer. This property is important to us for the design of the dispersal protocol in side blockchains and we show in Section \ref{sec:pcmt} that the PCMT also satisfies the repetition property.

\vspace{-0.45cm}
\subsection{DA attacks in blockchains with light nodes}\label{sec:DA_attack_LN}
\vspace{-0.15cm}

\subsubsection{System Model}
\debb{To reduce the storage requirement, blockchain systems run \emph{light nodes}. In this case, light nodes act as accepting nodes and the validating nodes are full nodes that send fraud proofs to the light nodes \cite{dataAvailOrg,CMT,SSskewITW,TCOMLDPC,PCMT}. }
To analyze DA attacks in blockchains with light nodes, we consider the following simplified system model.  Consider a blockchain system that has a \emph{block producer}, light nodes, and a \emph{full node oracle}. The full node oracle and all the light nodes are honest. The block producer can be malicious. 
Each light node is connected to 
the full node oracle and the block producer. 
\bluetext{The above system description follows \cite{CMT, TCOMLDPC} where the full node oracle represents the fact that the network of honest full nodes is connected and each light node is connected to at least one honest full node.} 
The system functions in the following way:
\begin{enumerate}[wide, labelwidth=!, labelindent=10pt, label=\roman*)] 
    \item 
    When the block producer generates a block, it constructs its CMT.  
    It then sends the CMT to the full node oracle and the CMT root to the light nodes. On receiving sampling requests from the light nodes, the block producer returns the requested samples with their Merkle proofs.  
    \item The full node oracle, on receiving a CMT from the block producer, decodes the CMT 
    using the hash-aware decoder (as described in the general CMT framework). After decoding the base layer of the CMT that contains the transaction data, it verifies all transactions and sends a fraud proof to all light nodes if it finds invalid transactions. During decoding, if the full node oracle

    \noindent
     detects an IC attack, it sends out an IC proof to all the light nodes.
    \item Light nodes only store the CMT root corresponding to the block generated by the block producer. On receiving a CMT root, light nodes make sampling request for coded symbols of the CMT base layer from the block producer. They perform Merkle proof checks on the returned symbols and send the symbols that satisfy the Merkle proofs to the full node oracle. A light node accepts the block if the block producer returns back all the requested samples (that all pass the Merkle proof checks). On receiving fraud proofs or IC proofs sent out by the full node oracle, light nodes verify the proof and reject the CMT root if the proof is correct. \\[-4mm]
\end{enumerate}

\vspace{-0.4cm}
\subsubsection{Threat Model}

We consider an adversary that acts as a malicious block producer and conducts a DA attack by hiding coded symbols of the CMT. On receiving sampling requests from light nodes, it only returns coded symbols that it has not hidden and ignores other requests. The adversary causes a DA attack at layer $L_j$ of the CMT by i) correctly generating the CMT of a block according to the general CMT framework in Section \ref{sec:general-CMT-framework}, ii) hiding a small portion of the coded symbols of $L_j$ such that the full node oracle is unable to decode $L_j$. 

The light nodes must detect a DA attack that the adversary may perform on any layer of the CMT \cite{CMT, TCOMLDPC}. To do so, they randomly sample a few base layer coded symbols to check the availability of the base layer. Randomly sampling the base layer of the CMT results in the random sampling of all the intermediate layers of the CMT via the Merkle proof of the base layer samples \cite{CMT} that allows the light nodes to check the availability of the intermediate layers.\\[-2mm]

 \vspace{-0.45cm}
\subsubsection{System Specific Metric}\label{sec:p_f_calulation}
\debb{The system specific metric in the case of light nodes}
 is the probability of failure for a single light node to detect a DA attack $P_f(s)$, where $s$ is the total number of base layer samples requested by the light nodes. 
A light node fails to detect a DA attack if none of the samples requested from the base layer of the CMT or their Merkle proofs are hidden by the malicious node. 
Let $\alpha_{\min ,j}$ be the undecodable threshold of layer $L_j$ of the CMT. Then $P_f(s) = \max_{\closedN{j}{l}}\left(1 - \frac{\alpha_{\min ,j}}{N_j}\right)^s$.
\debb{Thus, large $\alpha_{\min,j}$ result in small $P_f(s)$.}

\vspace{-0.6cm}
\subsection{DA oracle in Side blockchains}\label{sec:SB_prelims}
\vspace{-0.15cm}
\subsubsection{System Model}
\debb{To improve the transaction throughput, some blockchains (also called the \emph{trusted} or \emph{main} blockchain) support a large number of \emph{side blockchains} (smaller blockchain systems) by storing the Merkle root of the side blockchain blocks in their ledger \cite{AceD}, \cite{DE-PEG}. A single trusted blockchain supports a large number of side blockchains, each of which makes transactions in parallel resulting in a higher transaction throughput \cite{AceD}. In this case, the nodes in the trusted blockchain act as the accepting nodes, and  the validating nodes are other nodes in the side blockchain.}
To mitigate DA attacks in side blockchains, the DA oracle was introduced in \cite{AceD}. The DA oracle acts as an interfacing layer between the side blockchain nodes and the trusted blockchain with the goal of storing chunks of the transaction block in order to ensure availability. Let the DA oracle have $\tNoracle$ nodes where the adversary is able to corrupt a maximum $\tbetaoracle$ fraction of them.
Similar to \cite{AceD, DE-PEG}, we assume $\tbetaoracle < \frac{1}{2}$. For side blockchains that use the DA oracle, there exists a dispersal protocol which is a rule that specifies which oracle node receives which base layer CMT symbols,  each receiving $\tkoracle$ of them along with the Merkle proofs of the received symbols and the CMT root.  The DA oracle functions in the following way:

\begin{enumerate}[wide, labelwidth=!, labelindent=10pt, label=\roman*)] 
    \item When the block producer generates a block, it constructs its CMT. 
    The block producer then uses the dispersal protocol to communicate the coded symbols of the CMT base layer and their Merkle proofs to the $\tNoracle$ oracle nodes. Each oracle node also receives the CMT root. 
    \item Each oracle node on receiving the $\tkoracle$ coded symbols of CMT as specified in the dispersal protocol performs Merkle proof checks on the received symbols. 
    Each oracle node accepts the CMT root and votes a \emph{yes} if all its received  symbols pass the Merkle proof checks.
    \item For a parameter $\tgammaoracle$ such that $\tgammaoracle \leq 1 - 2\beta$, the CMT root is sent to the trusted blockchain nodes if at least $\tgammaoracle + \tbetaoracle$ fraction of the oracle nodes vote \emph{yes}. In this case, each oracle node stores the received $\tkoracle$ coded symbols along with their Merkle proofs and the CMT root. 
    \end{enumerate}

\vspace{-0.1cm}
Note that in the above system, other side blockchain nodes (that are not the block producer) prevent IC attacks and invalid transactions by sending IC proofs and fraud proofs to the oracle nodes. The oracle nodes check the proof validity and forward it to the trusted blockchain nodes. 

\vspace{-0.05cm}
\subsubsection{Dispersal Protocol} The dispersal protocol satisfies the condition that whenever the CMT
root is sent by the DA oracle to the trusted blockchain nodes, any side blockchain node must be able to decode each layer of the CMT from the coded symbols stored at the oracle nodes. We define the dispersal protocol by the set $\tCdispersal = \{\tAdispersal_1, \ldots, \tAdispersal_{\tNoracle} \}$, where
$\tAdispersal_i$, $\closedN{i}{\tNoracle}$ denotes the set of base layer coded symbols sent to the $i$th oracle node and $\vert\tAdispersal_i\vert = \tkoracle$. Consider the following.

\vspace{-0.3cm}
\begin{definition}
Dispersal Protocol $\tCdispersal$ is $(j,\tteta)$-correct if every $\tgammaoracle$ fraction of oracle nodes collectively receives at least $ N_j -  \tteta+1$ distinct coded symbols from layer $L_j$ of the CMT. 
\end{definition}

\vspace{-0.3cm}
A dispersal protocol that is $(j,\alpha_{\min ,j})$-\emph{correct} ensures that if the CMT root is sent by the DA oracle to the trusted blockchain, any side blockchain node will be able to decode back layer $L_j$ of the CMT. Thus, we want a dispersal protocol that is $(j,\alpha_{\min ,j})$-\emph{correct}  for all $1 \leq j \leq l$. Consider the following lemma. The proofs of all lemmas in this paper are provided in the Appendix.

\vspace{-0.3cm}
\begin{lemma}\label{lemma:comm_cost_mu_min}
Let $\tmu_{\min} = \lfloor\min_{1 \leq j \leq l}\left(  \frac{\alpha_{\min ,j}-1}{N_j}\right) N_l \rfloor+1$.
For CMTs that satisfy the repetition property as mentioned in Section \ref{sec:general-CMT-framework}, if a dispersal protocol is $(l,\tmu_{\min})$-correct, then it is $(j,\alpha_{\min, j})$-\emph{correct}  for all $1 \leq j \leq l$.
\end{lemma}

\vspace{-0.32cm}
Thus, based on the above lemma, we would like to design a $(l,\tmu_{\min})$-correct dispersal protocol. First, consider the following definition.

\vspace{-0.32cm}
\begin{definition} 
(\cite[Definition 2]{DE-PEG}) Dispersal protocol $\tCdispersal = \{\tAdispersal_1, \tAdispersal_2, \ldots, \tAdispersal_{\tNoracle} \}$ is called a $\tkoracle$-dispersal if each $\tAdispersal_i$ is a $\tkoracle$ element subset chosen uniformly
at random with replacement from all the $\tkoracle$ element subsets of
the $N_l$ base layer coded symbols of the CMT. 
\end{definition}

\vspace{-0.31cm}
Now, we can show the following for  a $\tkoracle$-dispersal based on \cite[Lemma 2]{DE-PEG}. 

\vspace{-0.32cm}
\begin{lemma}\label{lemma:not_correct}
Let $H_e(p) = -p\ln(p) - (1-p)\ln(1-p)$. For a $\tkoracle$-dispersal protocol $\tCdispersal$, 
\begin{align*}
    \mathrm{Prob}(\tCdispersal \text{ is not $(l,\tmu_{\min})$-correct } ) \leq e^{\tNoracle  H_e(\tgammaoracle)}\left(\sum_{j=0}^{N_l-\tmu_{\min}} (-1)^{N_l-\tmu_{\min} -j} {N_l \choose j} {N_l - j - 1 \choose \tmu_{\min} -1}\left[\frac{{j \choose \tkoracle}}{{N_l \choose \tkoracle}}\right]^{\tgammaoracle \tNoracle}\right).
\end{align*}
\end{lemma}

\vspace{-0.3cm}
The RHS of the above inequality can be made smaller than an arbitrary threshold $p_{th}$ by using a sufficiently large $\tkoracle$. Let $\tkoracle^*(\tmu_{\min},\tNoracle,N_l,\tgammaoracle,p_{th})$ be the smallest $\tkoracle$ such that the RHS in Lemma \ref{lemma:not_correct} is less than $p_{th}$. In this paper, we use a $\tkoracle^*(\tmu_{\min},\tNoracle,N_l,\tgammaoracle,p_{th})$-dispersal protocol.

\vspace{-0.02cm}
\subsubsection{System Specific Metric}\label{sec:comm_cost_calculation}
\debb{Since the side blockchains use storage at oracle nodes to mitigate DA attacks, the system specific metric is the} communication cost associated with the dispersal which can be written as \tcomm$(\tmu_{\min},\tNoracle,N_l,\tgammaoracle,p_{th}) = \tNoracle\tkoracle^*(\tmu_{\min},\tNoracle,N_l,\tgammaoracle,p_{th})X$, where $X$, called the single sample download size, is the total size of one CMT coded symbol along with its Merkle proof. Large undecodable thresholds $\alpha_{\min ,j}$ result in a large $\tmu_{\min}$ and hence a smaller $\tkoracle^*(\tmu_{\min},\tNoracle,N_l,\tgammaoracle,p_{th})$, which inturn reduces the communication cost. 

\vspace{-0.57cm}
\subsection{Design objectives for CMT}\label{sec:performance_metrics}
\vspace{-0.18cm}
As mentioned earlier, in this paper, we focus on designing polar codes for the CMT at large code lengths. The different metrics that are of importance to a CMT at large code lengths are i) IC proof size; ii) decoding complexity; iii) undecodable thresholds $\alpha_{\min ,j}$; iv) threshold complexity of computing the undecodable thresholds; v) CMT root size. As seen before, the system specific metrics depend on the undecodable thresholds $\alpha_{\min ,j}$ of the different layers of the CMT.  Improved performance of these metrics requires large undecodable thresholds.  Additionally, the design complexity of the system depends on the threshold complexity. For example, in the case of light nodes, in order to find out the system design parameter of the number of samples that a light node should request to get a desired probability of failure, the undecodable thresholds for the different layers of the CMT need to be computed. Similarly, for in side blockchains, in order to determine the value of $\tkoracle^*(\tmu_{\min},\tNoracle,N_l,\tgammaoracle,p_{th})$ for the dispersal protocol design, the knowledge of the undecodable thresholds is required. Thus, for a low system design complexity, the threshold complexity must be small. At the same time, the CMT must also result in small IC proof sizes, small decoding complexity, and small CMT root size. In the next section, we provide a CMT construction using polar codes called the PCMT which performs 
well in all the above metrics when the size of the transaction blocks is large. 

\vspace{-0.5cm}
\section{Polar Coded Merkle Tree (PCMT)}\label{sec:pcmt}
\vspace{-0.15cm}
In this section, we first provide the necessary background about polar codes that we use in 
the PCMT construction. We then explain the construction method of the PCMT.

\vspace{-0.52cm}
\subsection{Polar Code Preliminaries}\label{sec:polar_prelims}
\begin{figure}[t]
    \centering
\begin{subfigure}{0.33\linewidth}
\begin{minipage}{0.99\linewidth}
\begin{tikzpicture}
  \node (img) {\includegraphics[scale=0.33]{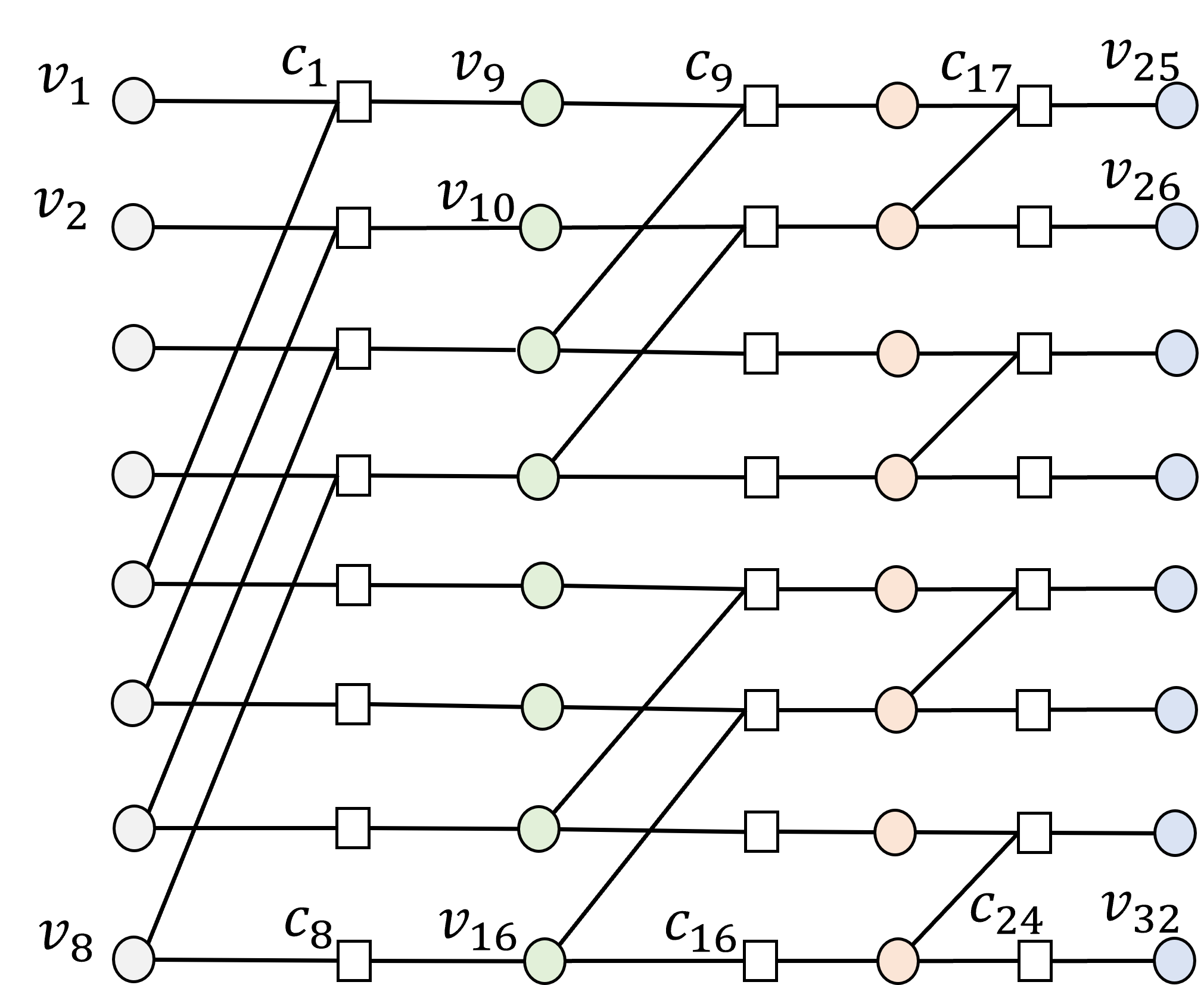}};
 \end{tikzpicture}
 \end{minipage}
\end{subfigure}%
    \begin{subfigure}{0.33\linewidth}
\begin{minipage}{0.99\linewidth}
\begin{tikzpicture}
  \node (img) {\includegraphics[scale=0.33]{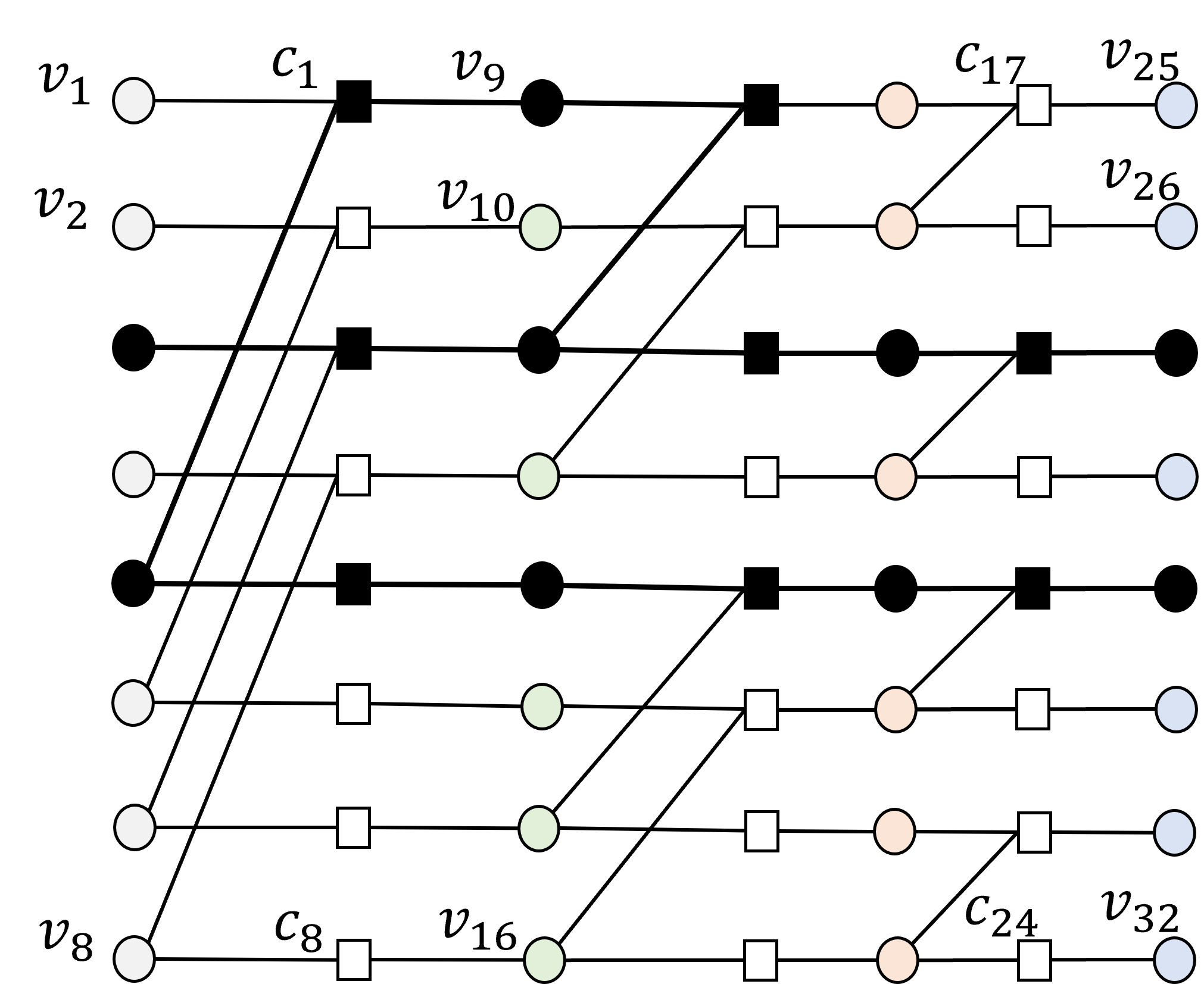}};
 \end{tikzpicture}
 \end{minipage}
    \end{subfigure}%
\begin{subfigure}{0.33\linewidth}
\begin{minipage}{0.99\linewidth}
\begin{tikzpicture}
  \node (img) {\includegraphics[scale=0.33]{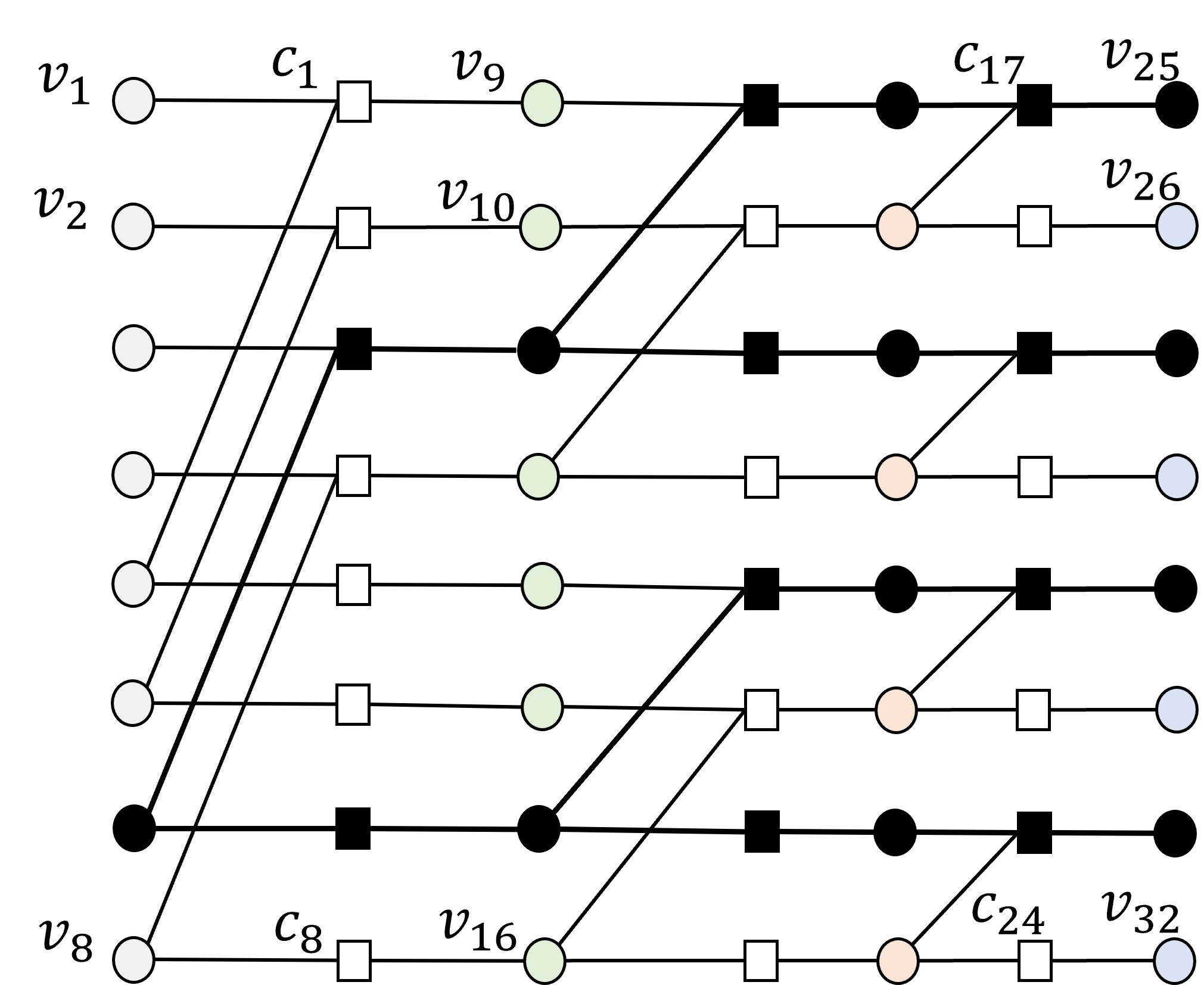}};
 \end{tikzpicture}
 \end{minipage}
\end{subfigure}
     \vspace{-10pt}
    \caption{Left panel: FG $\tFG_{8}$ where circles represent VNs and squares represent CNs; Middle panel: Stopping set; Right panel: Stopping tree. The black VNs and CNs in the middle and right panels represent the stopping set/tree.}
    \label{fig:FG_examples}
\end{figure}

\vspace{-0.18cm}
The transformation matrix $\tP_{2^n} = \tP^{\otimes n}_2$ is used to define an ($\tNmm,k$) polar code of codelength
$\tNmm = 2^n$ for some \redtext{positive} integer $n$ and information length $k$ \cite{PolarCodesErikan}. 
Each row of $\tP_{2^n}$ corresponds to either a data (information) symbol or a frozen symbol (zero symbol in this paper). The generator matrix of the  polar code is the submatrix of $\tP_{2^n}$ corresponding to the data symbols. The design of a polar code involves choosing which rows of $\tP_{2^n}$ should correspond to the data symbols (or equivalently which rows should correspond to frozen symbols). A polar code can also be represented using a factor graph (FG) \cite{Polar-SStree-TCOM}. 
For example, the FG representation of $\tP_{8}$ is shown in Fig. \ref{fig:FG_examples} left panel. In general, the FG of $\tP_{\tNmm}$, which we denote by $\tFG_{\tNmm}$, has $n+1$ columns of VNs and $n$ columns of CNs. For the variable node at VN column $m$ and row $i$ in FG $\tFG_{\tNmm}$, we define its VN index $\tVNi = (m-1)\tNmm + i$, $\closedN{i}{\tNmm}$, $\closedN{m}{n + 1}$ and refer to the VN as $v_{\tVNi}$. Similarly, for each check node at CN column $m$ and row $i$ in FG $\tFG_{\tNmm}$, we define its CN index $\tCNi = (m-1)\tNmm + i$, $\closedN{i}{\tNmm}$, $\closedN{m}{n}$ and refer to the CN as $c_{\tCNi}$. The VN and CN indexing is also shown in Fig. \ref{fig:FG_examples}.
Note that in the FG of polar codes, CNs have a small degree of either 2 or 3. We leverage this property in our PCMT construction to result in small IC proof sizes.

The construction of the CMT according to Section \ref{sec:general-CMT-framework} requires systematic encoding. Systematic  encoding of $(\tNmm, k)$ polar codes can be performed similar to  \cite{SystematicPOlarCodes}
by operating on the FG $\tFG_{\tNmm}$ of the code. 
While efficient systematic encoding of polar codes is presented in \cite{SystematicPOlarCodes}, we propose a new encoder known as the \emph{peeling encoder for polar codes (PEPC)}.
The main motivation for using a PEPC is it enables us to use a smaller FG obtained by FG pruning while still allowing successful encoding and decoding (using the pruned FG) as well as improved performance. The 

\noindent
encoder of \cite{SystematicPOlarCodes} does not allow this optimization.  

Given information and frozen index sets $\tImm \subset [\tNmm]$ and $\tFmm = [\tNmm] \setminus \tImm$, such that $\vert \tImm \vert = k$ (also, let $\tNmm = 2^n$), systematic encoding in the PEPC is performed as follows:
i) place the $k$ data symbols at the VNs $\{v_{n\tNmm + i} \;\vert\; i \in \tImm\}$ (i.e., VNs in the rightmost column of FG $\tFG_{\tNmm}$ at rows corresponding to $\tImm$) and set the VNs at $\{v_{i}\; \vert\; i \in \tFmm\}$ (i.e., VNs in the leftmost column of FG $\tFG_{\tNmm}$ at rows corresponding to $\tFmm$) to zero symbols; ii) determine (decode) the rest of the VNs from the check constraints of the FG $\tFG_{\tNmm}$ using a peeling decoder. By design, the coded symbols $\{v_{n\tNmm + i} \;\vert\; \closedN{i}{\tNmm}\},$ are systematic.
\greentext{Since we are relying on a peeling decoder for encoding, we need to verify the correctness of the encoding procedure, which we do in the following lemma.}

\vspace{-0.4cm}
\begin{lemma}\label{lemma:PDE_sucessfull}
\redtext{The PEPC always results in a valid codeword for all choices of information set $\tImm$.}
\end{lemma}

\vspace{-0.3cm}
For decoding polar codes, we again use a peeling decoder on the code FG. Similar to LDPC codes, the peeling decoder on the FG of polar codes fails if all VNs corresponding to a stopping set of the FG are erased. Mathematically, a stopping set is a set of VNs with the property that every CN connected to a VN in this set is connected to at least two VNs in the set. The set of VNs of a stopping set $\tssingle$ that are in the rightmost column of the FG is called the \emph{leaf set} \cite{Polar-SStree-TCOM} of $\tssingle$ which we denote as \tLS($\tssingle$). An important category of stopping sets in the FG of polar codes is called \emph{stopping trees} \cite{Polar-SStree-TCOM}. A stopping tree is a stopping set that only contains one VN from the leftmost column of the FG, which is called the root of the stopping tree.
An example of a stopping set and a stopping tree in the FG $\tFG_{8}$ is shown in Fig. \ref{fig:FG_examples}. For a given information index set $\tI$, let $\tSS^{\tI}$ denote the set of all stopping sets in the FG $\tFG_{N}$ that
do not have any frozen VNs from the leftmost column of the FG. 
The following lemma from \cite{Polar-SStree-TCOM} provides important properties of stopping sets in the FG of polar codes that help simplify its undecodable threshold. 

\vspace{-0.4cm}
\begin{lemma}\label{lemma:paper-ss-property}
(\cite{Polar-SStree-TCOM}) Consider a polar FG $\tFG_{N}$, where $N$ is a power of two and let $\tI$ be the information set. Each VN $v_{i}$, $\closedN{i}{\tNmm}$, in $\tFG_{\tNmm}$ is the root of an unique stopping tree. Let $\tssingletree^{\tNmm}_i$ be the unique stopping tree with root $v_{i}$ in $\tFG_{\tNmm}$. Then, $\vert\text{\tLS}(\tssingletree^{N}_i)\vert = \ttT_{\tNmm}(i)$ and 
$\min_{\tssingle \in \tSS^{\tI}} \vert\text{\tLS}(\tssingle) \vert  = \min_{i \in \tI} \vert\text{\tLS}(\tssingletree^{N}_i)\vert =  \min_{i \in \tI} \ttT_{\tNmm}(i)$. 
\end{lemma}

\vspace{-0.95cm}
\subsection{PCMT construction using polar codes}\label{sec:pcmt-construction-indexing}
\vspace{-0.15cm}
In this section, we provide the construction method for a PCMT within the general CMT framework provided in Section \ref{sec:general-CMT-framework}. We later show in Section \ref{sec:pruned_construction} how the PCMT construction provided in this section can be customized to use pruned polar FGs that improve the performance of the metrics mentioned in Section \ref{sec:performance_metrics}. For the purposes of clarity, we first provide the PCMT construction that uses the  entire FG of the polar codes as shown in Section \ref{sec:polar_prelims}. We later demonstrate how the PCMT construction can be modified when the full FG is not used.

We next describe the construction of a PCMT $\tCMT = (K, R, q, l)$. Consider the general layer of the PCMT with codelength $N$ and  information length $k$.
When the full FG of polar codes is used, $N = \frac{k}{R}$, but it need not be the case in general\footnote{\debb{In this paper, we use $K$ to represent the PCMT parameter and $k$ to represent the information length of the general layer.}}. For now, assume that $N$ is a power of 2. We later remove this assumption.
Let  $\tI$ and $\tF$ be the information and frozen index sets of the general layer.  We have $\vert \tI \vert = k$ and $\vert \tF \vert = N - k$. For notational ease, we re-index the row indices in FG $\tFG_{N}$ such that $\tI$ and $\tF$ are the indices $[1,k]$ and $(k,N]$, respectively. 
For FG $\tFG_{N}$, we define $\ttotVN$ as the total number of VNs in the FG. Additionally, for  FG $\tFG_{N}$, define an index called the \emph{dropped index} $\tdropindex$ which is the difference between $\ttotVN$ and $N$. For the full FG of polar codes, $\ttotVN = N(\log N + 1)$ and \tdropindex$ = N(\log N + 1) - N = N\log N$.    

For the PCMT, we also define certain intermediate coded symbols that are used to form the PCMT as $\ttN[\tVNi]$ which corresponds to VN $v_{\tVNi}$ in the FG  $\tFG_{N}$.  In the general CMT framework, we have $\tttS = \{\ttN[\tVNi] \;\vert\; \closed{\tVNi}{\tdropindex + 1}{\tdropindex + k}\}$, $\tttP = \{\ttN[\tVNi] \;\vert\; \openleft{\tVNi}{\tdropindex + k}{\tdropindex + N}\}$, and $\tttC = \tttS \cup  \tttP$. Next, we explain the different procedures involved in the

\noindent
general CMT framework for the PCMT construction.

\vspace{-0.05cm}
\subsubsection{Formation of PCMT coded symbols}
We first explain the $\text{\tparity}()$ procedure. For the data symbols $\tttS$, use a PEPC to find the parity symbols $\tttP$, where VNs corresponding to $\tfrozen =\{\ttN[\tVNi]\;\vert\; \openleft{\tVNi}{k}{N}\}$ in $\tFG_{N}$ are set as zero symbols. The PEPC also provides the set of symbols $\tdropped = \{\ttN[\tVNi] \;\vert\; \closedN{\tVNi}{\tdropindex}\}$ in FG $\tFG_{N}$ which are dropped from the PCMT, i.e., they are not included in $\tttC$. However, before dropping, we use their information in the  $\text{\tparent}()$ procedure \deb{which is as follows for a PCMT}. We have
\begin{align}\label{eqn:data_symbol_formation}
  \symaccent{\ttN}[\tVNi]= {\fontfamily{qcr}\selectfont \text{concat}}( \{ {\fontfamily{qcr}\selectfont \text{Hash}}(\ttN[x]) \;\vert\;  \closedN{x}{\ttotVN},\; &\tVNi = 1 + (x-1)_{\symaccent{k}} \}  ),\; \forall \closed{\tVNi}{\symaccent{\tdropindex} + 1}{\symaccent{\tdropindex} + \symaccent{k}},
\end{align}
where $\symaccent{\tttS} = \{\symaccent{\ttN}[\tVNi] \;\vert \; \closed{\tVNi}{\symaccent{\tdropindex} + 1}{\symaccent{\tdropindex} + \symaccent{k}}\}$. 
An example for the formation\footnote{The $\text{\tparent}$ procedure for the PCMT is the same as that provided in \eqref{eqn:formparent-CMT}. The only difference is that now we consider the modulo operation across all the VNs in the FG.} of the symbols

\begin{figure}[t]
    \centering
\begin{minipage}{0.99\linewidth}
\centering
\begin{tikzpicture}
  \node (img) {\includegraphics[scale=0.39]{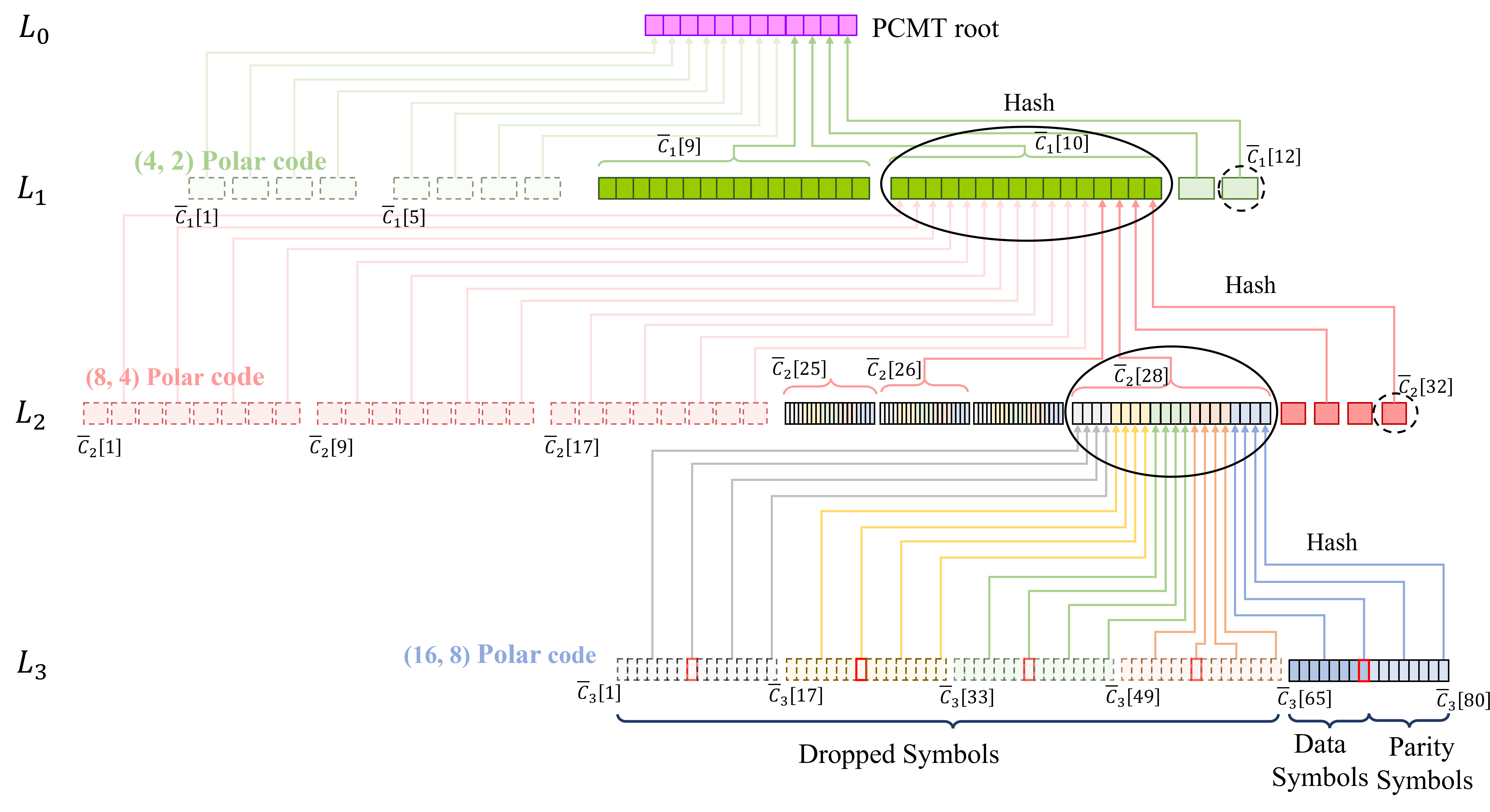}};
 \end{tikzpicture}
 \end{minipage}
     \vspace{-16pt}
    \caption{PCMT $\tT$
$= (K = 8, R = 0.5, q = 4 , l = 3)$. In the PCMT, the coded symbols of all the columns of the polar FG are hashed into the parent layer. 
The dropped symbols are shown in dotted. The symbols in $L_3$ are colored according to the column they belong to in FG $\tFG_{16}$. The circled symbols in $L_1$ and $L_2$ are the Merkle proof of the red symbols in $L_3$. The data (parity) symbols in the Merkle proofs are shown in solid (dashed) circles. }
    \label{fig:PCMT_construction}
\end{figure}

\noindent
$\ttN_{2}[28]$ and $\ttN_{1}[10]$ in the PCMT $\tCMT = (K=8, R = 0.5, q = 4, l = 3)$ is shown in Fig. \ref{fig:PCMT_construction}. 

    In the above $\text{\tparent}()$ procedure, the data symbols in $\symaccent{\tttS}$ are formed using the hashes of all the $\ttotVN$ intermediate coded symbols $\ttN[\tVNi]$ of the general layer, i.e., all the symbols in $\tdropped \cup \tttC$ or all the VNs in the FG $\tFG_{N}$. Each data symbol in $\symaccent{\tttS}$  is formed by combining $\symaccent{q} = \frac{\ttotVN}{\symaccent{k}}$ hashes (of the intermediate coded symbols of the general layer) together according to \eqref{eqn:data_symbol_formation}. The intuition behind using the hashes of all the intermediate coded symbols in the $\text{\tparent}()$ procedure is so that the symbols in $\tdropped$ also have Merkle proofs. Although dropped, the symbols in $\tdropped$ can be decoded back by a peeling decoder using the available (non-erased) symbols of $\tttC$. Once decoded, they can be used to build IC proofs of small size using the degree 2 and 3 CNs in the polar FG $\tFG_{N}$. \\[-2mm]

  \vspace{-0.4cm}  
 \subsubsection{Merkle Proof of PCMT symbols}  
 Due to the above $\text{\tparent}()$ procedure, the symbols in both $\tttC$ and $\tdropped$ of the general layer have Merkle proofs. Since $\tdropped \cup \tttC$ are all the intermediate coded symbols $\ttN[\tVNi]$, we \greentext{specify} the Merkle proof \tproof$(\ttN[\tVNi])$. 
 For $\closed{j}{2}{l}$,
 the Merkle proof of the symbol $\ttN_{j}[\tVNi]$, $\closedN{\tVNi}{\ttotVN_j}$, consists of a data symbol and parity symbol from each layer of the PCMT above $L_j$ similar to an LCMT in \cite{CMT, TCOMLDPC, AceD}. Precisely, the Merkle proof is given by the following. \greentext{For $\closed{j}{2}{l}$}
\begin{align}\label{eqn:pcmt_merkle_proof}
    &\text{\tproof}(\ttN_{j}[\tVNi]) = \{\ttN_{j'}[\tdropindex_{j'}+1 + (\tVNi-1)_{k_{j'}}],\;  \ttN_{j'}[\tdropindex_{j'}+1 + k_{j'} + (\tVNi-1)_{N_{j'} - k_{j'}}]\;\vert\; \closedN{j'}{j-1}\}. 
\end{align}

\vspace{-0.2cm}
An illustration of Merkle proof for different symbols in the PCMT $\tCMT = (K=8, R = 0.5, q = 4, l = 3)$ is shown in Fig. \ref{fig:PCMT_construction}. The Merkle proof in \eqref{eqn:pcmt_merkle_proof} is defined such that the data symbols from each layer in $\text{\tproof}(\ttN_{j}[\tVNi])$ lie on the path of $\ttN_{j}[\tVNi]$ to the PCMT root as shown in Fig. \ref{fig:PCMT_construction}. Similar to  an LCMT \greentext{(cf. \cite{CMT, TCOMLDPC, AceD})}, the data symbols in this path are used to check the integrity of $\ttN_{j}[\tVNi]$ in \tverify($\ttN_{j}[\tVNi]$,  \tproof($\ttN_{j}[\tVNi]$), \troot). 
Due to the definitions of $\text{\tparent}$() procedure in \eqref{eqn:data_symbol_formation} and Merkle proofs in \eqref{eqn:pcmt_merkle_proof}, we have the following. 

\vspace{-0.35cm}
 \begin{lemma}\label{lemma:PCMT-repetition}
 The PCMT satisfies the repetition property.
 \end{lemma}

\vspace{-0.32cm}
\debb{The proof of the above lemma is similar to the proof of \cite[Lemma 2]{AceD} and is omitted.}

\subsubsection{Hash-aware peeling decoder and IC proofs}

We decode the PCMT using a hash-aware
peeling decoder similar to an LCMT in \cite{CMT,TCOMLDPC,AceD}. In the general CMT framework, the \tdecodelayer() procedure for the general layer is as follows: it acts on the FG $\tFG_{N}$ that is used to encode the general layer. It takes as inputs the frozen symbols $\tfrozen$ and the non-hidden symbols of $\tttC$. Using a peeling decoder, it finds all symbols in $\tdropped \cup \tttC$ (i.e., the value of all the VNs in FG $\tFG_{N}$). The hash of every decoded symbol is matched with its hash provided by the parent layer. In the case that the hashes do not match, an IC proof is generated using the degree 2 or 3 CN of the FG $\tFG_{N}$ as per the general CMT framework.

\vspace{-0.43cm}
\subsection{System metrics for the PCMT}
\vspace{-0.15cm}
With the PCMT construction provided above, we now analyze the main system metrics. Note that the decoding complexity, IC proof size, and the PCMT root size have straightforward calculations and we delegate their discussion to Section \ref{sec:sims} where we also compare the performance
to other systems. In this subsection, we focus on the undecodable threshold of the PCMT.

    Recall that the undecodable threshold $\alpha_{\min,j}$ for layer $L_j$ of the PCMT is the minimum number

    \noindent
    of coded symbols that must be hidden (erased) from layer $L_j$ to prevent the peeling decoder from decoding the layer. Consider the general layer $L_j$ of the PCMT. %
Note that the VNs corresponding to $\tfrozen$ are set to zero symbols during the \tparity() procedure in the PCMT and hence cannot be erased. Thus, stopping sets in $\tSS^{\tI}$ are all the sets of VNs that, if erased, will prevent the hash-aware peeling decoder from decoding the general layer.  Since all the coded symbols except the rightmost column of $\tFG_{N}$ are dropped, i.e., they are not stored in the PCMT in \tttC, the hash-aware peeling decoder fails if the leaf set of a stopping set in $\tSS^{\tI}$ is hidden/erased. Thus, the undecodable threshold (for the general layer)
$\talpha = \min_{\tssingle \in \tSS^{\tI}} \vert\text{\tLS}(\tssingle) \vert 
 = \min_{i \in \tI} \ttT_{\tNmm}(i)$ (from Lemma \ref{lemma:paper-ss-property}).

Note that (from Lemma \ref{lemma:paper-ss-property}) $T_N(i)$ is the leaf set size of the stopping tree with root $v_i$, \greentext{$\closedN{i}{N}$}. 

\noindent
Thus based on the expression of the undecodable threshold, the best strategy for the adversary to result in the smallest undecodable threshold is to erase/hide the smallest leaf set amongst all stopping trees with non-frozen root. Clearly, the undecodable threshold depends on the choice of the information index sets $\tI$ used in the general layers of the PCMT. In the next section, we will provide a method called the \emph{Sampling Efficient Freezing (SEF)} algorithm to choose the information index sets in order to maximize the undecodable thresholds.

\vspace{-0.4cm}
\section{Polar code design for the PCMT: Sampling Efficient Freezing Algorithm }\label{sec:SEF}
\vspace{-0.2cm}
In this section, we provide a method to select the frozen index set $\tF$ (or equivalently the
information index set $\tI$) to be used in the general layer that results in large $\talpha$. Note that $\vert \tF \vert = N- k$ and $\tI = [N] \setminus \tF$, where $k$ is the message length. Since, $\talpha = \min_{i \in \tI}\ttT_{N}(i)$, a na\"\i ve frozen set selection method to maximize $\talpha$ would be to select the indices of $k$ VNs from the leftmost column of the FG $\tFG_{N}$ that have the smallest stopping tree leaf set sizes $\ttT_{N}(i)$.
We call the na\"\i ve frozen set selection as \emph{Na\"\i ve-Freezing (NF)} algorithm that satisfies  
$\talpha^{NF} = \min\left(\ttT_{N};N-k+1\right)$ (i.e., the $(N-k+1)$-th smallest value of $\ttT_{N}$). 

It should be noted that for an $(N,k)$ polar code, the NF algorithm results in the largest possible undecodable threshold. 
However, next, we show that it is possible to  prune the FG  of polar codes via a more informed method of selecting the frozen index sets using the SEF algorithm. 
The pruning via the SEF algorithm punctures the polar code (i.e., it reduces the code length at a fixed information length) without decreasing the undecodable threshold.  
Since for a fixed undecodable threshold, the performance of the system specific metrics is inversely related to the code length, the SEF algorithm results in better performance of the system specific metrics compared to the NF algorithm. Additionally, the SEF algorithm allows us to design $(N, k)$ polar codes where $N$ can be of any length and is not limited to be a power of two. 
For the remainder of this section, assume that for all FG $\tFG_{\tNmm}$, the rows in $\tFG_{\tNmm}$ are indexed $1$ to $\tNmm$ from top to bottom (as opposed to indexing mentioned in Section \ref{sec:pcmt-construction-indexing}). 
The SEF algorithm is based on the following lemma. 

\vspace{-0.4cm}
\begin{lemma}\label{lemma:last_few_frozen}
Consider FG $\tFG_{\tNmm}$ where $\tNmm$  is a power of two and let $\tFmm$ and $\tImm$ be the frozen and

\noindent
information index sets, respectively.  For a parameter $\tlast$, define the set of VNs $\tlastVN^{\tlast}_{\tNmm}[m] = \{v_{\tVNi}\; \vert\; \tVNi = (m-1)\tNmm + i, \closed{i}{\tNmm - \tlast +1}{\tNmm}\}$. If  $[\tNmm - \tlast +1,\tNmm] \subset \tFmm$, then: i) $\forall$  $\tssingle \in \tSS^{\tImm}$, $\tssingle$ does not have any VNs in $\tlastVN^{\tlast}_{\tNmm}[\log \tNmm + 1]$; ii) all VNs in $\{\tlastVN^{\tlast}_{\tNmm}[m] \;\vert \; \closedN{m}{\log \tNmm + 1} \}$ are zero symbols.
\end{lemma}
\vspace{-0.4cm}

\begin{figure}[t]
\begin{subfigure}{0.4\linewidth}
\begin{minipage}{0.99\linewidth}
\centering
\begin{tikzpicture}
  \node (img) {\includegraphics[scale=0.34]{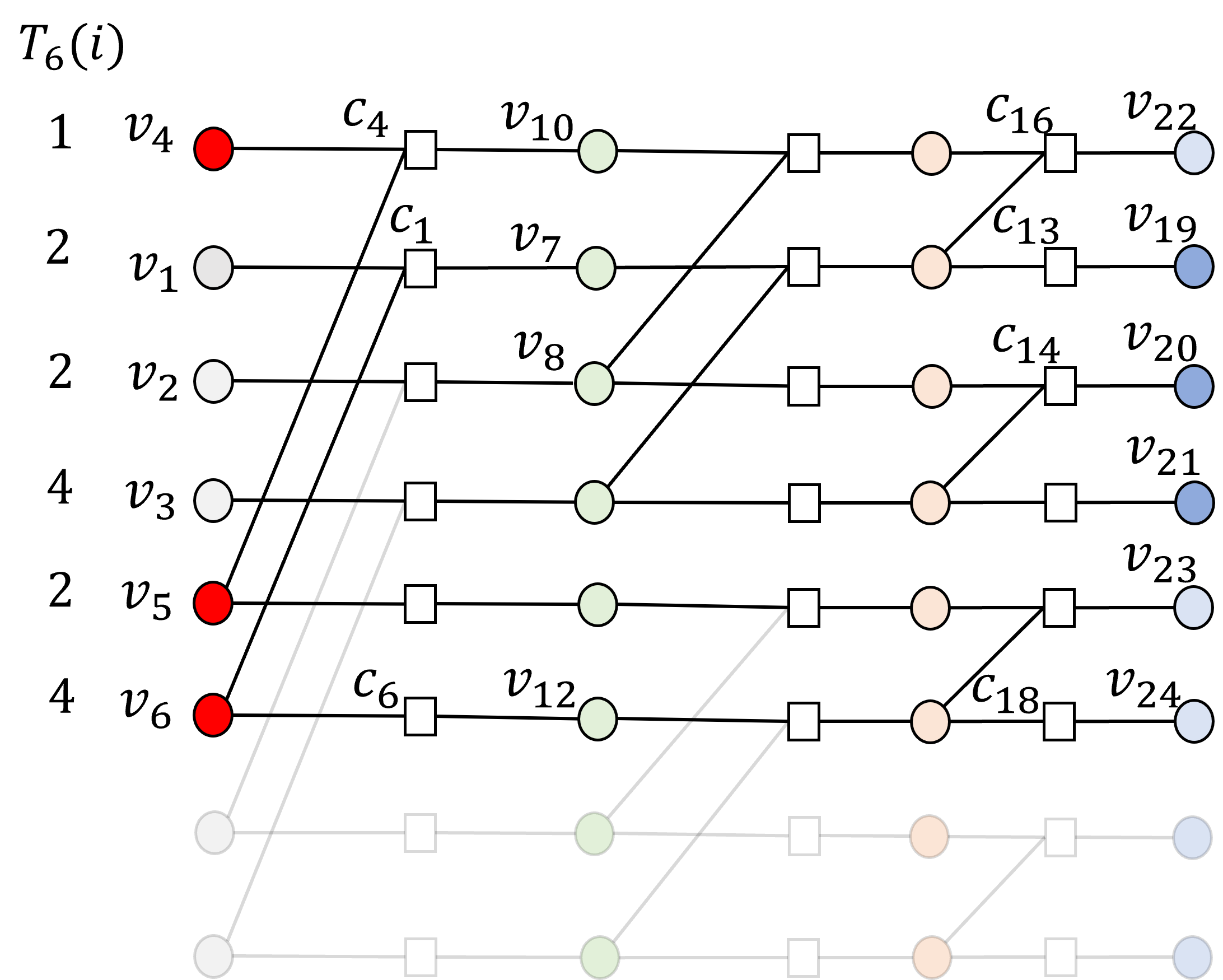}};
 \end{tikzpicture}
 \end{minipage}
\end{subfigure}%
\begin{subfigure}{0.3\linewidth}
\begin{minipage}{0.99\linewidth}
\centering
\begin{tikzpicture}
  \node (img) {\includegraphics[scale=0.36]{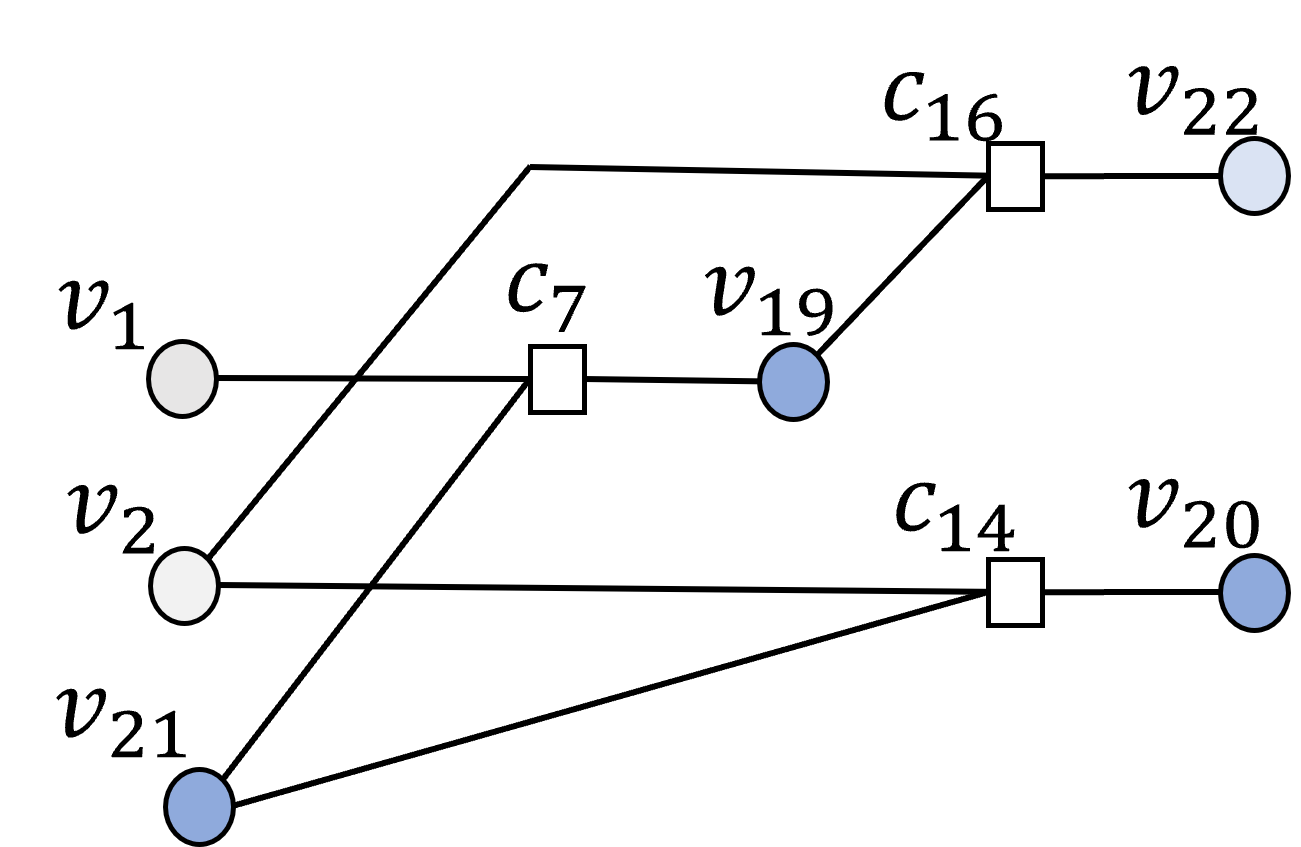}};
 \end{tikzpicture}
 \end{minipage}
    \end{subfigure}%
\begin{subfigure}{0.33\linewidth}
\begin{minipage}{0.99\linewidth}
    \centering
\begin{tikzpicture}
  \node (img) {\includegraphics[scale=0.36]{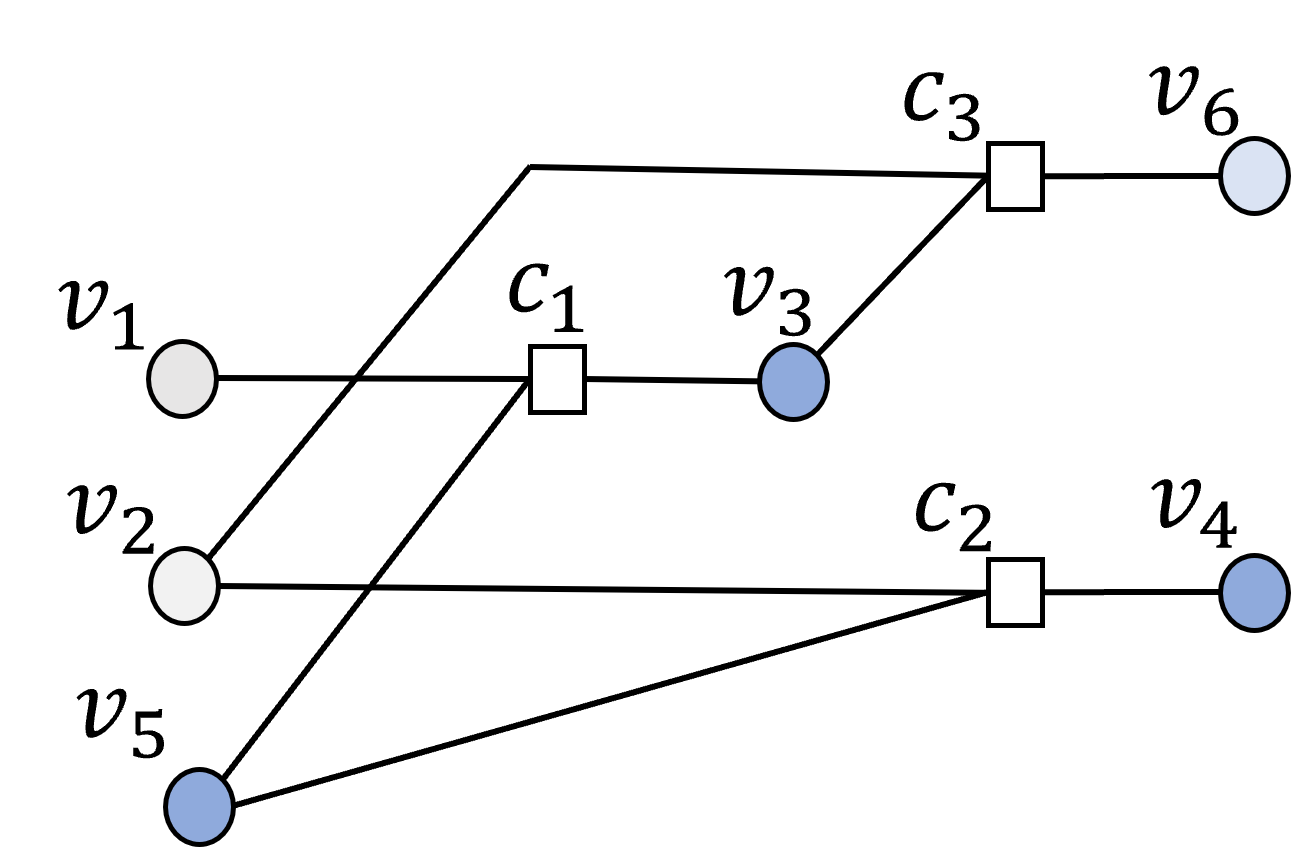}};
 \end{tikzpicture}
 \end{minipage}
\end{subfigure}    
     \vspace{-5pt}
    \caption{Left panel: FG $\tFG_{6}$ obtained by removing the VNs from the last 2 rows of $\tFG_{8}$ (removed VNs are shown in low opacity). The values of the stopping tree size for each VN in the leftmost column on FG $\tFG_{6}$, i.e., $\ttT_6(i)$ is provided in the figure. The VNs marked in red are the frozen VNs $\tF$ selected using the SEF algorithm.
    The rows in FG $\tFG_{6}$ are numbered such that  the information and frozen indices $\tI$ and $\tF$, respectively, are the indices $[1,3]$ and $(3,6]$ as required for the PCMT construction mentioned in Section \ref{sec:pcmt}. The non-dropped VNs in FG $\tFG_{6}$ (in the PCMT construction) are marked in blue where the darker blue circles represent the information symbols and the light blue circles represent the parity symbols. Note that since the last two rows in FG $\tFG_{6}$ are frozen, the VNs in the last two rows from the FG can be removed and we get $N_{\tname} = 4$. Thus, the input to the pruning algorithm is the FG $\tFG_{4}$ which is obtained by removing the last two rows from the FG $\tFG_{6}$ in the left panel; Middle panel: Pruned FG $\tpFG_{4}$ obtained by using the pruning algorithm with input FG $\tFG_{4}$ obtained from left panel. The VN indexing corresponds to the index of the corresponding VNs in the unpruned FG; Right panel: Pruned FG $\tpFG_{4}$ same as the middle panel but with VNs and CNs re-indexed in ascending order according to their index in middle panel.}
    \label{fig:SEF}
\end{figure}

According to the above lemma, if VNs corresponding to the last $\tlast$ rows from the bottom in the leftmost column of FG $\tFG_{\tNmm}$ are all frozen, then no stopping set in $\tSS^{\tImm}$ can have a VN from the last $\tlast$ rows in the rightmost column of FG $\tFG_{\tNmm}$. This property helps puncture the polar code while keeping the undecodable threshold constant, which improves the performance of the system specific metrics since i)  the VNs in $\tlastVN^{\tlast}_{\tNmm}[\log \tNmm + 1]$ do not need to be sampled thus improving the probability of failure; ii) the VNs in $\tlastVN^{\tlast}_{\tNmm}[\log \tNmm + 1]$ do not need to be dispersed to the oracle nodes thus reducing the communication cost. We formally calculate the undecodable threshold of the SEF algorithm in Lemma \ref{lemma:undecodable_sef}.
\deb{Lemma \ref{lemma:last_few_frozen} also allows us to reduce the size of FG $\tFG_{\tNmm}$ by noting that all the VNs in the last $\tlast$ rows of $\tFG_{\tNmm}$ are zero symbols and hence can be pruned along with their associated edges.}
This removal will give us the FG $\tFG_{\tNmm - \tlast}$ of a polar code of length $\tNmm - \tlast$. We leverage this property to design polar codes of lengths that are not powers of two. An example of FG $\tFG_{6}$ is shown in Fig. \ref{fig:SEF}. The detailed SEF algorithm to select the frozen index set $\tF$ of an $(N,k)$ polar code where $N$ is not necessarily a power of two, is provided in Algorithm \ref{alg:SEF}. The algorithm is used to design polar codes for all the PCMT layers.

\begin{algorithm}[t]
{\fontsize{11pt}{13.2pt}\selectfont 
\caption{SEF Algorithm}\label{alg:SEF}
\begin{algorithmic}[1]
\State \textbf{Inputs:} $N$, $k$ \textbf{Output:} $N_{\tname}$, $\tFG_{N_{\tname}}$, $\tF_{\tname}$ %
\State \textbf{Initialize:} $\tNm = 2^{\lceil \log N \rceil}$, polar FG $\tFG_{\tNm}$, $\tlast_{1} = \tNm - N$, $i = N$

\State $t_N$ = $\ttT_{\tNm}$ with last $\tlast_{1}$ entries removed

\State $\tF =  \{e \;|\; \closedN{e}{N} ,\;  t_{N}(e) < \min(t_{N}; N - k+1)\}$

\While{$\vert \tF \vert < N- k$} 
\If{$i \not\in \tF$} 
$\tF  = \tF \cup i$ \textbf{end if};
 $i = i-1$
\EndIf
\EndWhile

\State $\tlast_2 = \max\left(\{\tlast \; \vert \;[N - \tlast +1, N]  \subset \tF\}\right)$; $N_{\tname}  = N - \tlast_2$

\State $\tFG_{N_{\tname}}$ =  FG obtained by removing all VNs in $\{\tlastVN^{\tlast_{1}+\tlast_{2}}_{\tNm}[m]\;\vert\; \closedN{m}{\log \tNm+1}\}$ and their connected edges from $\tFG_{\tNm}$ (also remove any CNs that have no connected edges);  $\tF_{\tname} = \{e \;|\; e \in \tF, e \leq N_{\tname}\}$

\end{algorithmic}
}
\end{algorithm}

The SEF algorithm takes as input information length $k$ and target code length $N$ (for a target rate of $R = \frac{k}{N}$). The outputs of the algorithm  are the actual code length $N_{\tname}$ where $N_{\tname} \leq N$, FG $\tFG_{N_{\tname}}$ to be used for the PCMT construction where $\tFG_{N_{\tname}}$ has $N_{\tname}$ coded symbols, and the frozen index set $\tF_{\tname}$ such that $\vert \tF_{\tname}\vert = N_{\tname} - k$. The actual rate of the code is $\frac{k}{N_{\tname}} \geq R$.

In the SEF algorithm, $\tNm$ denotes the smallest power of two larger than $N$. We derive the FG for the $(N,k)$ polar code from the FG $\tFG_{\tNm}$.
The vector $\ttT_{\tNm}$ stores the stopping tree sizes for VNs $v_{\tVNi}, \closedN{\tVNi}{\tNm}$ in the FG $\tFG_{\tNm}$ (these are the VNs in the leftmost column of the FG).
We start the algorithm by implicitly removing the  last $\tlast_1 = \tNm - N$ rows from FG $\tFG_{\tNm}$ to obtain a FG with code length $N$ (i.e., $\tFG_{N}$) where the vector $t_N$ stores the stopping tree sizes of the corresponding VNs in the leftmost column of the FG (step 3). Then
in steps 4-6, we select the frozen index set $\tF$  that contains the indices in $[N]$ to be frozen for the FG $\tFG_{N}$ such that $\vert \tF \vert = N - k$ (output $\tF_{\tname}$ is derived from $\tF$). For the selection of $\tF$, we first select all the indices $e$ in $[N]$ such that the VNs $v_{e}$ have their stopping tree sizes less than $\min(t_N; N-k+1)$ (step 4). 
Then, the remaining indices in $\tF$ (so that $\vert \tF \vert = N - k$) are selected as the VN indices from the bottom row of FG $\tFG_{N}$ that are not already present in $\tF$ (steps 5-6). The variable $\tlast_2$ (step 7) represents the largest  number $\tlast$ such that the last $\delta$ rows from the bottom of FG $\tFG_{N}$ are frozen.
Thus, $\tlast_2$ represents the rows of FG $\tFG_{N}$, the VNs corresponding to which can be removed from the FG $\tFG_{N}$ without affecting the undecodable threshold. We achieve the removal in step 8 that gives us the output FG $\tFG_{N_{\tname}}$. The removal also results in $N_{\tname} = N - \tlast_2$ (step 7).  The output frozen index set $\tF_{\tname}$ is keeping the indices from $\tF$ that are less than or equal to $N_{\tname}$ (step 8). 
An example of the application of the SEF algorithm is provided in Fig. \ref{fig:SEF}.  Let $\tNm = 2^{\lceil \log N \rceil}$. For the FG $\tFG_{N_{\tname}}$, we denote the total number of VNs in the factor graph by \tptotVN$(\tFG_{N_{\tname}})$.

It is important to note that in the SEF algorithm, we freeze the bottom rows of FG $\tFG_{\tNm}$ which allows us to completely prune the VNs and CNs in these rows from the FG. However, for conventional channels (e.g. the BEC) this type of pruning is not possible since the last few rows contain the most reliable VNs and are rarely frozen \cite{PolarCodesErikan}. 
In the next section, we explain how the SEF algorithm is used for the PCMT construction provided in Section \ref{sec:pcmt-construction-indexing}. 

\vspace{-0.4cm}
\begin{remark}
    In the SEF algorithm, we first freeze all rows whose corresponding VNs have stopping tree sizes less than $\min(t_{N};N - k+1)$. Then the remaining indices to freeze are chosen from the bottom of the FG. Alternatively, we can first freeze all rows with stopping tree sizes less than $\tau$ for some $\tau < \min(t_{N};N - k+1)$ and then freeze the remaining indices from the bottom of the FG. However, since the stopping tree sizes are a power of 2 \cite{Polar-SStree-TCOM}, it is easy to see that the alternative approach cannot provide a larger ratio $\frac{\talpha}{N_{\tname}}$. Thus, the SEF algorithm optimizes to get the smallest probability of failure.  
    \bluetext{For convenience, we use the same SEF algorithm for the DA oracle in side blockchains.}
\end{remark}
\vspace{-0.4cm}

\vspace{-0.6cm}
\subsection{Building the PCMT using the SEF Algorithm}\label{sec:SEF_PCMT}
\begin{figure}[t]
    \centering
\begin{subfigure}{0.6\linewidth}
\begin{minipage}{0.99\linewidth}
\begin{tikzpicture}
  \node (img) {\includegraphics[scale=0.36]{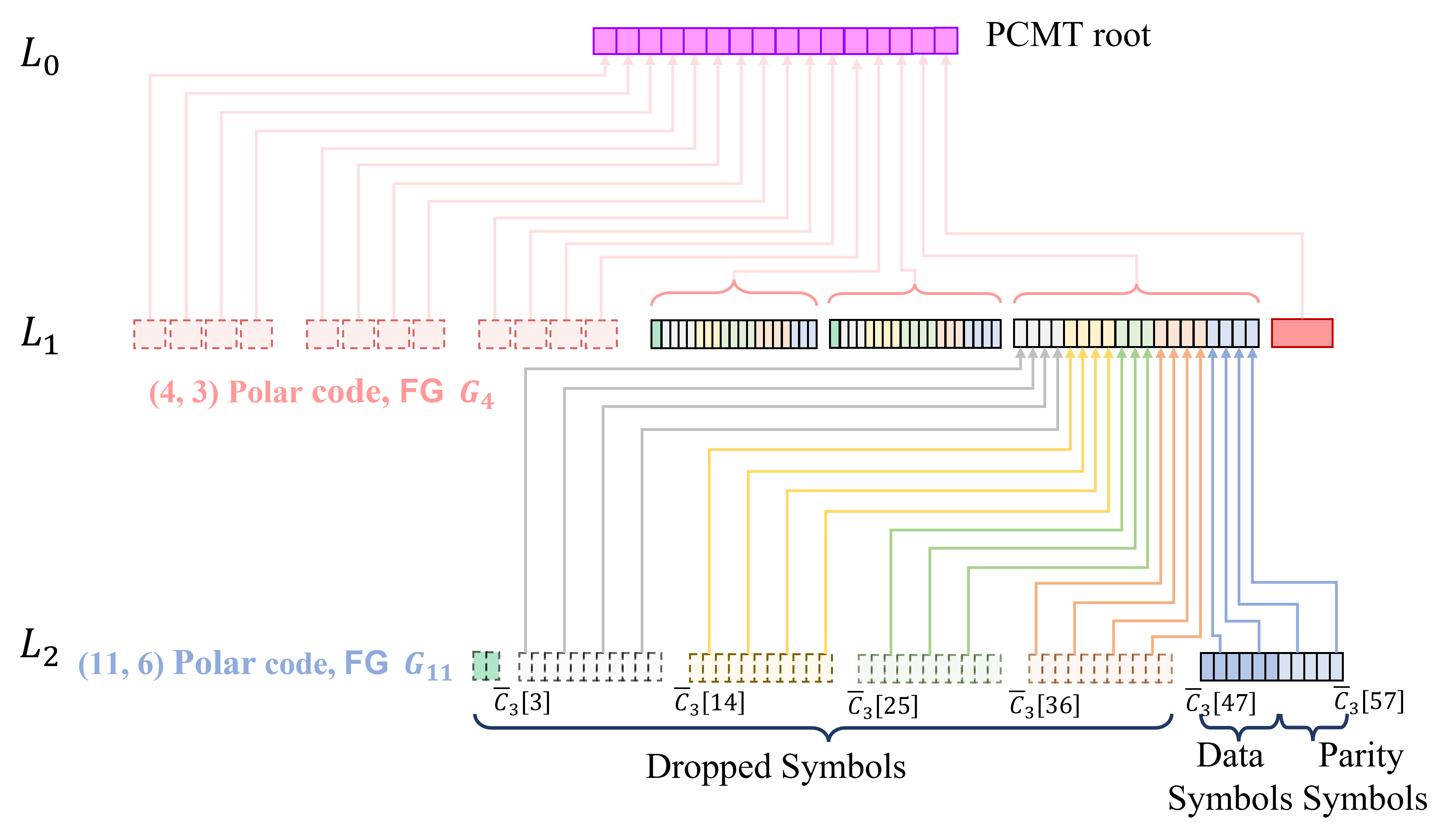}};
 \end{tikzpicture}
 \end{minipage}
\end{subfigure}%
\begin{subfigure}{0.4\linewidth}
\begin{minipage}{0.99\linewidth}
\begin{tikzpicture}
  \node (img) {\includegraphics[scale=0.30]{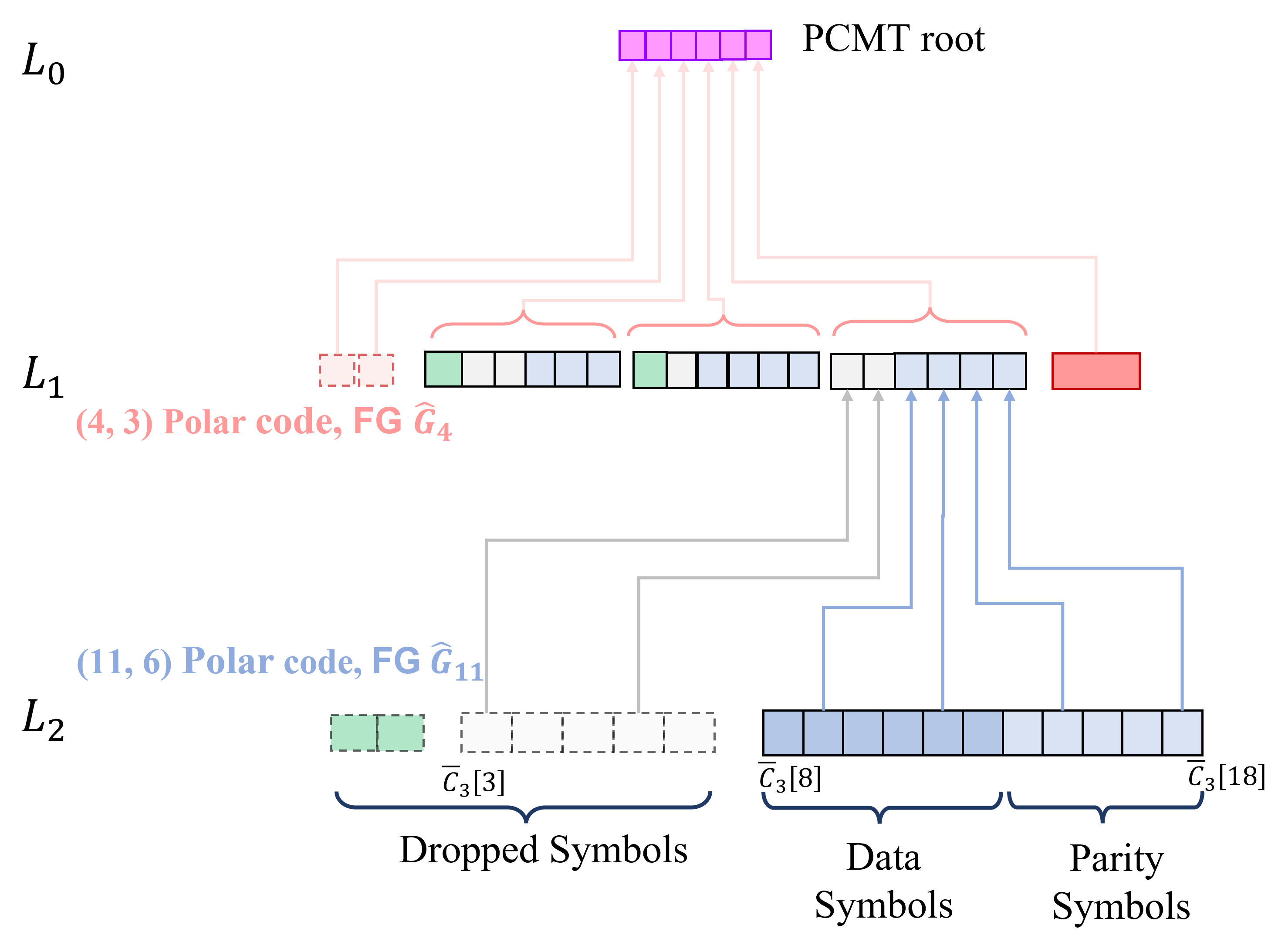}};
 \end{tikzpicture}
 \end{minipage}
    \end{subfigure}%
    \vspace{-14pt}
    \caption{Left panel: PCMT $\tCMT = (K = 6, R = 0.5, q = 2,  l = 2)$ constructed using the FGs output by the SEF algorithm. Here, $N_{\tname,2} = 11$ and $N_{\tname,1} = 4$; Right panel: PCMT $\tCMT = (K = 6, R = 0.5, q = 2,  l = 2)$ constructed using the pruned FGs output by the pruning algorithm. Note that the $N_{\tname}$ values for different layers are the same as that of the left pannel. In both panels, the zero padded VNs are shown in green.}
    \label{fig:SEF_PCMT_pruning}
\end{figure}

\vspace{-0.15cm}
\deb{Consider the construction of a  PCMT $\tCMT = (K, R, q, l)$. 
\debb{For the general layer $L_j$ with information length $k$}, we use the SEF algorithm with inputs $(\frac{k}{R},k)$ to get the outputs $(N_{\tname}, \tFG_{N_{\tname}}, \tF_{\tname})$ which are used for the construction of the PCMT as described in Section \ref{sec:pcmt-construction-indexing}. In particular, the set of VNs $\mathcal{V} = \{v_{\tVNi} \;\vert\; \closedN{\tVNi}{\tptotVN(\tFG_{N_{\tname}})} \}$ of FG $\tFG_{N_{\tname}}$ is used in \tparent$()$ procedure of the PCMT. 
Recall that the \tparent$()$ procedure groups the hashes of the symbols of the general layer into $\symaccent{k}$ symbols (that form the information symbols of the parent layer). Thus to make an even partition into $\symaccent{k}$ groups, we zero pad the set of VNs $\mathcal{V}$. Let $\symaccent{q} = \lceil \frac{ \tptotVN(\tFG_{N_{\tname}})}{\symaccent{k}} \rceil$. \debb{We zero pad $\symaccent{q} \cdot \symaccent{k} - \tptotVN(\tFG_{N_{\tname}})$ VNs to $\mathcal{V}$ and set $\ttotVN  = \symaccent{q}\cdot \symaccent{k}$ and dropped index $\tdropindex = \ttotVN - N_{\tname}$.} Using these parameters, we construct the PCMT as explained in Section \ref{sec:pcmt-construction-indexing}. An example of the PCMT construction using the above procedure is shown in Fig. \ref{fig:SEF_PCMT_pruning} left panel.  We have the following lemma for a PCMT constructed using SEF polar codes. }

\vspace{-1.3cm}
\begin{lemma}\label{lemma:undecodable_sef}
For the general layer $L_j$ of a PCMT $\tCMT = (K, R, q, l)$ constructed using the SEF algorithm as explained above, where $\tI_{\tname} =[N_{\tname}] \setminus  \tF_{\tname}$, the undecodable threshold is $\talpha = \min_{i \in \tA_{\tname}} \ttT_{\frac{k}{R}}(i)$. Additionally, the threshold complexity \greentext{(see Section \ref{sec:performance_metrics})} is the complexity of the SEF algorithm (applied on all layers of the PCMT) and is $\sum_{j=1}^{l}O(\frac{K}{(qR)^{l-j}})$.
\end{lemma}
\vspace{-0.5cm}

Note that due to step 4 of the SEF algorithm, $\min_{i \in \tI} \ttT_{\frac{k}{R}}(i) \geq \min(t_N; N-k+1)$ and 
hence the undecodable threshold of the SEF algorithm is always as big as that of the NF algorithm.
In Section \ref{sec:sims}, we demonstrate the performance of the PCMT built using SEF polar codes with respect to the performance metrics mentioned in  Section \ref{sec:performance_metrics}. In the next lemma, we analyze the asymptotic performance of the PCMT constructed using the SEF algorithm for large block sizes and compare it to regular Merkle trees that do not utilize channel coding.

\vspace{-0.4cm}
\begin{lemma}\label{lemma:scaling-law}
    Consider a PCMT $\tCMT = (K, R, q, l)$ built using the SEF algorithm. Let the block size and the hash size be $b$ and $y$, respectively.
    Let $P^{p}_f(s)$ and $P^{u}_f(s)$ be the probability of failure  to detect a DA attack on the base layer of the PCMT and uncoded Merkle tree, respectively. 
    \browntext{The sample size $s$ is selected in both cases to be the largest while ensuring that the total download size of the $s$ samples (i.e., the size of $s$ base layer symbols and their Merkle proofs) is 
    at most $\frac{b}{D_r}$, where $D_r > 1$ is a fixed constant.}     
    For \greentext{$b \gg y K$} (case of large block sizes), $\ln \left (\frac{ P^{u}_f(s)}{ P^{p}_f(s)}\right) = \Omega(\sqrt{K})$.  
\end{lemma}
\vspace{-0.4cm}

According to Lemma \ref{lemma:scaling-law},
the PCMT has an exponentially better probability of failure compared
to the uncoded Merkle tree and the exponent improves with the information length. \browntext{Note that the above asymptotic improvement is true for all rates $R$ used in the PCMT.}
Additionally, note that it is difficult to provide a similar scaling law for deterministic LDPC codes due to the NP-hardness of calculating the undecodable threshold $\talpha$ for LDPC codes. 
In the next section, we provide techniques to further prune the FG of polar codes (compared to the pruning in the SEF algorithm) to further improve the performance of the metrics mentioned in  Section \ref{sec:performance_metrics}.

 \vspace{-0.6cm}
 \section{Pruning the Factor Graph of Polar codes for the PCMT construction}\label{sec:pruned_construction}
 \vspace{-0.2cm}
 For inputs $(\frac{k}{R}, k)$ to the SEF algorithm, let the output be the FG $\tFG_{N_{\tname}}$ and let $\tNm = 2^{\lceil \log (k/R) \rceil}$. 
 The FG $\tFG_{N_{\tname}}$ contains $N_{\tname}(\log \tNm+1)$ VNs. In the PCMT, the hashes of all these VNs are stored in the parent layer. Let us compare this \greentext{construction} to a CMT where an $(\frac{k}{R}, k)$ channel code is used in the general layer. In this case, the hashes of only $\frac{k}{R}$ VNs are stored in the parent layer. More hashes in the case of a PCMT imply that each symbol in a PCMT is of a larger size than the corresponding symbol of a CMT with the same parameters $(K, R, q, l)$. 
 Large symbol sizes increase the Merkle proof sizes which can increase the IC proof sizes, limit the total number of samples for a fixed download size budget increasing the probability of failure, and increase the communication cost in the case of DA oracles. A higher number of VNs in the PCMT FG also results in higher decoding complexity. Thus in this section, we provide a pruning algorithm to remove VNs from the FG of polar codes so as to reduce the number of  VNs hashed together in the PCMT (to reduce the Merkle proof sizes) and the decoding complexity.

 We now explain the pruning algorithm. It takes as input the FG $\tFG_{N}$ of a polar code, where $N$ is the number of coded symbols (also the number of VNs in the rightmost column of the FG) and the frozen index set $\tF$. We call the $N$ VNs in the rightmost column of the FG as the non-dropped VNs (since they are the VNs that are actually stored in the PCMT and are not dropped as per Section \ref{sec:pcmt-construction-indexing}). The remaining VNs  are called the dropped VNs. We keep track of which VNs are the non-dropped VNs in the algorithm. 
 The output of the algorithm is the pruned FG $\tpFG_{N}$ (note that the number of coded symbols remains the same, hence subscript $N$). 
Our algorithm is similar to \cite{sparsepolar} which was designed for the belief propagation decoder. However, our algorithm is designed for the peeling decoder and ensures that the input and output FGs have the same decoding output under a peeling decoder. This property also ensures that the undecodable threshold does not change due to pruning. Additionally, unlike \cite{sparsepolar}, the maximum degree of CNs is not increased by our pruning algorithm which ensures that the IC proof size does not increase due to pruning.  
 Note that, in our case, a CN of degree $d$ connected to coded symbols $\tsymbol_1, \tsymbol_2, \ldots, \tsymbol_d$ satisfies the parity check constraint $\sum_{i = 1}^{d} \tsymbol_i = 0$. We remove VNs and CNs from the FG while ensuring that the CN constraints are not affected.

 Our algorithm has the following main components:
 \subsubsection{Frozen VNs} In the encoding process using polar codes as mentioned in Section \ref{sec:polar_prelims}, the frozen VNs are set to zero symbols. Thus, the frozen VNs do not affect the CN constraints and can be removed from the FG. We call the procedure that acts on the FG $\tFG_N$ and removes all the frozen VNs as \tfrozenprune$(\tFG_{N})$. 
 \subsubsection{Degree 1 CNs} Due to the removal of VNs, degree 1 CNs can be formed in the FG. The parity check constraint of a degree 1 CN is satisfied iff the connected VN is a zero symbol. Thus, the degree 1 CN and the connected VN can be removed from the FG.  We call the procedure that removes all the degree 1 CN and its connected VN from $\tFG_{N}$ as \tremovedegreoneCN$(\tFG_{N})$. 
 \subsubsection{Degree 2 CNs} The two VNs that are connected to a degree 2 CN must have the same value for the parity check constraint to be satisfied. Thus, these two connected VNs can be merged into a single VN and the degree 2 CN can be removed from the FG. Here, we distinguish the following cases based on the type of VNs connected to the degree 2 CN. The first case is when the connected VNs are dropped VNs. In this case, we merge the two VNs and drop the degree 2 CN. The new merged VN takes place (for VN indexing purposes) of the VN with the smaller VN index in the FG. The second case is when one of the connected VN is a dropped VN and the other one is a non-dropped VN. In this case, we again merge the two VNs and remove the degree 2 CN. However, the newly merged VN takes place of the non-dropped VN in the FG and is now a non-dropped VN. The third case is when the two connected VNs are non-dropped VNs. In this case, we do not perform any action, i.e., the two VNs are not merged and the degree 2 CN is not dropped. This step is to ensure that the number of non-dropped VNs i.e., the number of coded symbols of the polar code remains fixed.   We call the procedure that performs the above actions on FG $\tFG_{N}$ as \tmergedegretwoCN$(\tFG_{N})$. 
 \subsubsection{Empty CNs} CNs that are not connected to any VNs get created due to the above operations which we remove from the FG. We call the corresponding procedure \tremoveemptyCN$(\tFG_{N})$.
  
 The pruning algorithm is as follows. We first perform step 1 on $\tFG_{N}$. We then repeat steps 2,3, and 4 until the size of the FG does not change anymore, at which point we terminate and output the pruned FG as $\tpFG_{N}$. The size of the FG is defined as the sum of the number of VNs and CNs in the graph. Note that the complexity of the pruning algorithm is at most $O(N\log N)$ since each VN is touched at most once by the algorithm.  

An example of the output of the pruning algorithm with input $\tFG_{4}$ (which is obtained from Fig. \ref{fig:SEF} left panel by removing VNs from the last two rows) is shown in Fig. \ref{fig:SEF} right panel where the non-dropped VNs are marked in blue. 
The above pruning algorithm ensures that the FG $\tFG_{N}$ and the pruned FG $\tpFG_{N}$ are equivalent in terms of the peeling decoder, i.e., given the values of the $N$ coded symbols (some of which may be erased due to a DA attack), the peeling decoder produces the same decoding output for VNs in FG $\tpFG_{N}$ as their corresponding values  in FG $\tFG_{N}$. This property also implies that the FGs $\tFG_{N}$ and $\tpFG_{N}$ have the same undecodable threshold.   Let the total number of VNs in the pruned FG $\tpFG_{N}$ be \tptotVN$(\tpFG_{N})$. Also, let the VNs in $\tpFG_{N}$ be indexed in ascending order according to the indices of the VNs in the original FG $\tFG_{N}$. 
 Fig. \ref{fig:SEF} right panel shows the indexing of the VNs in the output FG $\tpFG_{4}$. According to the indexing, the VNs with the $N$ (here $N = 4$ in the example) largest indices are the non-dropped VNs. The remaining VNs are the dropped VNs. The VNs in FG $\tpFG_{N}$ are $\{v_{\tVNi} \;\vert\; \closedN{\tVNi}{\tptotVN(\tpFG_{N})} \}$ where the VNs $\{v_{\tVNi} \;\vert\; \closed{\tVNi}{\tptotVN(\tpFG_{N}) - N +1}{\tptotVN(\tpFG_{N})} \}$ are the non-dropped VNs and contain the coded symbols. Note that the pruning algorithm does not increase the maximum degree of the CNs in the factor graph, i.e., the maximum CN degree of a CN in $\tpFG_{N}$ is still 3. Next, we explain how we use the polar code FGs output by the pruning algorithm for the PCMT construction.
 
 For the general layer $L_j$ and SEF algorithm output $\tFG_{N_{\tname}}$, we first use the pruning algorithm  with input $\tFG_{N_{\tname}}$ to get the FG $\tpFG_{N_{\tname}}$. Now for the PCMT construction, we use the procedure mentioned in Section \ref{sec:SEF_PCMT} with the pruned FG $\tpFG_{N_{\tname}}$.
 An example of a PCMT built using the pruned FGs is shown in Fig. \ref{fig:SEF_PCMT_pruning} right panel.  We call the PCMT built using the pruned FGs as the PrPCMT.
Note that the asymptotic performance of the PrPCMT is the same as the PCMT in Lemma \ref{lemma:scaling-law} since the PrPCMT performance is upper bounded by the PCMT performance.

 \setlength{\extrarowheight}{3pt}
\setlength{\tabcolsep}{2.8pt}
  \begin{table*}[t]
  \caption{\footnotesize Comparison of various performance metrics of 2D-RS codes, an LCMT, and a PCMT/PrPCMT. The LCMT and PCMT/PrPCMT have the same $(K, R, q, l)$ parameters. The maximum degree of the CNs  in the LDPC codes and polar FG used on the base layer of the CMTs are $d_c$ and $d_p = 3$, respectively. The size of the transaction block is $b$. 2D-RS has $K$ data symbols and $\lceil \log \sqrt{N_l}\rceil$ layers in the Merkle tree where $N_l = \frac{K}{R}$.  
  For a PCMT/PrPCMT, $\ttotVN_j$ is the total number of VNs in the FG used to encode layer $L_j$.  Note that the system specific performance depends on the single sample download size and the undecodable threshold $\talpha$.
    }\label{figtable:Performance_comp}
    \vspace{-10pt}
    \centering
\resizebox{!}{4.2cm}{
\begin{tabular}[b]{| L | M | N | O|}

\hline
& 2D-RS & LCMT  & PCMT, PrPCMT  \\
\hline
Root size & $2y\big\lceil\sqrt{N_l}\;\big\rceil$ &$y N_1$ & $ y\ttotVN_1$\\
\hline
Single sample download size $X$ & $\frac{b}{K} + y\big\lceil \log \sqrt{N_l} \;\big\rceil$ & $\frac{b}{K} + y(2q-1)(l-1)$ & $\frac{b}{K} + y\sum_{j=1}^{l-1}\left(2\lceil \frac{\ttotVN_{j+1}}{k_j}\rceil-1\right)$\\
\hline
IC proof size &  $(\frac{b}{K} + y\big\lceil \log \sqrt{N_l} \;\big\rceil)\big\lceil \sqrt{K}\;\big\rceil$& $\frac{(d_c-1)b}{K} + d_cy(q-1)(l-1)$ & $ \frac{(d_p-1)b}{K} +  d_py\sum_{j=1}^{l-1}\left(\lceil \frac{\ttotVN_{j+1}}{k_j}\rceil-1\right)$\\
\hline
Decoding complexity & $O(N_l^{1.5})$ & $O(N_l)$& $O(\ttotVN_l) \leq O(N_l\lceil\log N_l\rceil)$\\
\hline
$\talpha$ & Analytical expression in \cite{dataAvailOrg} &NP-hard to compute & Lemma \ref{lemma:undecodable_sef}\\
\hline
Threshold complexity & $O(1)$ & NP-hard &$\sum_{j=1}^{l}O(\frac{K}{(qR)^{l-j}})$\\
\hline
\end{tabular}
} 
  \end{table*}

 \vspace{-0.8cm}
  \section{Simulation Results and Performance Comparison }\label{sec:sims}
\begin{figure*}[t]
    \centering
    \begin{minipage}{0.99\linewidth}
    \begin{subfigure}{0.5\linewidth}
\begin{minipage}{0.99\linewidth}
\centering
\begin{tikzpicture}
  \node (img)
  {\includegraphics[scale=0.25]{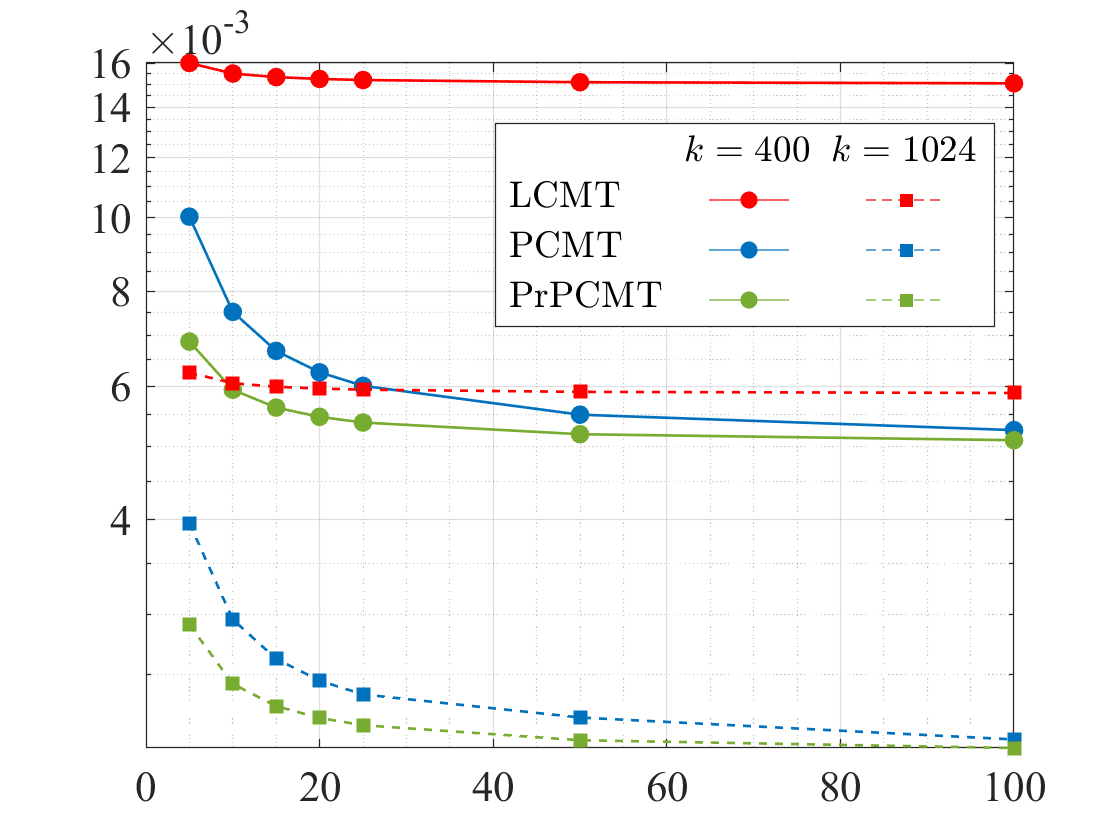}};
  \node[below=of img, node distance=0cm, yshift=1.4cm,font=\color{black}] {Data symbol  size $c$ (in KB)};
  \node[left=of img, node distance=0cm, rotate=90, anchor=center,yshift=-1.35cm,font=\color{black}] {IC proof size / block size};
 \end{tikzpicture}
 \end{minipage}
    \end{subfigure}%
\begin{subfigure}{0.5\linewidth}
\begin{minipage}{0.99\linewidth}
\centering
\begin{tikzpicture}
  \node (img)
  {\includegraphics[scale=0.25]{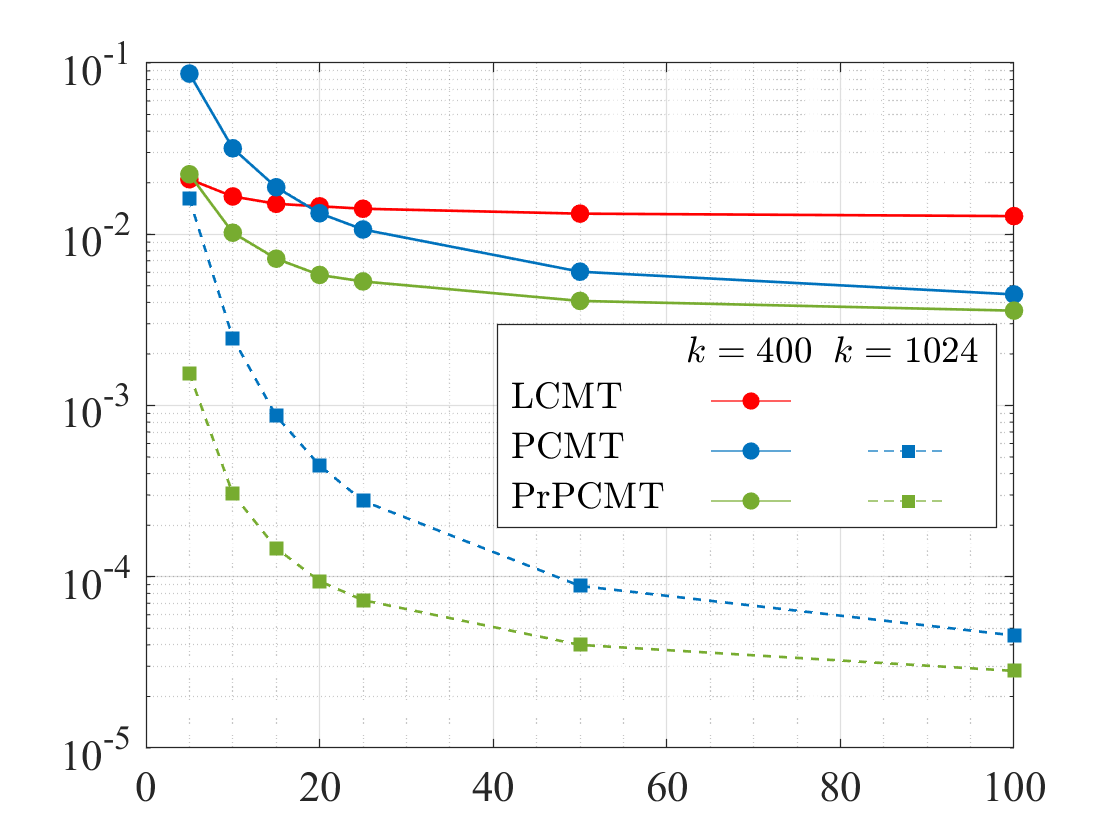}};
  \node[below=of img, node distance=0cm, yshift=1.4cm,font=\color{black}] {Data symbol size $c$ (in KB)};
  \node[left=of img, node distance=0cm, rotate=90, anchor=center,yshift=-1.2cm,font=\color{black}] {$P_f(s)$};
 \end{tikzpicture}
 \end{minipage}
\end{subfigure}
\vspace{-8pt}
 \end{minipage}
\begin{subfigure}{0.5\linewidth}
\begin{minipage}{0.99\linewidth}
\centering
\begin{tikzpicture}
  \node (img)
  {\includegraphics[scale=0.25]{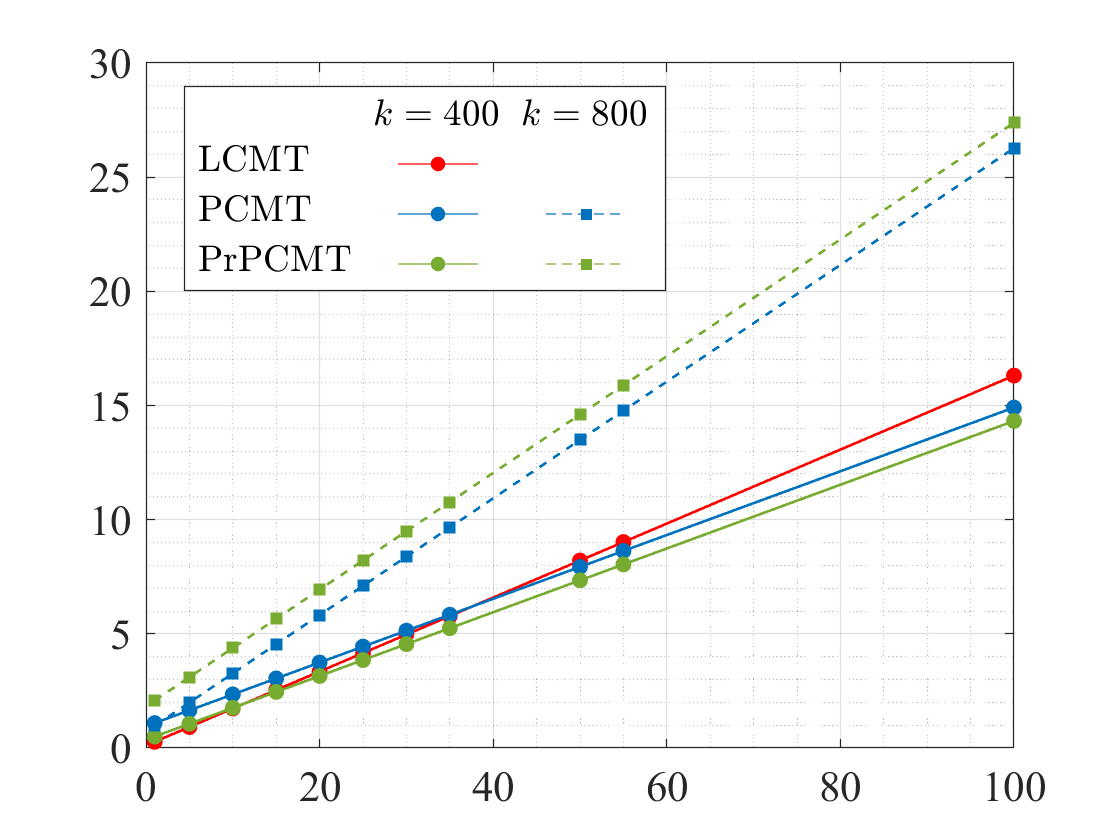}};
  \node[below=of img, node distance=0cm, yshift=1.4cm,font=\color{black}] {Data symbol size $c$ (in KB)};
  \node[left=of img, node distance=0cm, rotate=90, anchor=center,yshift=-1.35cm,font=\color{black}] {Communication cost (in GB)};
 \end{tikzpicture}
 \end{minipage}
\end{subfigure}%
\begin{subfigure}{0.5\linewidth}
\begin{minipage}{0.99\linewidth}
\centering
\begin{tikzpicture}
  \node (img)
  {\includegraphics[scale=0.25]{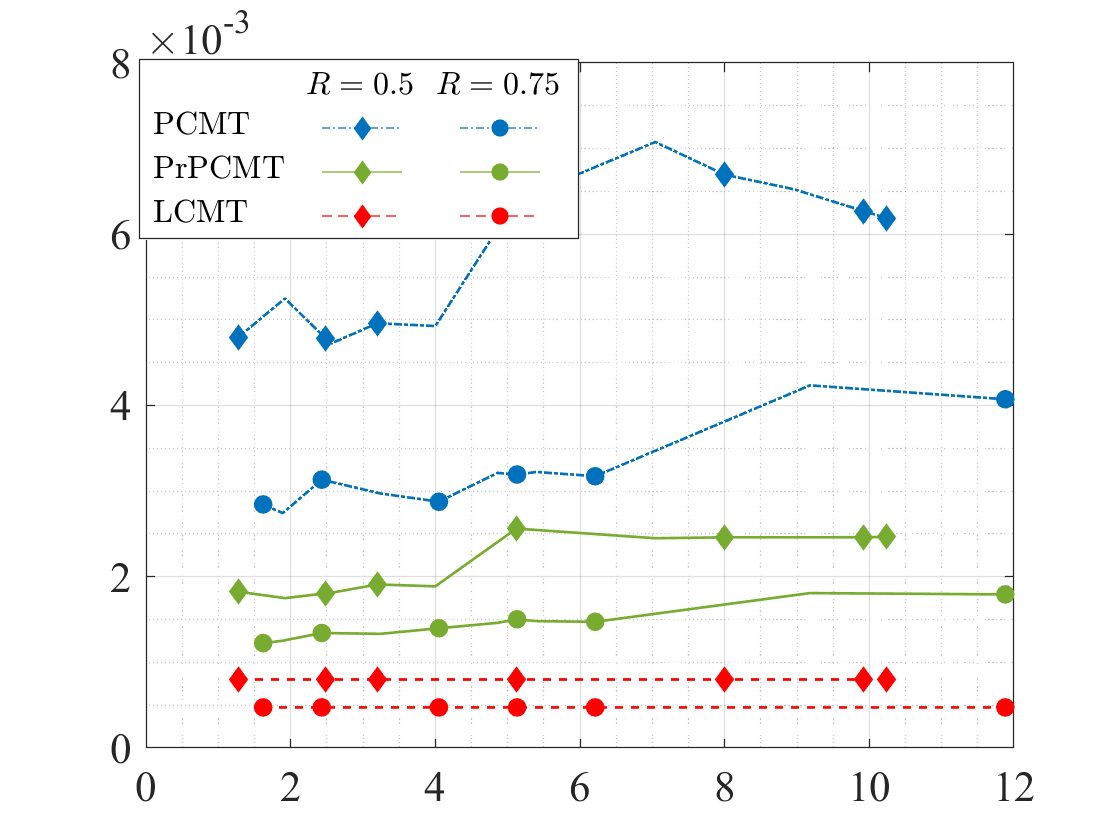}};
  \node[below=of img, node distance=0cm, yshift=1.4cm,font=\color{black}] {Block size (in MB)};
  \node[left=of img, node distance=0cm, rotate=90, anchor=center,yshift=-1.35cm,font=\color{black}] {CMT root size / block size};
 \end{tikzpicture}
 \end{minipage}
\end{subfigure}
     \vspace{-9pt}
    \caption{ Comparison of various CMT performance metrics for different coding methods.
     Top panels and bottom left panel use CMT parameters $\tCMT = (K, R = 0.5, q = 4, l = 4)$. Top left panel: IC proof size normalized by the block size for different data symbol sizes $c$. Top right panel: Probability of light node failure $P_f(s)$ for different data symbol sizes $c$. Bottom left panel: Communication cost associated with DA oracle for different data symbol sizes $c$. Bottom right panel: CMT root size normalized by the block size as the block size is varied. For rates $R = 0.5$ and $0.75$ we use $(q = 4,l = 4)$ and $(q = 4,l=3)$, respectively. }
    \label{fig:metrics_as_c_varies}
\end{figure*}
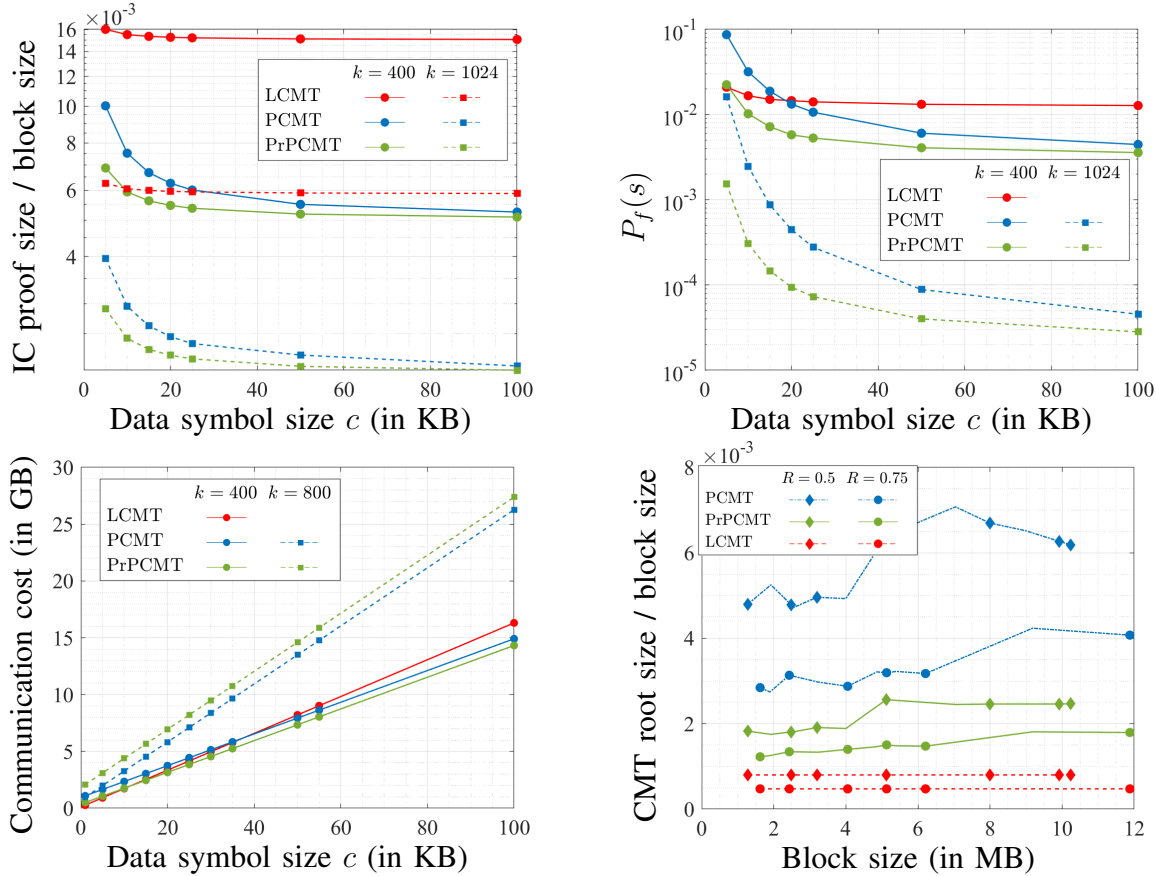

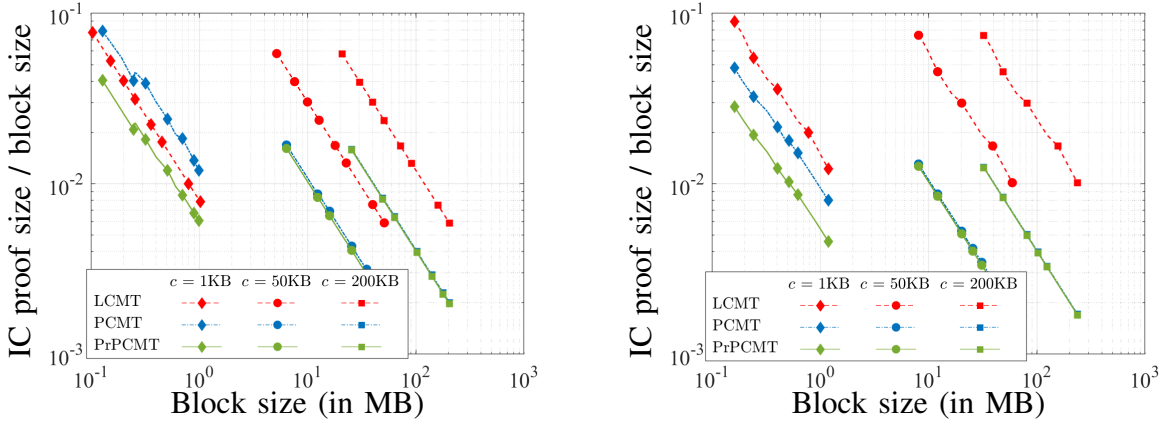
\begin{figure*}[t]
    \centering
    \begin{subfigure}{0.5\linewidth}
\begin{minipage}{0.99\linewidth}
\centering
\begin{tikzpicture}
  \node (img)
  {\includegraphics[scale=0.25]{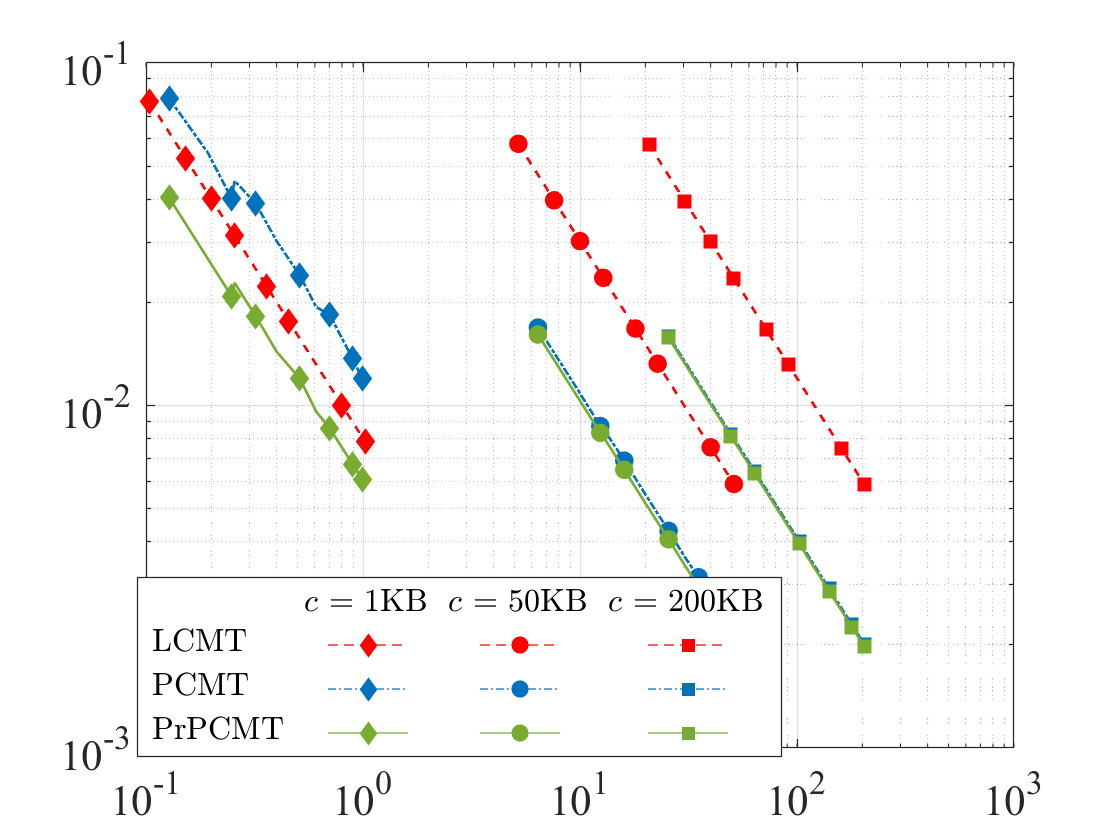}};
  \node[below=of img, node distance=0cm, yshift=1.4cm,font=\color{black}] {Block size  (in MB)};
  \node[left=of img, node distance=0cm, rotate=90, anchor=center,yshift=-1.2cm,font=\color{black}] {IC proof size / block size};
 \end{tikzpicture}
 \end{minipage}
    \end{subfigure}%
\begin{subfigure}{0.5\linewidth}
\begin{minipage}{0.99\linewidth}
\centering
\begin{tikzpicture}
  \node (img)
  {\includegraphics[scale=0.25]{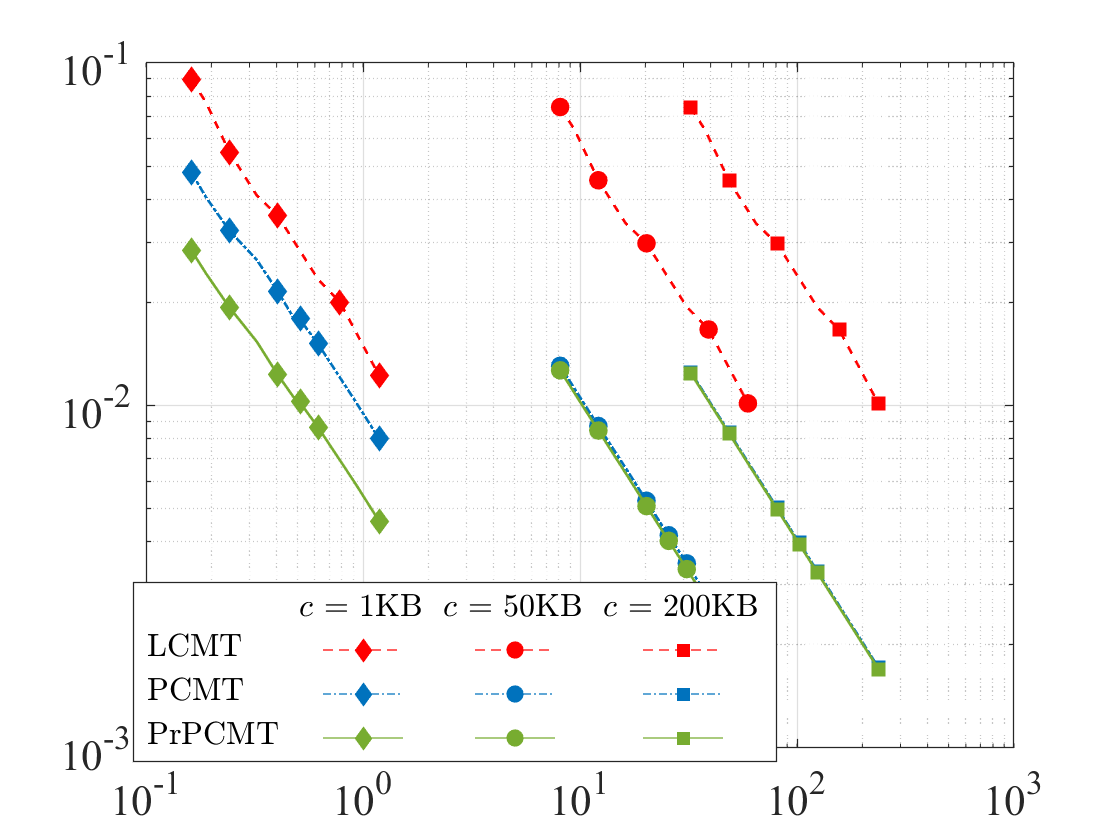}};
  \node[below=of img, node distance=0cm, yshift=1.4cm,font=\color{black}] {Block size  (in MB)};
  \node[left=of img, node distance=0cm, rotate=90, anchor=center,yshift=-1.2cm,font=\color{black}] {IC proof size / block size};
 \end{tikzpicture}
 \end{minipage}
\end{subfigure}
     \vspace{-10pt}
    \caption{\deb{IC proof size normalized} by block size as the block size is varied for different data symbol sizes $c$. Left panel: Rate $R = 0.5$, $\tCMT = (K, R = 0.5, q = 4, l = 4)$; Right panel: Rate $R = 0.75$, $\tCMT = (K, R = 0.75, q = 4, l = 3)$. }
    \label{fig:Ic_proof_size_vs_block_size}
     \vspace{-25pt}
\end{figure*}

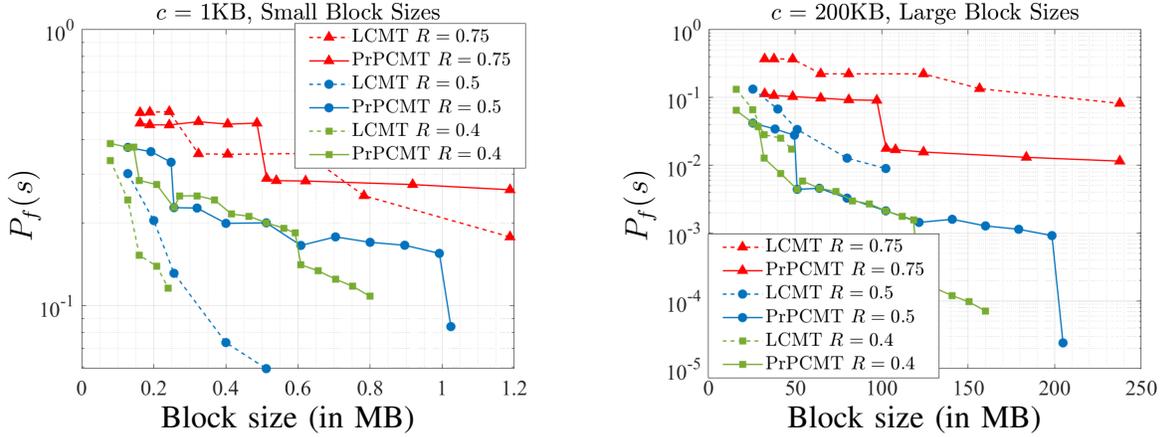
\begin{figure*}[t]
    \centering
    \begin{subfigure}{0.5\linewidth}
\begin{minipage}{0.99\linewidth}
\centering
\begin{tikzpicture}
  \node (img)
  {\includegraphics[scale=0.25]{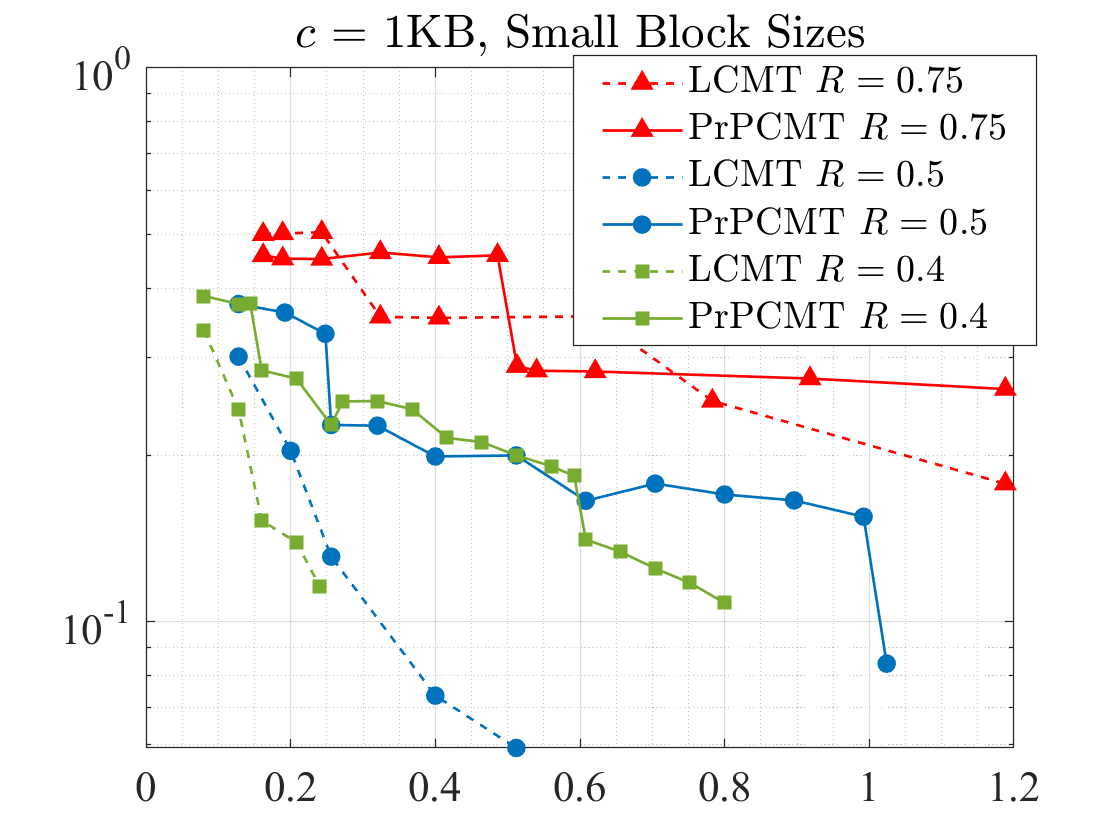}};
  \node[below=of img, node distance=0cm, yshift=1.4cm,font=\color{black}] {Block size  (in MB)};
  \node[left=of img, node distance=0cm, rotate=90, anchor=center,yshift=-1.35cm,font=\color{black}] {$P_f(s)$};
 \end{tikzpicture}
 \end{minipage}
    \end{subfigure}%
\begin{subfigure}{0.5\linewidth}
\begin{minipage}{0.99\linewidth}
\centering
\begin{tikzpicture}
  \node (img)
  {\includegraphics[scale=0.25]{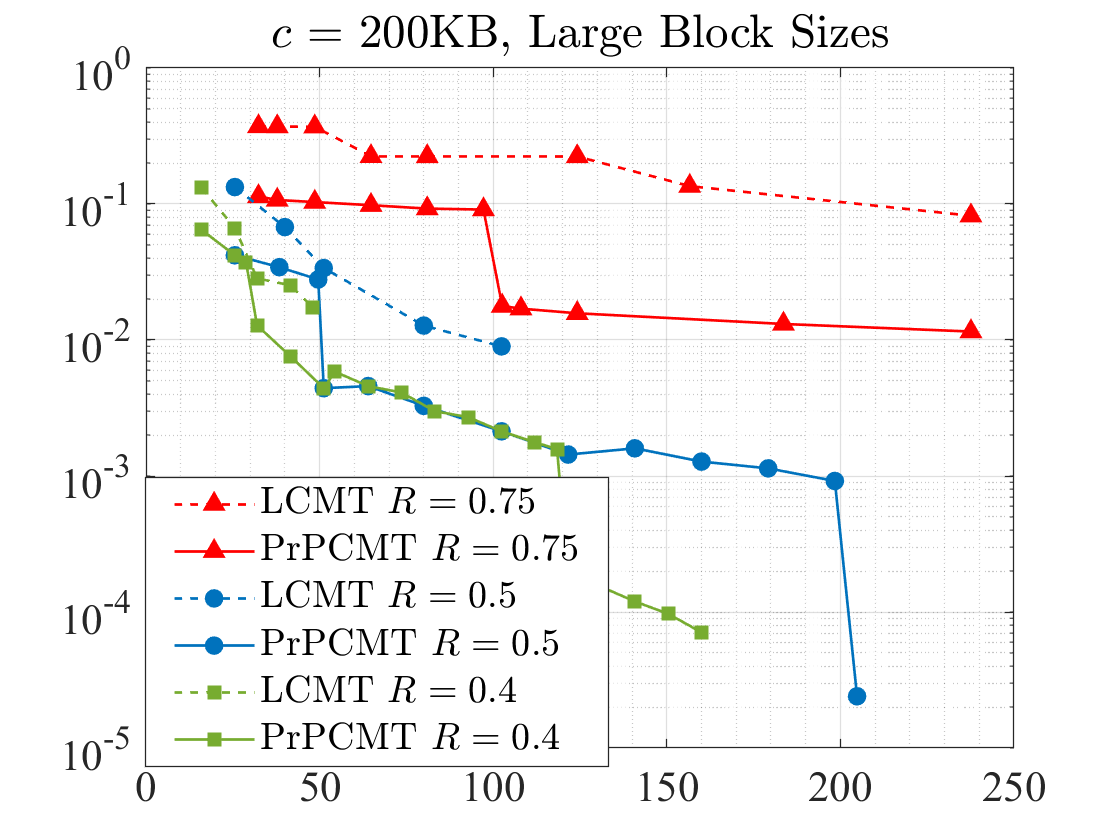}};
  \node[below=of img, node distance=0cm, yshift=1.4cm,font=\color{black}] {Block size  (in MB)};
  \node[left=of img, node distance=0cm, rotate=90, anchor=center,yshift=-1.2cm,font=\color{black}] {$P_f(s)$};
 \end{tikzpicture}
 \end{minipage}
\end{subfigure}
     \vspace{-10pt}
    \caption{ $P_f(s)$ vs. block size $b$ for the LCMT and PrPCMT. The two panels use $(R, q, l) = (0.4, 5, 4)$, $(0.5, 4, 4)$, and $(0.75, 4, 3)$ and a constant data symbol size c. Sample size $s$ for the PCMT and LCMT are selected such that the total sample download size is $\frac{b}{3}$. Left Panel: $c = 1$KB; Right Panel: $c = 200$KB.}
    \label{fig:P_f_vs_block_size}
\end{figure*}

\begin{figure*}[t]
    \centering
    \begin{subfigure}{0.5\linewidth}
\begin{minipage}{0.99\linewidth}
\centering
\begin{tikzpicture}
  \node (img)
  {\includegraphics[scale=0.25]{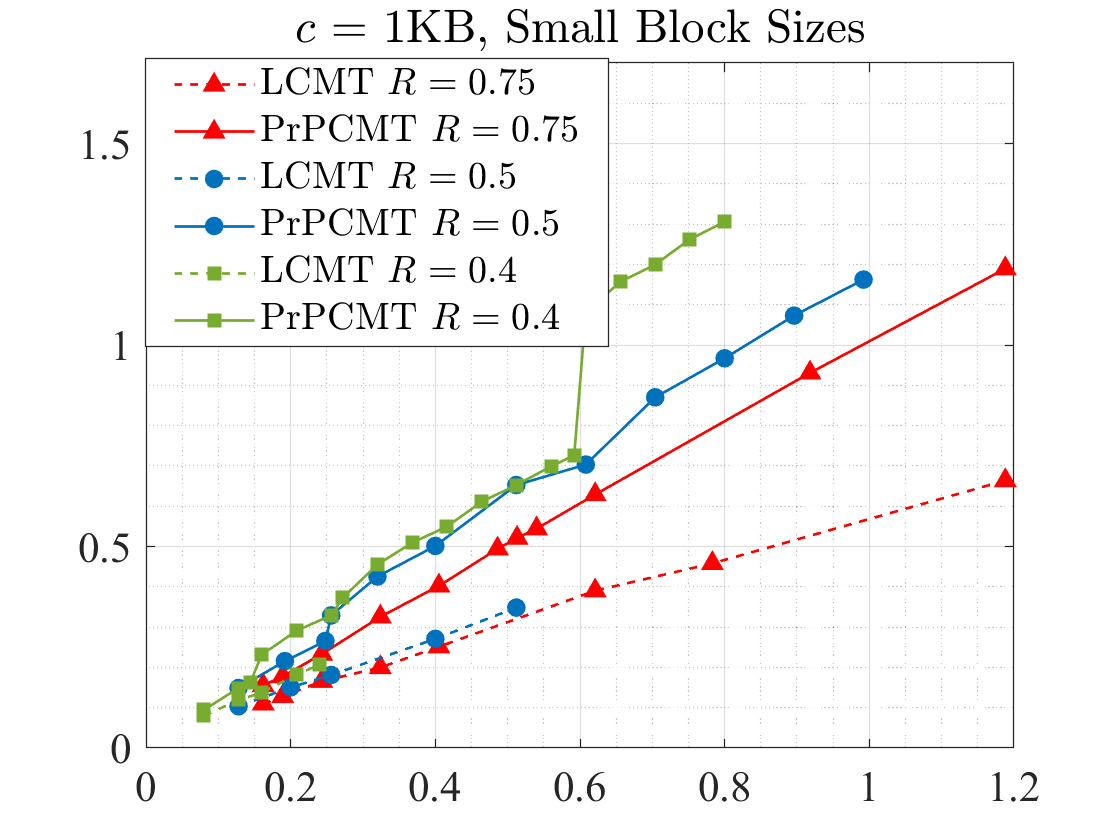}};
  \node[below=of img, node distance=0cm, yshift=1.4cm,font=\color{black}] {Block size  (in MB)};
  \node[left=of img, node distance=0cm, rotate=90, anchor=center,yshift=-1.35cm,font=\color{black}] {Communication cost (in GB)};
 \end{tikzpicture}
 \end{minipage}
    \end{subfigure}%
\begin{subfigure}{0.5\linewidth}
\begin{minipage}{0.99\linewidth}
\centering
\begin{tikzpicture}
  \node (img)
  {\includegraphics[scale=0.25]{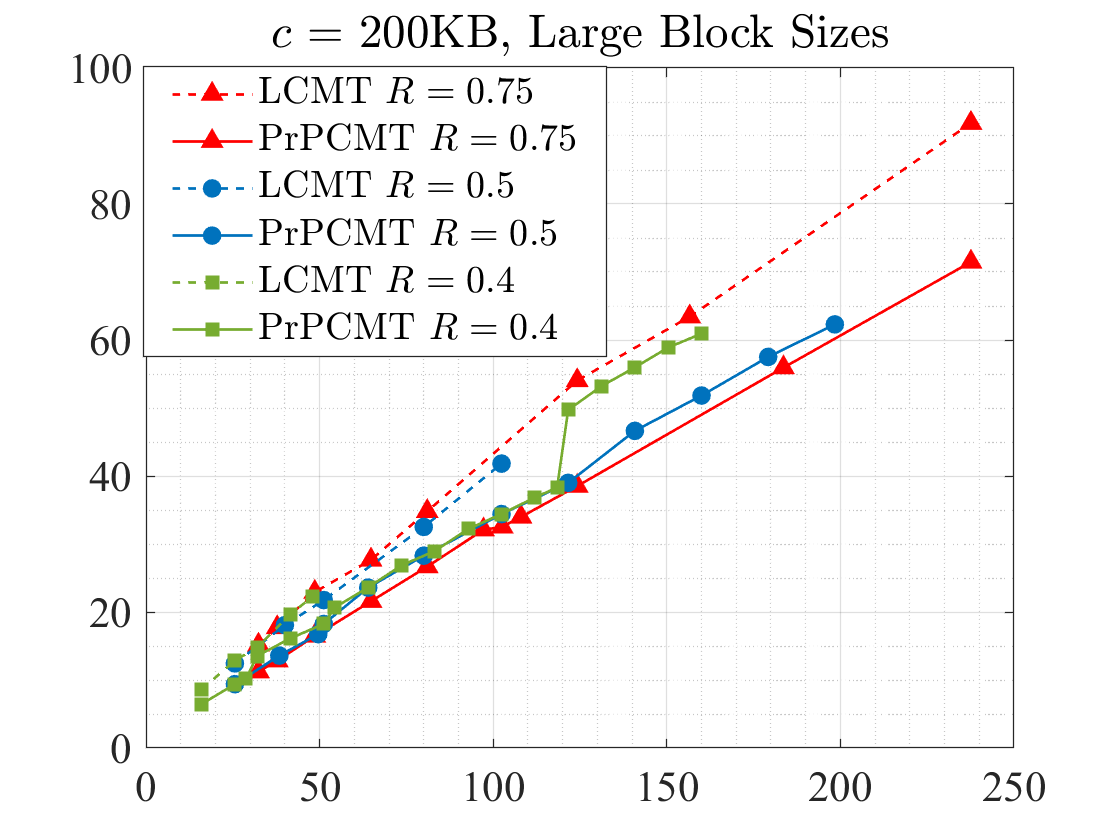}};
  \node[below=of img, node distance=0cm, yshift=1.4cm,font=\color{black}] {Block size  (in MB)};
  \node[left=of img, node distance=0cm, rotate=90, anchor=center,yshift=-1.35cm,font=\color{black}] {Communication cost (in GB)};
 \end{tikzpicture}
 \end{minipage}
\end{subfigure}
     \vspace{-10pt}
    \caption{ Communication cost vs. block size $b$ for the LCMT and PrPCMT. The two panels use $(R, q, l) = (0.4, 5, 4)$, $(0.5, 4, 4)$, and $(0.75, 4, 3)$ and a constant data symbol size c. Left Panel: $c = 1$KB; Right Panel: $c = 200$KB. }
    \label{fig:X_c_vs_block_size}
\end{figure*}
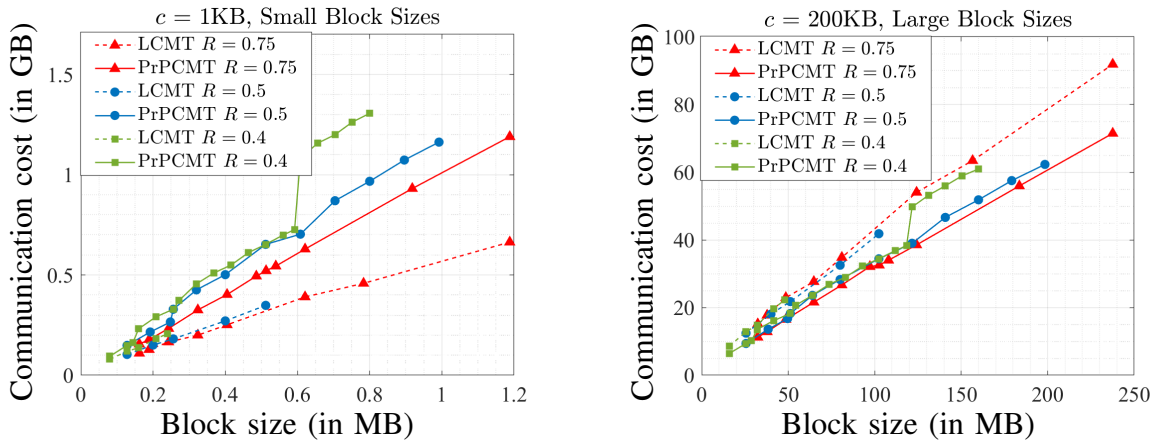

 \vspace{-0.15cm} 
In this section, we demonstrate the benefits of a PCMT and a PrPCMT when the size of the
transaction block $b$ is large. We demonstrate the improvements with respect to the performance metrics mentioned in Section \ref{sec:performance_metrics}. We also compare the performance with an LCMT (a CMT built using the parity check matrices of LDPC codes) \cite{TCOMLDPC} and 2D-RS codes \cite{dataAvailOrg}. 
Although 2D-RS codes offer a high undecodable threshold and hence a very good performance on the system specific metrics, they have a very high IC proof size and decoding complexity. %
Thus, we first compare the performance of a PCMT with an LCMT in Figs. \ref{fig:metrics_as_c_varies}, \ref{fig:Ic_proof_size_vs_block_size}, \ref{fig:P_f_vs_block_size}, and \ref{fig:X_c_vs_block_size}. Finally in Table \ref{figtable:numerical_comp}, we compare the performance to 2D-RS codes.
For CMT parameter $K$, we use the block size $b = cK$, where $c$ is the data symbol size of the base layer. We denote the output size of the {\fontfamily{qcr}\selectfont \text{Hash}} function as $y$ and use $y = 256$ bits in our simulations. All the PCMTs and PrPCMTs are built using polar codes designed by the SEF algorithm described in Section \ref{sec:SEF}. All LCMTs are built (as described in Section \ref{sec:general-CMT-framework}) using LDPC codes constructed using the PEG algorithm
\cite{PEG} where we set the degree of all VNs to 3. For PEG LDPC codes, the undecodable threshold $\alpha_{\min, j}$ for each LCMT layer  is calculated by
solving an Integer Linear Programs (ILP) as described in \cite{ILPsearch}
and is computationally infeasible for larger code lengths. Due to complexity issues of calculating the undecodable threshold (and hence the system specific metrics) for an LCMT, we compute the system specific metrics for an LCMT in Figs. \ref{fig:metrics_as_c_varies}, \ref{fig:P_f_vs_block_size}, and \ref{fig:X_c_vs_block_size} only for feasible code lengths and, thus, for feasible block sizes.  To calculate the IC proof size of an LCMT, we use the maximum CN degree $d_c$ (which is found from the parity check matrix of the LDPC code used in the base layer of the LCMT). For a PCMT and PrPCMT, the maximum CN degree $d_p = 3$.
The probability of light node failure $P_f(s)$ for different coding methods is calculated based on the equation provided in Section \ref{sec:p_f_calulation}. 
In the probability of failure calculation, the sample size $s$ for an LCMT, a PCMT, and  a PrPCMT are selected such that the total sample download size is $b/3$ in all cases. The total sample download size is equal to $Xs$, where $s$ is the total number of samples and $X$ is the single sample download size that depends on the channel code used. \bluetext{The equation to calculate the single sample download size $X$ is provided in Table \ref{figtable:Performance_comp}.} 
The communication cost associated with the DA oracle in side blockchains is calculated using the equation provided in Section \ref{sec:comm_cost_calculation} where we again calculate $X$ based on Table \ref{figtable:Performance_comp}. Note that we  use the equation provided in Section \ref{sec:comm_cost_calculation} for the communication cost for the PCMT/PrPCMT since the PCMT/PrPCMT satisfies the repetition property as shown in Lemma \ref{lemma:PCMT-repetition}.
For the DA oracles, we use the parameters $\beta = 0.49$, $\tgammaoracle = 1 - 2\beta$, $p_{th} = 10^{-8}$ and the number of oracle nodes $\tNoracle = 400$.
Table \ref{figtable:Performance_comp} provides a comparison of the various performance metrics for 2D-RS codes, an LCMT and a PCMT. Derivation of the formulae in Table \ref{figtable:Performance_comp} is provided in the appendix. We use the equations for different quantities provided in Table \ref{figtable:Performance_comp} to generate Figs. \ref{fig:metrics_as_c_varies}, \ref{fig:Ic_proof_size_vs_block_size}, \ref{fig:P_f_vs_block_size}, and \ref{fig:X_c_vs_block_size}.  

In Fig. \ref{fig:metrics_as_c_varies} top and bottom left panels, we compare the performance of different CMT metrics as the size of the data symbol $c$ varies. We compare results for different values of $K$ where the block size  $b = cK$. In Fig. \ref{fig:metrics_as_c_varies} top left panel, we compare the IC proof normalized by the block size for different coding methods. We see that for different values of $c$, the LCMT has a larger IC proof size compared to the PCMT  and PrPCMT. The low value of the IC proof size for the PCMT and PrPCMT is due to a low CN degree of 3 in the polar FGs. In the figure, we also see that the IC proof size for the PrPCMT is lower than the PCMT. Looking at the expression for the IC proof size for the  PCMT and PrPCMT in Table \ref{figtable:Performance_comp}, we can see that the lower value of the IC proof size for the PrPCMT is due to a lower value of the total number of VNs $\ttotVN_j$, which is a result of FG pruning. In Fig. \ref{fig:metrics_as_c_varies} top right and bottom left panels, we plot the performance of the system specific metrics as the value of $c$ varies. Note that in these plots, we do not have  curves corresponding to $k = 800$ and $k = 1024$ for the LCMT due to an infeasible complexity of calculating the undecodable thresholds $\alpha_{\min, j}$ at $k = 800$ and $k = 1024$. For $k = 400$,  we see from Fig.  \ref{fig:metrics_as_c_varies} top right and bottom left panels that the PCMT has a higher probability of failure and communication cost compared to the LCMT at small data symbol sizes $c$ and gets smaller than the LCMT as $c$ increases. The reason for a smaller value of the system specific metrics for the PCMT as $c$ gets larger is due to a small penalty in the single sample download size $X$ for the PCMT compared to the LCMT at large $c$. In the figures, we also see that the PrPCMT always has a lower probability of failure and communication cost compared to the PCMT. Note that the PrPCMT and  PCMT have the same undecodable threshold and the lower value of the system specific metrics in the PrPCMT is due to a smaller single sample download size as a result of factor graph pruning.

In Fig. \ref{fig:metrics_as_c_varies} bottom right panel, we compare the CMT root size normalized by the block size for different coding methods as the block size varies. In the figure, we fix the data symbol size $c=10$KB and vary the parameter $K$ such that the block size is  $b = cK$. For each case, we calculate the CMT root size using the equations provided in Table \ref{figtable:Performance_comp}.  From the figure, we see that the PCMT has a significantly larger CMT root size compared to the LCMT. The root size gets reduced in the PrPCMT as can be seen by comparing the green and the blue curves. The reduction is due to the reduction in the number of VNs in the FG of polar codes due to pruning. We see that the CMT root size for the PrPCMT is slightly more than that of the LCMT. However, since the size of the CMT root is very small compared to the actual block size (root size/block size is in the order of $10^{-3}$- $10^{-2}$), a slight increase in root size is out weighted by the significant improvements in the IC proof size and system specific metrics. \debb{Additionally, note that since the CMT root size and decoding complexity are both proportional to $\ttotVN$, an improvement in the CMT root size also translates to a similar improvement in the decoding complexity.} 

In Fig. \ref{fig:Ic_proof_size_vs_block_size}, we plot the normalized IC proof size vs. block size $b$
 for an LCMT, a PCMT, and a PrPCMT and different data symbol sizes $c$ and rate $R$. 
 Similar to Fig. \ref{fig:metrics_as_c_varies} bottom right panel, we vary the values of $K$ and set $b = cK$. In Fig. \ref{fig:Ic_proof_size_vs_block_size} left panel, we see that for $c = 200$ and $50$KB which correspond to large block sizes, the IC
proof size for the PCMT and PrPCMT is smaller compared to the LCMT. At $c = 1$KB which corresponds to small block sizes, the IC proof size for the PCMT is larger than that of the LCMT. However, the IC proof size of PrPCMT is still smaller than that of the LCMT. For a larger rate of $R = 0.75$, we can see from Fig. \ref{fig:metrics_as_c_varies} right panel that for all values of $c$, the IC proof size for the PCMT and PrPCMT is always smaller than that of the LCMT. Note that the maximum CN degree is always $3$ in the polar FG irrespective of the rate $R$. However, for PEG LDPC codes (used in the LCMT), the maximum CN degree increases with an increase in the rate which results in a larger IC proof size compared to the PCMT even for small values of $c$ as seen in Fig. \ref{fig:Ic_proof_size_vs_block_size} right panel.

\vspace{-0.05cm}
In Fig. \ref{fig:P_f_vs_block_size}, we compare $P_f(s)$ for the PrPCMT and LCMT and different rates $R$. We compare $P_f(s)$ for small and large block sizes in the left and right panels, respectively. From Fig. \ref{fig:P_f_vs_block_size} left panel, we see
that the PrPCMT has a worse probability of failure compared to the
LCMT for small block sizes. However in Fig. \ref{fig:P_f_vs_block_size} right panel, we see that for large block sizes, the
PrPCMT has a significantly lower probability of failure compared to the LCMT across all rates $R$ and block sizes $b$. The reason for a lower $P_f(s)$ at large block sizes for the PrPCMT is due to a higher $\talpha$ for SEF polar codes and negligible penalty in the single sample download size. 

\vspace{-0.05cm}
In Fig. \ref{fig:X_c_vs_block_size}, we compare the communication cost associated with the DA oracle for the PrPCMT and LCMT \footnote{We remark that the plots in Figs. \ref{fig:P_f_vs_block_size} and \ref{fig:X_c_vs_block_size} corresponding to PrPCMT are not smooth due to the sudden increase in the undecodable threshold as the code length increases.}. Similar to $P_f(s)$, we compare the communication cost for small and large block sizes (in left and  right panels, respectively). From Fig. \ref{fig:X_c_vs_block_size}, we see that for small block sizes, the PrPCMT has a larger communication cost compared to the LCMT. However, for large block sizes, the PrPCMT has a lower communication cost compared to the LCMT for all rates $R$ and block sizes $b$ (due to the same reason as Fig. \ref{fig:P_f_vs_block_size} right panel).

 \setlength{\extrarowheight}{2.8pt}
\setlength{\tabcolsep}{2.8pt}
  \begin{table}[t]
      \caption{Comparison of various performance metrics for 2D-RS, an LCMT, and a PrPCMT. The table uses $\mathcal{T}_1 = (K,R,q,l) = (512,0.5,4,6)$, $\mathcal{T}_2 = (2048,0.5,4,8)$, $\mathcal{T}_3 = (4096, 0.5,4,9)$, $c = 20$KB, and $b = cK$. 
    The total sample download size is calculated such that $P_f(s)$ is $0.01$. Due to the high threshold complexity for the LCMT, we do not have a corresponding total sample download size and communication cost value for $\mathcal{T}_2$ and $\mathcal{T}_3$.%
    }\label{figtable:numerical_comp}
    \vspace{-8pt}
    \centering
\resizebox{!}{2.9cm}{
\begin{tabular}[b]{| c |c  c c |c c c |c c c|}

\hline
& \multicolumn{3}{c|}{2D-RS} & \multicolumn{3}{c|}{LCMT}  & \multicolumn{3}{c|}{PrPCMT}  \\

& $\mathcal{T}_1$ & $\mathcal{T}_2$ & $\mathcal{T}_3$ &
 $\mathcal{T}_1$ & $\mathcal{T}_2$ & $\mathcal{T}_3$ &$\mathcal{T}_1$ & $\mathcal{T}_2$ & $\mathcal{T}_3$\\
\hline
Root size (KB) & 2.05 &4.10&	5.82&	1.02&	1.02&	1.02&	2.34&	2.34&	2.34
\\
\hline
IC proof size (KB) & 456.2&	913.9&	1294.3&	123.7&	124.7&125.4&	46.1&	49.8&	52.2
\\
\hline
Total sample download size (MB) &0.85	&0.95&	0.97&	3.27& -& -&			3.06&	6.89&	8.31
\\
\hline
Communication Cost (GB) &2.52 &	8.95&	17.28&	4.40&-&-&			4.16&	17.97&	40.87
\\
\hline
\end{tabular}
}  
  \end{table}
 
A comparison across various performance metrics for 2D-RS codes, an LCMT, and a PrPCMT is provided in Table \ref{figtable:numerical_comp} (\greentext{see Table \ref{figtable:Performance_comp} for threshold and decoding complexity}). 
 We first note that the PrPCMT outperforms the LCMT with respect to the IC-proof size, total sample download size, and communication cost with a small increase in root size and decoding complexity. Additionally, the PrPCMT has a lower threshold complexity as opposed to the LCMT where the threshold complexity of the LCMT is NP-hard. 
 \greentext{On the other hand,} the PrPCMT outperforms 2D-RS codes significantly in terms of the root size, IC proof size, and decoding complexity while having a higher total sample download size and communication cost.
Overall, the PrPCMT simultaneously performs well across all the different performance metrics relevant to this application.

 \vspace{-0.65cm}
 \section{Conclusion}\label{sec:conclusion}
 \vspace{-0.25cm}
 In this paper, we considered the problem of designing polar codes to mitigate DA attacks in
 two major blockchain systems. We first provided a novel construction of a Merkle tree using polar codes called a PCMT that can be used to mitigate DA attacks. Then, we provided a specialized polar code design algorithm for the PCMT called the SEF algorithm and a graph pruning algorithm to reduce the size of the polar FGs. We demonstrated that a PCMT built using SEF polar codes with pruned FGs performs well in mitigating DA attacks and outperforms an LCMT and 2D-RS codes that were previously used in literature across a variety of metrics.

\vspace{-0.5cm}
\appendix
\vspace{-0.25cm}
\subsubsection{Proof of Lemma \ref{lemma:comm_cost_mu_min}} 
Since the dispersal protocol is $(l,\tmu_{\min})$\emph{-correct}, every $\tgammaoracle$ fraction of oracle nodes collectively receive at least $ N_l - \tmu_{\min} + 1$ distinct coded symbols  or at least $\frac{N_l - \tmu_{\min} + 1}{N_l}$ fraction of distinct coded symbols from the base layer of the CMT. Since the CMT satisfies the repetition property, it implies that every $\tgammaoracle$ fraction of oracle nodes receives at least $\frac{N_l - \tmu_{\min} + 1}{N_l}$ fraction of distinct coded symbols from layer $L_j$ of the CMT for $1 \leq j \leq l$. Thus, every $\tgammaoracle$ fraction of oracle nodes receives at least $\lceil \left(\frac{N_l - \tmu_{\min} + 1}{N_l}\right)N_j\rceil$ distinct coded symbols from $L_j$. Now, $\lceil \left(\frac{N_l - \tmu_{\min} + 1}{N_l}\right)N_j\rceil 
= N_j - \lceil\left(\frac{\tmu_{\min}-1}{N_l}\right)N_j \rceil 
\geq N_j - \lceil\frac{N_j}{N_l}\left(\lfloor\left(  \frac{\alpha_{\min, j}-1}{N_j}\right) N_l \rfloor\right)\rceil 
\geq N_j - \lceil\frac{N_j}{N_l}\left(\left(  \frac{\alpha_{\min, j}-1}{N_j} \right) N_l\right)\rceil 
= N_j -\alpha_{\min, j}+ 1$. Thus, every $\tgammaoracle$ fraction of oracle nodes receives at least $N_j -\alpha_{\min, j}+1$ coded symbols implying that the dispersal protocol is $(j,\alpha_{\min, j})$-\emph{correct}.

\subsubsection{Proof of Lemma \ref{lemma:not_correct}}

We prove the lemma using \cite[Lemma 4]{DE-PEG}. 
Let $\chi(n,l,s,T,m)= \sum_{j=0}^{n}(-1)^{n-j}{l \choose j}{l - j - 1 \choose l-n-1}\left[\frac{{s-l+j \choose m}}{{s \choose m}}\right]^T$. Note that $\mathrm{Prob}(\vert \cup_{i \in S}A_i \vert \leq N_l - \tmu_{\min}) = \chi(N_l - \tmu_{\min}, N_l,N_l,\gamma \tNoracle,\tkoracle)$ due to \cite[Lemma 4]{DE-PEG}. Additionally, ${\tNoracle \choose \tgammaoracle \tNoracle} \leq e^{\tNoracle H_e(\tgammaoracle)}$. We have

\vspace{-0.65cm}
\begingroup
\allowdisplaybreaks
\begin{align*}
    &\mathrm{Prob}(\mathcal{C} \text{ is not } (l,\tmu_{\min})\text{-correct}) = \mathrm{Prob}(\exists S \text{ such that } \vert S \vert = \gamma \tNoracle, \vert \cup_{i \in S}A_i \vert \leq N_l - \tmu_{\min})\\& \leq  \sum_{S \subseteq [N_l] : \vert S \vert = \tgammaoracle \tNoracle}\mathrm{Prob}(\vert \cup_{i \in S}A_i \vert \leq N_l - \tmu_{\min})
   = \sum_{S \subseteq [N_l] : \vert S \vert = \tgammaoracle \tNoracle}\chi(N_l - \tmu_{\min}, N_l,N_l,\gamma \tNoracle,\tkoracle) \\
   &= {\tNoracle \choose \tgammaoracle \tNoracle}\chi(N_l - \tmu_{\min}, N_l,N_l,\tgammaoracle \tNoracle,\tkoracle) 
   \leq e^{\tNoracle H_e(\tgammaoracle)}\chi(N_l - \tmu_{\min}, N_l,N_l,\tgammaoracle \tNoracle,\tkoracle)\\
    &= e^{\tNoracle  H_e(\tgammaoracle)}\left(\sum_{j=0}^{N_l-\tmu_{\min}} (-1)^{N_l-\tmu_{\min} -j} {N_l \choose j} {N_l - j - 1 \choose \tmu_{\min} -1}\left[\frac{{j \choose \tkoracle}}{{N_l \choose \tkoracle}}\right]^{\tgammaoracle \tNoracle}\right).
\end{align*}  
\endgroup

\subsubsection{Proof of Lemma \ref{lemma:PDE_sucessfull}}
We prove Lemma \ref{lemma:PDE_sucessfull} by proving the following property of stopping sets in the FG of SEF polar codes. \browntext{To the best of our knowledge}, we have not seen the following result before in the literature and, hence, it may be of independent interest. 
Let $n = \lceil \log N \rceil$.

\vspace{-0.35cm}
\begin{lemma}\label{lemma:full_row}
 Consider a polar FG $\tFG_{N}$ produced by the SEF algorithm. 
 Every stopping set of $\tFG_{N}$ must contain the VNs of at least one full row from the FG i.e., every stopping set contains all VNs in the set $\{v_{\tVNi} \;\vert \; \tVNi = (m-1)N + i,\; \closedN{m}{\;n + 1\;}  \}$ for some $ \closedN{i}{N}$. 
\end{lemma}
\vspace{-0.55cm}
\begin{proof}
    Let $\tssingle$ be a stopping set of $\tFG_{N}$. Let $\tFG^{\tssingle}_N$ be the induced subgraph of $\tFG_{N}$ corresponding to the set of VNs in $\tssingle$.  Observe that the FG $\tFG_N$ has two types of edges (see Fig. \ref{fig:FG_examples}): horizontal edges and slanted edges (which connect a degree 3 VN to a degree 3 CN). 
    We consider two cases: i) $\tFG^{\tssingle}_N$ does not have any slanted edges; ii) $\tFG^{\tssingle}_N$  has at least one slanted edge.

    For case i), it can be easily seen that the stopping set $\tssingle$ must include a full row of VNs. For case ii), since $\tFG^{\tssingle}_N$  has at least one slanted edge, it implies that $\tssingle$ has at least one VN of degree 3. Thus, define the set 
$\Delta_{\tssingle} = \{(i,m) \;\vert\; \closedN{i}{N},\; \closedN{m}{n}, \tVNi = (m-1)N + i, v_{\tVNi} \in \tssingle, \text{degree of } v_{\tVNi} = 3 \}$. Also define  $i_{\max} = \max(\{i \vert (i,m) \in \Delta_{\tssingle} \text{ for some }\closedN{m}{n}\})$. $\Delta_{\tssingle}$ contains the indices of all the degree 3 VNs of $\tssingle$ and $i_{\max}$ denotes the largest row index such that $\tssingle$ has a degree 3 VN from that row. Due to the definition of case ii), $\Delta_{\tssingle}$ is nonempty. We now show that $\tssingle$ has all the VNs in the row $i_{\max}$ of FG $\tFG_{N}$, i.e., $\tssingle$  contains all the VNs in $\{v_{\tVNi}\;\vert\;\tVNi = (m-1)N + i_{\max}; \closedN{m}{n+1}\}$. Let $\overline{m}$, $\closedN{\overline{m}}{n}$, be such that $(i_{\max}, \overline{m}) \in \Delta_{\tssingle}$. By the definition of a stopping set, the CNs to the right and left of $v_{(\overline{m}-1)N + i_{\max}}$ must belong to the induced subgraph graph of the stopping set. In other words, $c_{(\overline{m}-2)N + i_{\max}} \in \tFG^{\tssingle}_N $ and  $c_{(\overline{m}-1)N + i_{\max}} \in \tFG^{\tssingle}_N$ (unless $v_{(\overline{m}-1)N + i_{\max}}$ is the rightmost or the leftmost VN in which case we will have only one CN neighbor).   Now, to satisfy the stopping set property, for both these CNs, their corresponding VNs to their left and right in the same row  $i_{\max}$ must belong to the stopping set $\tssingle$.
If not, then to satisfy the stopping set property, the CN must be connected to a VN $v_{(m-1)N + i}\in \tssingle$  by a slanted edge. Note that a slanted edge connects a CN to a degree 3 VN in a lower row. In other words,  a slanted edge connects a CN from row $i_{\max}$ to a degree 3 VN in a row with index greater than $i_{\max}$. This condition violates the definition of $i_{\max}$. Thus, $v_{(\overline{m}-2)N + i_{\max}} \in \tssingle$ and $v_{(\overline{m})i_{\max}} \in \tssingle$. 
Now, considering $v_{(\overline{m}-2)N + i_{\max}} \in \tssingle$ and $v_{(\overline{m})i_{\max}} \in \tssingle$ as the starting VN (similar to $v_{ (\overline{m}-1)+i_{\max}}$), we can apply the above logic to show that $v_{(\overline{m}-3)N+i_{\max}} \in \tssingle$ and $v_{(\overline{m}+1)N + i_{\max}}\in \tssingle$. Repeatedly applying the same argument, we can show that all the VNs in $\{v_{\tVNi}\;\vert\;\tVNi = (m-1)N+i_{\max},\;\closedN{m}{n+1}\}$ belong to $\tssingle$. 
\end{proof}

\vspace{-0.4cm}
We now use the above result to prove Lemma \ref{lemma:PDE_sucessfull}. Since $\tI \cup \tF$, the information and frozen indices form a partition of all row indices. Now, due to the above lemma, every stopping set either contains a VN from the leftmost column of the FG belonging to the frozen indices or a VN from the rightmost column of the FG belonging to an information index. Thus for every stopping set, at least one VN of the stopping set is not erased at the start of the peeling decoding in the PEPC. Hence, the PEPC will always be successful and will result in a valid codeword.

\subsubsection{Proof of Lemma \ref{lemma:last_few_frozen}} 
Firstly, it is easy to see that when all the VNs in $\tlastVN^{\tlast}_{\tNm}[1]$ (i.e., the VNs 
in the last $\tlast$ rows from the leftmost column of FG $\tFG_{\tNm}$) are set to zero symbols, all the VNs in the last $\tlast$ rows of all the columns of the FG will be zero symbols. 
\browntext{This result proves claim ii) of the lemma.} 
\browntext{For claim i)}, let $\tssingle \in \tSS^{\tImm}$ and let $\tFG^{\tssingle}_{\tNmm}$ be the induced subgraph of $\tFG_{\tNmm}$ corresponding to the set of VNs in $\tssingle$. From the definition of $\tSS^{\tImm}$,  $\tssingle$ does not have any frozen VNs from the leftmost column of the FG $\tFG_{\tNmm}$. Now, since  $[\tNmm - \tlast +1, \tNmm] \subset \tFmm$, $\tssingle$ does not have any VNs in $\tlastVN^{\tlast}_{\tNmm}[1]$. We now prove the first claim of the lemma by contradiction. Assume that $\tssingle$ has a VN from $\tlastVN^{\tlast}_{\tNmm}[\log \tNmm + 1]$. In particular, assume that $v_{nN + i_1} \in \tssingle$, where $n = \log \tNmm$ and $\closed{i_1}{\tNmm -\tlast+1}{\tNmm}$. Now, by the property of stopping sets, $c_{(n-1)N + i_1} \in \tFG^{\tssingle}_{\tNmm}$. To satisfy the stopping set property, either $v_{(n-1)N + i_1} \in  \tssingle$ or $v_{(n-1)N+i_2} \in \tssingle$ where
$i_1 < i_2 \leq \tNmm$ and $v_{(n-1)N + i_2}$ and $c_{(n-1)N + i_1}$ are connected in $\tFG_{\tNmm}$. Thus, we have at least one index ${i}$, $\closed{i}{\tNmm - \tlast +1 }{\tNmm}$  such that $v_{(n-1)N + i} \in \tssingle$. Proceeding in a similar manner as above, we have at least one index ${i}$, $\closed{i}{\tNmm - \tlast +1 }{\tNmm}$  such that $v_{(n-2)N + i} \in \tssingle$. Repeating the same process until we reach the leftmost column, we can find at least one index ${i}$, $\closed{i}{\tNmm - \tlast +1 }{\tNmm}$  such that $v_{i} \in \tssingle$ which is a contradiction of the fact that $\tssingle$ does not have any VNs in set $\tlastVN^{\tlast}_{\tNmm}[1] = \{v_{i} \;|\; \closed{i}{\tNmm - \tlast +1 }{\tNmm}\}$.

\subsubsection{Proof of Lemma \ref{lemma:undecodable_sef}}
 The SEF algorithm produces an $(N_{\tname}, k)$ polar code with a FG $\tFG_{N_{\tname}}$. Let $\tNm = 2^{\lceil \log \frac{k}{R} \rceil}$. FG $\tFG_{N_{\tname}}$ is obtained from freezing (and hence removing) the last $\tlast_1 + \tlast_2$ rows of $\tFG_{\tNm}$.
For the output $\tF_{\tname}$, let  $\tI_{\tname} = [N_{\tname}]\setminus\tF_{\tname}$. Also define $\tFm = \tF_{\tname} \cup [N_{\tname} + 1, \tNm]$, $\tIm = [\tNm]\setminus \tFm$. Clearly, the sets $\tIm$ and $\tI_{\tname}$ are the same. Thus, the $(N_{\tname}, k)$ polar code can be seen as a code defined on the FG $\tFG_{\tNm}$ with frozen index set $\tFm$ and information index set $\tI_{N_{\tname}}$.
We now apply Lemma \ref{lemma:paper-ss-property} on FG $\tFG_{\tNm}$. The smallest leaf set size of all stopping sets in $\tSS^{\tI_{\tname}}$ is given by $\min_{\tssingle \in \tSS^{\tI_{\tname}}}\vert\text{\tLS}(\tssingle)\vert 
     = \min_{i \in \tI_{\tname}}\ttT_{\tNm}(i)
     = \min_{i \in \tI_{\tname}}\ttT_{\frac{k}{R}}(i)$.

     The threshold complexity is the complexity of the SEF algorithm which is at most linear in the input $k_j = \frac{K}{(qR)^{l-j}} $. Overall, the threshold complexity of the entire  PCMT is $\sum_{j=1}^{l}O(\frac{K}{(qR)^{l-j}})$.

\subsubsection{Proof of Lemma \ref{lemma:scaling-law}}

Since $b \gg y K$, we can ignore the size of the Merkle proofs in the total sample download size.
Thus, we can write the total sample download size as $\frac{bs}{K}$. \browntext{As such, $\frac{bs}{K} \leq \frac{b}{D_r}$ or $s = \lfloor\frac{K}{D_r}\rfloor$}. For uncoded Merkle tree, $P^u(s) = \left(1 - \frac{1}{K} \right)^s = \left(1 - \frac{1}{K} \right)^{\lfloor\frac{K}{D_r}\rfloor}$. 
Now, let $N = \frac{K}{R}$, $n = \lceil \log N\rceil$, and $\tNm = 2^n$.  Based on Lemma \ref{lemma:undecodable_sef}, for the base layer we have $\talpha = \min_{i \in \tI_{\tname}}\ttT_{\tNm}(i)$ where $\tI_{{\tname}} = [N_{\tname}] \setminus \tF_{\tname}$, and $\tF_{N_{\tname}}$ is the output of the SEF algorithm with inputs $(\frac{K}{R}, K)$. 
Now, due to step 5 of the SEF algorithm, $\talpha = \min(t_N ; N - K + 1)$ where $t_N$ is obtained from $\ttT_{\tNm}$ by removing the last $\tNm - N$ entries from the bottom. Additionally note that, $\min(t_N ; N - K + 1) \geq \min(\ttT_{\tNm} ; N - K + 1)$. Thus, $\talpha \geq \min(\ttT_{\tNm} ; N - K + 1)$.
The vector $\ttT_{\tNm}$ has the following property \cite{Polar-SStree-TCOM}: $\ttT_{\tNm}$ has exactly $n \choose q$ entries with value $2^q$ for $\closed{q}{0}{n}$. Using this property, we have a simple algorithm to lower bound $\talpha$.  Let $q^*$ be the largest $\closed{q}{0}{n}$ such that $\sum_{r = 0}^{q-1}{n \choose r} \leq N-K$. Then, $\talpha \geq \min(\ttT_{\tNm} ; N - K + 1) = 2^{q^*}$.

Now, for $0< q - 1 \leq \frac{n}{2}$,  $\sum_{r = 0}^{q-1}{n \choose r}  \leq 2^{nH_2(\frac{q-1}{n})}$ \deb{(bound on the volume of a hamming ball)}. Let $q_1$ be largest $\closed{q}{0}{1+\frac{n}{2}}$ such that $2^{nH_2(\frac{q-1}{n})} \leq N-K$. Then from the definitions of $q^*$ and $q_1$, $q^* > q_1$.  Now $2^{nH_2(\frac{q-1}{n})} \leq N-K \implies
q \leq 1 + nH_2^{-1}(\frac{\log(N - K)}{n})$. Note that $\frac{\log (N - k)}{n} \leq 1$. Thus, $1 + nH_2^{-1}(\frac{\log(N - K)}{n}) \leq 1 + \frac{n}{2}$. Hence, $q_1 = 1 + nH_2^{-1}(\frac{\log(N - K)}{n}) )$ and $\talpha = 2^{q^*} \geq 2^{1 + nH_2^{-1}(\frac{\log(N - K)}{n})} \geq 2^{1 + (\log N)\cdot H_2^{-1}(\frac{\log(N - K)}{n})}.$ As such, $P^{p}_f(s) = \left ( 1 - \frac{\talpha}{N_{\tname}}\right)^{\lfloor\frac{K}{Dr}\rfloor} \leq \left ( 1 - \frac{2^{1 + (\log N)\cdot H_2^{-1}(\frac{\log(N - K)}{n})}}{N}\right)^{\lfloor\frac{K}{D_r}\rfloor}$. 
To compute the asympotic growth rate of $\ln \frac{P^{u}_f(s)}{P^{p}_f(s)}$, we compute the following limit: $\Delta = \underset{K \rightarrow \infty}{\lim}\frac{1}{\sqrt{K}}\ln\frac{P^{u}_f(s)}{ \left ( 1 - \frac{2^{1 + (\log N)\cdot H_2^{-1}(\frac{\log(N - K)}{n})}}{N}\right)^{\lfloor\frac{K}{D_r}\rfloor}}$. 
As $K \rightarrow \infty$, $2^{1 + (\log N)\cdot H_2^{-1}(\frac{\log(N - K)}{n})} \rightarrow 2^{1+ \frac{(\log N)}{2}}$. Hence, $\Delta = \underset{K \rightarrow \infty}{\lim}\frac{1}{\sqrt{K}}\ln\frac{(1 - \frac{1}{K})^{\lfloor\frac{K}{D_r}\rfloor}}{ \left ( 1 - 2\cdot2^{-\frac{(\log N)}{2}}\right)^{\lfloor\frac{K}{D_r}\rfloor}} = \underset{K \rightarrow \infty}{\lim}\frac{1}{\sqrt{K}} \ln\frac{(1 - \frac{1}{K})^{\lfloor\frac{K}{D_r}\rfloor}}{ \left ( 1 - 2\cdot\frac{\sqrt{R}}{\sqrt{K}}\right)^{\lfloor\frac{K}{D_r}\rfloor}} = \underset{K \rightarrow \infty}{\lim}\frac{1}{\sqrt{K}}  \ln \left ( \frac{e^{-\frac{1}{D_r}}}{e^{-\frac{2\cdot\sqrt{KR}}{D_r}}}\right) 
= \frac{2\sqrt{R}}{D_r} $, where we have utilized the fact that $\underset{x \rightarrow \infty}{\lim} \frac{1}{x}\lfloor x\rfloor = 1$. 
\deb{Thus, noting that $\Delta$ uses an upper bound on $P^{p}_f(s)$, we get $\ln \frac{P^{u}_f(s)}{P^{p}_f(s)} = \Omega(\sqrt{K})$.}

 \subsubsection{Derivation of formulae in Table \ref{figtable:Performance_comp}}
The metrics for 2D-RS codes are calculated as described in \cite{dataAvailOrg}. We now derive the  corresponding metrics for the general CMT framework provided in Section \ref{sec:general-CMT-framework}. In the general CMT framework, let the hashes of $q_j$ symbols of $L_{j+1}$ be concatenated into a data symbol of $L_j$, $1 \leq j < l$. For an LCMT, $q_j = q$, $1 \leq j < l$. For a PCMT and PrPCMT, $q_j = \lceil \frac{\ttotVN_{j+1}}{k_j} \rceil$, $1 \leq j < l$ (see description in Section \ref{sec:SEF_PCMT}).

\textbf{Root size}: The root consists of hashes of all the symbols of $L_1$. Thus the root size for the LCMT is $yN_1$, and the PCMT and PrPCMT is $y\ttotVN_1$.

\textbf{Single sample download size $X$:} Each sample consists of a base layer symbol and the Merkle proof for the base layer symbol. Additionally, the Merkle proof of a base layer symbol consists of a data and a parity symbol from each layer above the base layer. The Merkle proof (for both the LCMT and PCMT/PrPCMT) satisfies the property that the data symbol in the proof from layer $L_j$ consists of the hash of the data symbol in the proof from layer $L_{j+1}$, $\closedN{j}{l-1}$. Thus, of the $q_j$ hashes present in the data symbol of the Merkle proof from layer $L_j$, $\closedN{j}{l-1}$, the hash corresponding to the data symbol
of the Merkle proof from layer $L_{j + 1}$ is not communicated in the Merkle proof. Thus, there are only $(q_j-1)$ hashes from each layer $L_j$, $\closedN{j}{l-1}$ for the data part in the Merkle proofs.
Thus, the size of the Merkle proof of a base layer symbol is $y\sum_{j = 1}^{l-1}(2q_j-1)$ (since the size of each parity symbol is $yq_j$). Hence, the download size for a single sample is $\frac{b}{K} + y\sum_{j = 1}^{l-1}(2q_j-1)$, where $\frac{b}{K}$ is the size of the base layer symbol. Substituting the values of $q_j$ for the LCMT, PCMT, and PrPCMT, we get the equations in Table \ref{figtable:Performance_comp}.

\textbf{IC proof size}: The IC proof for a failed parity check equation with $d$ symbols consists of $d-1$ symbols and the Merkle proofs of the $d$ symbols. Note that the proof size is the largest when the failed parity check equation is in the base layer. Thus, we provide the IC proof size when the $d$ symbols are from the base layer.   Also, in IC proofs, the Merkle proof only consists of the data symbols from each layer above the base layer \cite{CMT,AceD}.
Thus, the IC proof size is $\frac{(d-1)b}{K}  + dy\sum_{j=1}^{l-1}(q_j-1)$.
Substituting the values of $q_j$ and $d$,  we get the equations in Table \ref{figtable:Performance_comp}. 

\textbf{Decoding complexity}: Since the LCMT, PCMT, and  PrPCMT are decoded using a peeling decoder, the decoding complexity is proportional to the total number of VNs in the FG. Thus the decoding complexity is $O(N_l)$ for an LCMT and $O(\ttotVN_l)$ for the PCMT and PrPCMT.

\end{document}